\documentclass[%
 reprint,
 amsmath,amssymb,
 aps,
]{revtex4-2}

\usepackage{graphicx}
\usepackage{dcolumn}
\usepackage{bm}

\usepackage[hidelinks,colorlinks=true,linkcolor=blue,filecolor=blue,urlcolor=blue,citecolor=blue]{hyperref}

\usepackage{xcolor}
\usepackage[caption=false]{subfig} 

\newcommand{\density}[1][]{\@ifempty{#1}{P}{P_{#1}}}
\newcommand{\densityRight}[1][]{\@ifempty{#1}{\density_{+}}{\density_{#1\,+}}}
\newcommand{\densityLeft}[1][]{\@ifempty{#1}{\density_{-}}{\density_{#1\,-}}}
\makeatother

\newcommand{\ABindex}[1]{[#1]}

\begin{document}

\preprint{APS/123-QED}

\title{Run-and-tumble motion in a linear ratchet potential: Analytic solution, power extraction, and first-passage properties}

\author{Connor Roberts}
\email{connor.roberts16@imperial.ac.uk}
\affiliation{%
Department of Mathematics, Imperial College London, 180 Queen's Gate, London, SW7 2AZ, United Kingdom
}%
\affiliation{%
Centre for Complexity Science, Imperial College London, SW7 2AZ, United Kingdom
}%

\author{Zigan Zhen}
\email{z.zhen19@imperial.ac.uk}
\affiliation{%
Department of Mathematics, Imperial College London, 180 Queen's Gate, London, SW7 2AZ, United Kingdom
}%
\affiliation{%
Centre for Complexity Science, Imperial College London, SW7 2AZ, United Kingdom
}%

\date{\today}



\begin{abstract}
We explore the properties of run-and-tumble particles moving in a piecewise-linear ``ratchet'' potential by deriving analytic results for the system's steady-state probability density, current, entropy production rate, extractable power, and thermodynamic efficiency. The ratchet's broken spatial symmetry rectifies the particles' self-propelled motion, resulting in a positive current that peaks at finite values of the diffusion strength, ratchet height, and particle self-propulsion speed. Similar nonmonotonic behaviour is also observed for the extractable power and efficiency. We find the optimal apex position for generating maximum current varies with diffusion, and that entropy production can have nonmonotonic dependence on diffusion. In particular, for vanishing diffusion, entropy production remains finite when particle self-propulsion is weaker than the ratchet force. Furthermore, power extraction with near-perfect efficiency is achievable in certain parameter regimes due to the simplifications afforded by modelling ``dry" active particles. In the final part, we derive mean first-passage times and splitting probabilities for different boundary and initial conditions. This work connects the study of work extraction from active matter with exactly solvable active particle models and will therefore facilitate the design of active engines through these analytic results.
\end{abstract}

\maketitle



\section{Introduction}\label{sec:intro}

Active matter is composed of agents that consume energy from their environment to exert mechanical forces \cite{Marchetti2013Jul}. These systems break detailed balance and are thus impervious to the tools of equilibrium statistical mechanics \cite{tailleur2008statistical}. Because of the rich phenomenology they capture, even at the single-particle scale \cite{elgeti2015physics}, active matter models have become the most favoured route for exploring the broader field of nonequilibrium systems \cite{gompper20202020}.

A paradigmatic active matter model is the run-and-tumble (RnT) particle, which idealises the piecewise-ballistic locomotion of \textit{E.\ coli} bacteria \cite{Schnitzer1993Oct}. RnT particles are arguably the simplest of the canonical active particle models, which also includes active Ornstein-Uhlenbeck particles \cite{Fodor2016Jul,Martin2021Mar,Maggi2015May} and active Brownian particles \cite{Wagner2017Apr,Das2018Jan,caraglio2022analytic}. This has led to extensive study of their nonequilibrium steady-state distributions in confining potentials \cite{Tailleur2009Jun,zhen2022optimal,roberts2022exact,Angelani2017Jul,garcia2021run,razin2020entropy,dhar2019run,basu2020exact,frydel2022positing,Ezhilan2015Oct,Elgeti2015Mar}. Notably, most results for these models are obtained through simulations or under approximations \cite{Maggi2015May,Ezhilan2015Oct,Elgeti2015Mar,Fodor2016Jul,Martin2021Mar, fox1986uniform, farage2015effective, marconi2016velocity, jung1987dynamical, marconi2015towards}, including recent work concerning RnT particles subject to a differentiable ratchet potential \cite{derivaux2022rectification}. Only a handful of exact results for the steady-state distributions of these models are documented in the literature, and most of these are attributed to RnT particles in one dimension \cite{Angelani2017Jul, Malakar2018Apr,garcia2021run,razin2020entropy,dhar2019run,basu2020exact,frydel2022run,breoni2022one}.

When immersed in asymmetric environments, active particles display unidirectional motion \cite{Doering1992Oct,astumian1994fluctuation, Pavliotis2005Sep, Galajda2007Dec, Wan2008Jul,Angelani2009Jan, di2010bacterial, angelani2011active, Reichhardt2017Mar, zhen2022optimal, koumakis2014directed, muhsin2023inertial}. The emergence of such rectified motion has generated significant interest in active engines, whose constituent active agents perform work on an external load to store potential energy \cite{Julicher1997Oct,Parrondo1998Aug,pietzonka2019autonomous,Sekimoto1997May}. The net current observed in these systems is the hallmark of broken time-reversal symmetry, indicating a system is operating far from equilibrium. Exactly ``how far'' is quantified by the entropy production rate \cite{Fodor2016Jul,garcia2021run,cocconi2020entropy,Martin2021Mar}, which was recently calculated for RnT particles confined to an infinite-square well \cite{razin2020entropy}, but has yet to be calculated for asymmetric potentials.

The nonequilibrium behaviour of active matter results in first-passage properties that are significantly different to that of equilibrium systems. For instance, the mean escape time of a Brownian particle from a potential well depends only on the ratio of the relative height of the potential barrier to the particle's thermal energy \cite{Kramers1940Apr}. In contrast, the mean escape time of active particles can depend on the detailed shape of the potential \cite{Woillez2019Jun}. Studying the first-passage properties of active matter is particularly relevant to biology, which is often concerned with the time it takes motile agents to reach certain locations, such as in chemotaxis \cite{angelani2014first}. For the specific case of RnT particles, first-passage properties have been studied both with \cite{Malakar2018Apr} and without \cite{angelani2014first,le2019noncrossing, angelani2015run, Angelani2017Jul} the presence of additive diffusive noise. The noiseless case allows analytic results for run-and-tumble particles to be obtained in the case of partially absorbing boundaries \cite{angelani2014first, Angelani2017Jul}, encounter-based absorption \cite{bressloff2022encounter, bressloff2023encounter}, and in the presence of symmetric confining potentials \cite{dhar2019run}. However, even for the more general case of nonnegligible diffusive noise, it is possible to obtain analytic expressions for the first-passage time density on a semi-infinite line, as well as the splitting probabilities and mean first-passage times on a finite interval \cite{Malakar2018Apr}. It is therefore of interest to see how the combination of asymmetric potentials and diffusion, both of which will be considered here, generalise these known results.

In the present work, we obtain exact analytic results for RnT particles in the presence of an external piecewise-linear ``ratchet" potential on a one-dimensional ring. In addition, we consider the application of an external counterforce, assuming the role of an external load, to the resulting rectified motion. The counterforce allows for the study of work extraction and is therefore relevant to the operation of active engines. This model of RnT motion is implemented through the addition of active telegraphic noise to the motion of an overdamped Brownian particle in a ratchet. This can describe several physically equivalent nonequilibrium systems, discussed further in Sec.~\ref{sec:Model}.

In what follows, the derivation and discussion of new results begin in Sec.~\ref{sec:ParticleDensities} with a calculation of the state-dependent probability densities at stationarity, which are obtained by solving the governing coupled Fokker-Planck equations of the system and then confirmed through Monte Carlo simulations of the Langevin equation. From these densities, in Sec.~\ref{sec:Currents}, we obtain the corresponding currents and observe nonmonotonic dependencies on the system parameters, similarly to other nonequilibrium systems \cite{Semeraro2023Jan, Rizkallah2023May, Bijnens2021Mar, Ai2017Mar}. Then, following the formalism outlined in Ref.~\cite{seifert2005entropy}, we derive the steady-state entropy production rate in Sec.~\ref{sec:EntropyProduction}, which paves the way for the study of the power output and thermodynamic efficiency in Sec.~\ref{sec:OptimalWork}. Finally, in Sec.~\ref{sec:FirstPassage}, we study the transport properties of this system, namely,  the mean first-passage times for the separate cases of absorbing and reflecting boundary conditions and the splitting probabilities for the particle to exit from either end of the ratchet.

Given the large number of quantities of interest, the aim of this work is not to provide a full characterisation of the high-dimensional phase space in each case, but to give an overview of the most interesting and relevant features.



\section{Model}\label{sec:Model}

\begin{figure}
    \centering
    \includegraphics[width=0.32\textwidth, trim=4.7cm 1.9cm 5cm 0.7cm, clip]{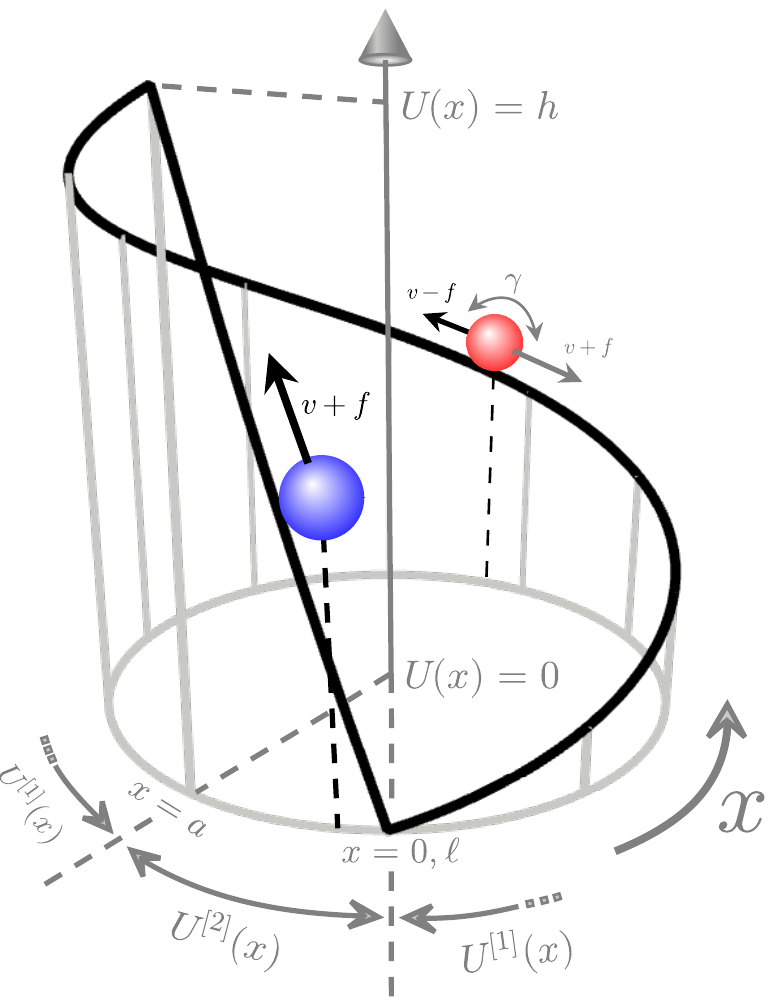}\caption{Schematic representation of RnT particles subject to a periodic ratchet potential $U(x) = U(x+\ell)$, Eq.~(\ref{eq:RatchetPotential}), with maximum amplitude $U(a)=h$. Due to the contribution imparted by the external force $f$, the overall drift of an RnT particle alternates, with Poissonian rate $\gamma$, between $v-f$, corresponding to the right-moving state $\sigma(t) = 1$ (red sphere, background), and $-(v+f)$, corresponding to the left-moving state $\sigma(t) = -1$ (blue sphere, foreground).}\label{fig:Schematic}
\end{figure}
\begin{figure}
    \centering
    \includegraphics[width=0.37\textwidth, trim=1cm 0.2cm 0cm 0cm, clip]{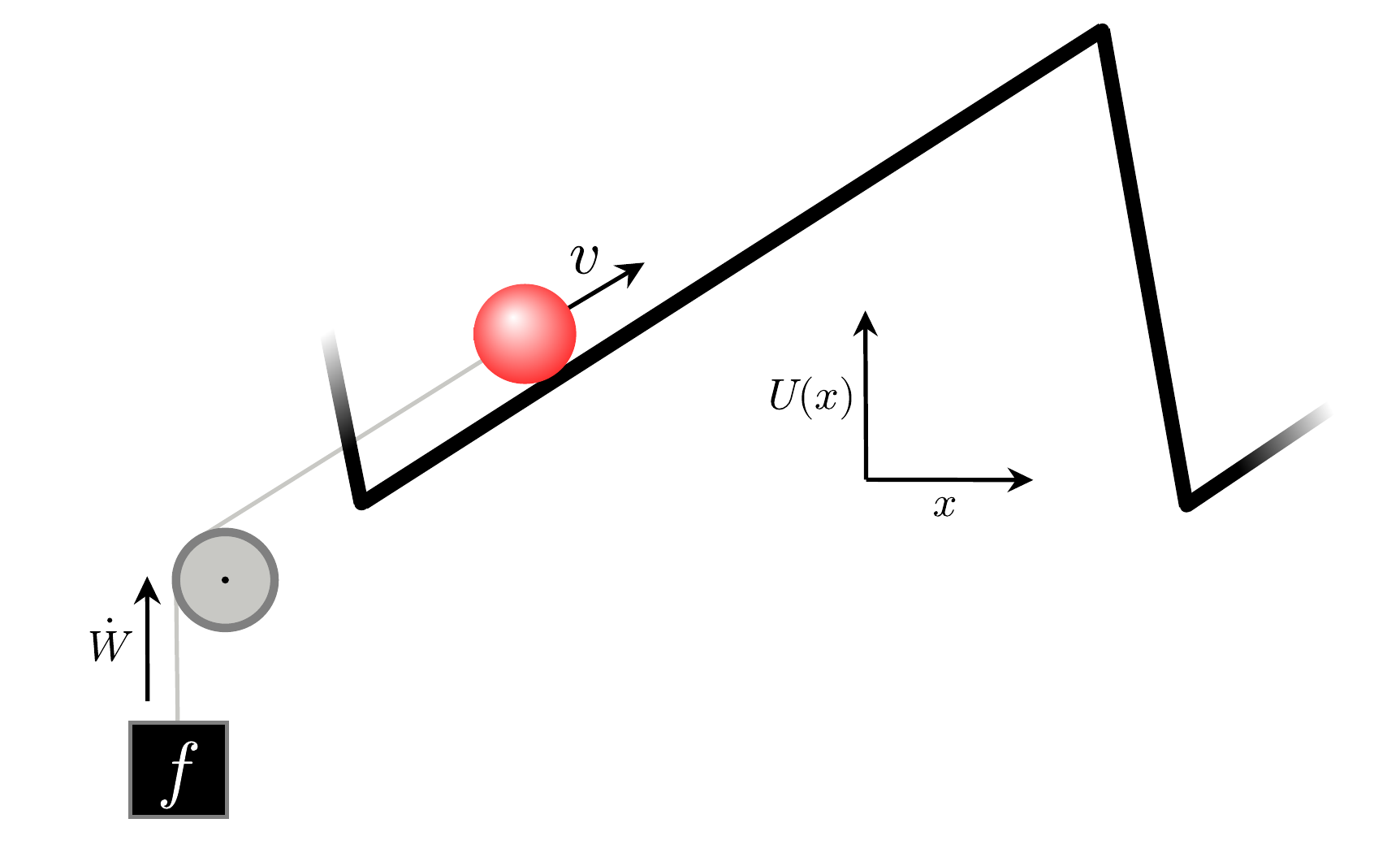}
    \caption{By applying an external counterforce of magnitude $f$, work $W$ can be extracted from the rectified motion of an RnT particle. Here, the counterforce is represented schematically as an external load being raised on a pulley. However, more practical implementations could include rotatable ratchet gears \cite{di2010bacterial} or passive movable obstacles \cite{pietzonka2019autonomous, derivaux2022rectification}.}\label{fig:engine}
\end{figure}

We consider an RnT particle, whose self-propulsion speed $w(t) \in \{v,-v\}$, where $v>0$, alternates or ``tumbles'' between two values with Poissonian rate $\gamma$, while moving at position $x \in \left[ 0, \ell \right)$ in a piecewise-linear ratchet potential of length $\ell$. The ratchet has maximum amplitude $h$ at the apex $x=a$, and periodic boundary conditions at $x=0$ and $x=\ell$, see Fig.~\ref{fig:Schematic}. The particle is therefore subject to the potential,
\begin{equation}\label{eq:RatchetPotential}
    U(x) = 
    \begin{cases}
		U^{[1]}(x) \equiv \frac{h}{a}x, & \quad 0 \leq x < a\\
        U^{[2]}(x) \equiv \frac{h}{\ell-a}(\ell-x), & \quad a \leq x < \ell
	\end{cases}.
\end{equation}
Henceforth, any reference to the position $x=0$ can be assumed to also refer to $x=\ell$, due to the periodicity, and we will thus refer only to the former. Without loss of generality, in Secs.~\ref{sec:ParticleDensities}-\ref{sec:OptimalWork}, we take $h \geq 0$ since replacing $h \rightarrow -h$ simply generates the mirror-image of the ratchet, i.e.\ $a \rightarrow \ell-a$, up to an (irrelevant) constant offset $-h$ after translating space by $x \rightarrow x - a$. For $a \neq \ell/2$, the ratchet is asymmetric and can therefore rectify the motion of the RnT particle. The rectified motion results in a current which can be exploited to perform useful work by applying a counterforce of magnitude $f$, such as an external load, see Fig.~\ref{fig:engine}. In the following, we take this counterforce to act to the left, i.e.\ in the negative $x-$direction, corresponding to a potential $fx$, so that work is done \textit{by} the particle when both $f$ and the current are positive, and there is thus still a net drift of particles to the right. We will consider the overdamped regime where forces, such as from the ratchet or the external counterforce, are equivalent to velocities (or external drifts) up to a factor of the particle mobility, which we set to unity.

The present system has several physically equivalent interpretations depending primarily on whether the breaking of detailed balance is attributed to external forcing or particle motility. For instance, $f$ has been introduced as a counterforce, but it could also be an asymmetry parameter of a biased RnT motion with self-propulsion $\bar{w}(t) \in \{v - f,-(v+f)\}$, Fig.~\ref{fig:Schematic}. Alternatively, $f$ could be considered a property of the ratchet by defining the potential $V(x) = U(x) + fx$, so that $-f$ is the tilt parameter of the tilted ratchet potential $V(x)$ \cite{Reimann2001Jun,Lindner2001Mar,berx2023reentrant}. Similarly, the tumble rate $\gamma$ could be viewed as the switching rate of a time-fluctuating ratchet $V(x,t) = \{U(x) -(v-f)x,~U(x)+(v+f)x\}$ acting on a Brownian particle. This is the case for Brownian motors, whose underlying principle is the rectification of thermal fluctuations through time-varying external forces \cite{makhnovskii2004flashing, astumian1994fluctuation, Parrondo1998Aug, magnasco1993forced}. Other interpretations also exist or can be devised from suitable combinations of the above.

We will occasionally draw on different interpretations when it offers additional insight, but the focus of the present work is on the single perspective of an RnT particle with symmetric self-propulsion $w(t) \in \{v,-v\}$ moving in the potential $U(x)$, Eq.~(\ref{eq:RatchetPotential}), whilst subject to an additional external counterforce $f$. Regardless, the results presented in this paper apply to all of the aforementioned interpretations as, in any case, the motion of the particle is governed by the same Langevin equation in the overdamped limit, i.e.\
\begin{equation}\label{eq:Langevin}
    \dot{x}(t) = -U'(x) + v \sigma(t) - f + \sqrt{2D}\xi(t),
\end{equation}
where $t$ is time, $U'(x)$ is the $x$-derivative of $U(x)$, the telegraphic noise $\sigma(t) \in \{1,-1\}$ switches sign with rate $\gamma$, the diffusion constant is $D$, and $\xi(t)$ is Gaussian white noise with correlator $\langle\xi(t)\xi(t')\rangle=\delta(t-t')$. As is the case for thermal fluctuations, the diffusion considered here is translational and therefore superimposed on the active component of the particle's motion. Henceforth, we will refer to particles as being ``right movers" or in the ``right-moving state" when $\sigma(t) = 1$ and as ``left movers" or in the ``left-moving state" when $\sigma(t) = -1$. 

The corresponding Fokker-Planck equation to Eq.~(\ref{eq:Langevin}) reads
\begin{widetext}
\begin{subequations}\label{eq:FokkerPlanckEquation}
\begin{alignat}{2}
\frac{\partial P^{[i]}_R(x,t)}{\partial t} &= D\frac{\partial^{2} P^{[i]}_R(x,t)}{\partial x^{2}}-\left(v-f-U'^{[i]}\right)\frac{\partial P^{[i]}_R(x,t)}{\partial x}-\gamma\left(P^{[i]}_R(x,t)-P^{[i]}_L(x,t)\right), \label{eq:FokkerPlanckEquationRightMover} \\
\frac{\partial P^{[i]}_L(x,t)}{\partial t} &= D \frac{\partial^{2}P^{[i]}_L(x,t)}{\partial x^{2}} -\left(-v-f-U'^{[i]}\right)\frac{\partial P^{[i]}_L(x,t)}{\partial x}-\gamma\left(P^{[i]}_L(x,t)-P^{[i]}_R(x,t)\right), \label{eq:FokkerPlanckEquationLeftMover}
\end{alignat}
\end{subequations}
\end{widetext}
where $P_{R}^{[i]}(x,t)$ and $P_{L}^{[i]}(x,t)$ are the probability densities at position $x$ and time $t$ of right movers and left movers respectively. The superscript $[i] \in \{[1], [2]\}$ indicates the section of the ratchet on which the probability densities $P_{R,L}^{[i]}(x,t)$ are valid. For the uphill section (from the perspective of a right mover) $x \in \left[0,a\right)$, the densities are given by $P_{R,L}^{[1]}(x,t)$, and for the downhill section $x \in \left[a,\ell\right)$, the densities are given by $P_{R,L}^{[2]}(x,t)$. Hence, each $U'^{[i]}$ of Eq.~(\ref{eq:RatchetPotential}) is in fact constant, specifically $U'^{[1]} = h/a$ and $U'^{[2]} = -h/(\ell-a)$, which has allowed the constants $(\pm v -f - U'^{[i]})$ to be brought in front of $\partial_{x} P_{R,L}^{[i]}(x,t)$ in Eq.~(\ref{eq:FokkerPlanckEquation}).

Unless specified otherwise, the default parameter values used in the figures throughout this paper are $\ell=1$, $a = 0.9$, $h = 4$, $D = 1$, $v = 1$, $\gamma = 1$ and $f=0$. We have chosen these parameters as they correspond to the case where the particle is confined to the potential minimum at $x=0$ when $D=0$, since $v - f < h/a$ and $ v + f < h/(\ell -a)$ represent the conditions for right movers and left movers to be confined respectively. In this ``confined'' case, or ``locked" state \cite{muhsin2023inertial, lindner1999inertia}, an overall current can be generated only by diffusive fluctuations that allow the particle to cross the top of the potential barrier at $x=a$. We will treat this confined case extensively because of its nontrivial behaviour that becomes apparent for $D>0$ (c.f.\ the trivial $D=0$ solution for the confined case in Appendix~\ref{app:DerivationParticleDensity_D=0} and Ref.~\cite{angelani2011active}), as well as its greater relevance, such as to the activated escape of particles via noise-induced transitions \cite{Woillez2019Jun}. Finally, we have opted to keep the discussion of all results in terms of physically meaningful dimensionful parameters, $\ell,a,h,D,v,\gamma$ and $f$, rather than reducing the description of the system to five dimensionless parameters whose role may be difficult to ascertain.


\section{Probability densities}\label{sec:ParticleDensities}

The probability densities at stationarity $ P_{R,L}^{[i]}(x)\equiv\lim\limits_{t\rightarrow \infty} P_{R,L}^{[i]}(x, t)$ are derived from solving the coupled Fokker-Planck equations (\ref{eq:FokkerPlanckEquation}) with the left-hand sides set to zero, i.e.\ $\partial_{t}P_{R}^{[i]}(x) = \partial_{t}P_{L}^{[i]}(x) = 0$. For the specific case of $f = 0$, the steady-state solution to Eq.~(\ref{eq:FokkerPlanckEquation}) has been derived in the context of molecular motors \cite{astumian1994fluctuation}, and was used principally to investigate how the steady-state current varies with the switching rate (equivalent to $\gamma$ here) of the potential. However, this solution has never been studied in the context of RnT motion. The current of an RnT particle moving in a piecewise-linear potential has been studied for the less general case of $D = 0$ \cite{angelani2011active}. However, to the best of our knowledge, a detailed study of the more general $D > 0$ solution for RnT particles does not exist in the literature. Furthermore, we believe our subsequent derivations of the entropy production, power output, thermodynamic efficiency and first-passage properties, as well as the generalisation to $f \neq 0$, to all be novel results.

To obtain the steady-state probability densities, we make the ansatz $P_{R,L}^{[i]}(x) = \mathcal{Z}_{R,L}^{[i]} \exp( \lambda^{[i]}x)$ \cite{astumian1994fluctuation}. Substituting this ansatz into Eq.~(\ref{eq:FokkerPlanckEquation}) for $\partial_{t}P_{R,L}(x,t) = 0$ results in four linearly independent solutions for each of the particle species, right movers and left movers, in each section of the ratchet, $i=1$ and $i=2$, totalling 16 degrees of freedom. Each set of eigenvalues $\lambda^{[i]}$, $i \in \{1,2\}$, is found by solving a quartic characteristic equation. One of the eigenvalues in each set vanishes and so the probability densities have the general solution
\begin{equation}\label{eq:ParticleDensityGeneralSoln}
P_{R,L}^{\ABindex{i}}(x)=
 \mathcal A_{R,L}^{\ABindex{i}}e^{\lambda_{\mathcal{A}}^{\ABindex{i}}x}
+ \mathcal B_{R,L}^{\ABindex{i}}e^{\lambda_{\mathcal{B}}^{\ABindex{i}}x}
+ \mathcal C_{R,L}^{\ABindex{i}}e^{\lambda_{\mathcal{C}}^{\ABindex{i}}x}
+ \mathcal D_{R,L}^{\ABindex{i}},
\end{equation}
with the 16 coefficients $\mathcal{A}_{R}^{[1]}, \mathcal{A}_{L}^{[1]}, \mathcal{A}_{R}^{[2]}, \dots, \mathcal{D}_{L}^{[2]}$ to be determined.

These coefficients are fixed by demanding each linearly independent term in Eq.~(\ref{eq:ParticleDensityGeneralSoln}) individually satisfies Eq.~(\ref{eq:FokkerPlanckEquation}) in the steady state, i.e.\ the resulting prefactors after substitution of each $\exp(\lambda^{[i]}x)$ must be zero \cite{astumian1994fluctuation}. This fixes half of the 16 degrees of freedom in Eq.~(\ref{eq:ParticleDensityGeneralSoln}). The remaining half is fixed through continuity of the probability densities and currents for each species across $x=0$ and $x=a$, as well as the normalisation condition $\int_{0}^{\ell}\mathrm{d}x~P(x) = 1$, where $P(x) = P_{R}(x) + P_{L}(x)$ is the total probability density. Hence, in principle, the steady-state densities for RnT particles in a piecewise-linear ratchet potential can be obtained in closed form. However, solving 16 coupled equations unsurprisingly results in unwieldy expressions for the coefficients $\mathcal{A}_{R}^{[1]}, \mathcal{A}_{L}^{[1]}, \mathcal{A}_{R}^{[2]}, \dots, \mathcal{D}_{L}^{[2]}$, so the bulk of the calculations was performed using Mathematica. The resulting solutions are too long to conveniently display as equations and so most of our results are displayed as graphs alongside descriptions of their qualitative behaviour in the main text. Nevertheless, we emphasise that all results presented in this paper originate from exact expressions. In fact, any observable that can be written exclusively in terms of the system parameters and the coefficients $\mathcal{A}_{R}^{[1]}, \mathcal{A}_{L}^{[1]}, \mathcal{A}_{R}^{[2]}, \dots, \mathcal{D}_{L}^{[2]}$, such as the steady-state current, Eq.~(\ref{eq:OverallCurrentExplicit}), and the steady-state entropy production rate, Eq.~(\ref{eq:EPR_FinalExpression}), can be written in closed form. The full details of how the expressions for the densities were obtained are left to Appendix~\ref{app:DerivationParticleDensity}. We also derive the steady-state probability densities for the nontrivial $D = 0$ case in Appendix~\ref{app:DerivationParticleDensity_D=0}. This simplification ultimately leads to closed-form expressions for the corresponding particle current that are compact enough to be written down, see Eqs.~(\ref{app:eq:D=0_CurrentExactExpression_Unconfined})-(\ref{app:eq:app:eq:D=0_current_leftconfined_parameters}).

\begin{figure}
        \centering
       \includegraphics[width=0.42\textwidth, trim=0.25cm 1.8cm 1.6cm 3.2cm, clip]{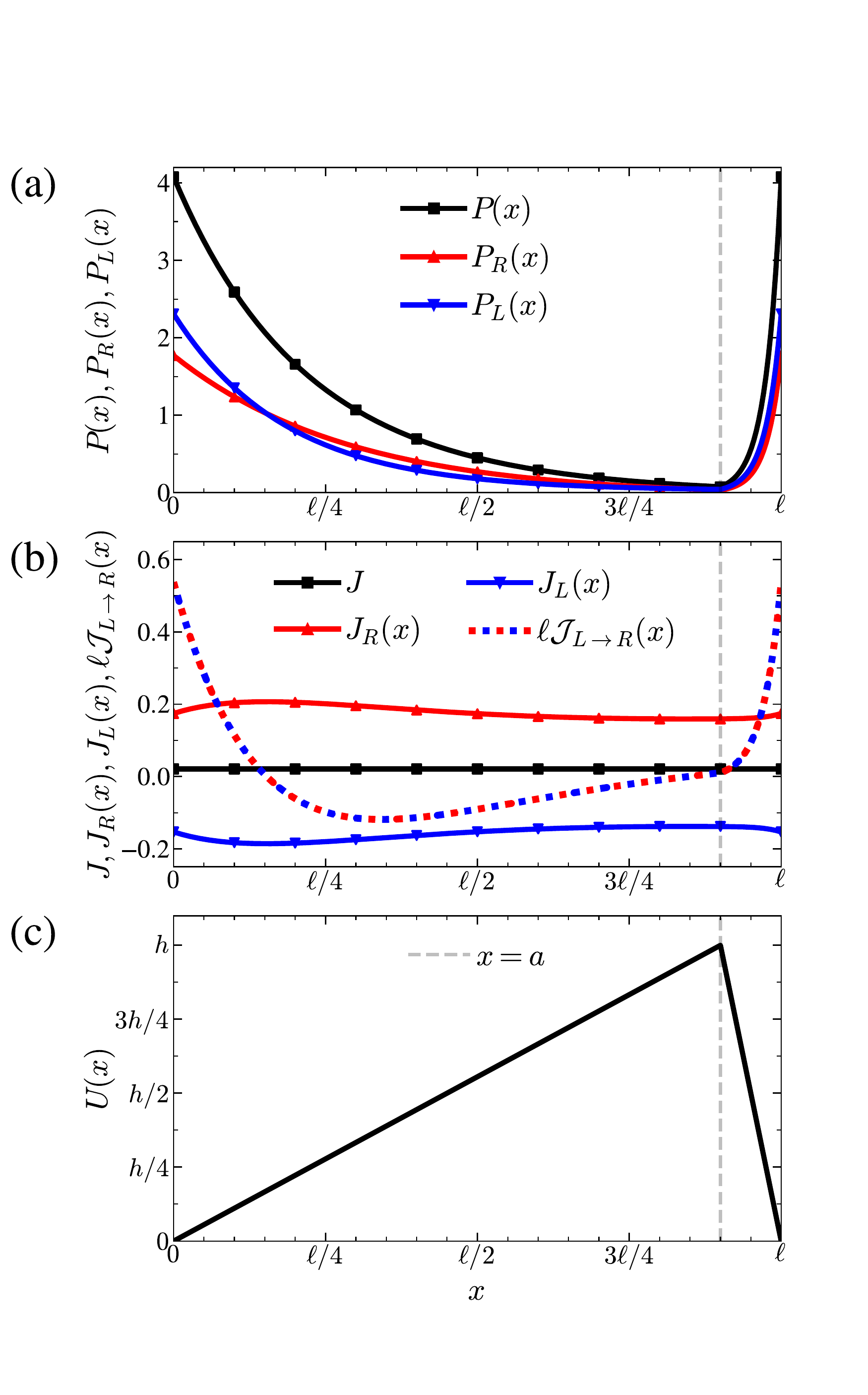}\caption{(a) Steady-state probability densities $P(x), P_{R}(x), P_{L}(x)$, Eq.~(\ref{eq:ParticleDensityGeneralSoln}), (b) and corresponding currents $J, J_{R}(x), J_{L}(x), \mathcal{J}_{L\rightarrow R}(x)$, Eqs.~(\ref{eq:ContinuityEquations})-(\ref{eq:OverallCurrentExplicit}), (c) for a particular form of the ratchet potential $U(x)$, Eq.~(\ref{eq:RatchetPotential}). The vertical dashed lines indicate the position of the ratchet apex, $x = a$. }\label{fig:DensityCurrentPotential}
    \end{figure}

\begin{figure}
        \centering
       \includegraphics[width=0.43\textwidth, trim=0.4cm 0.5cm 0.3cm 0.3cm, clip]{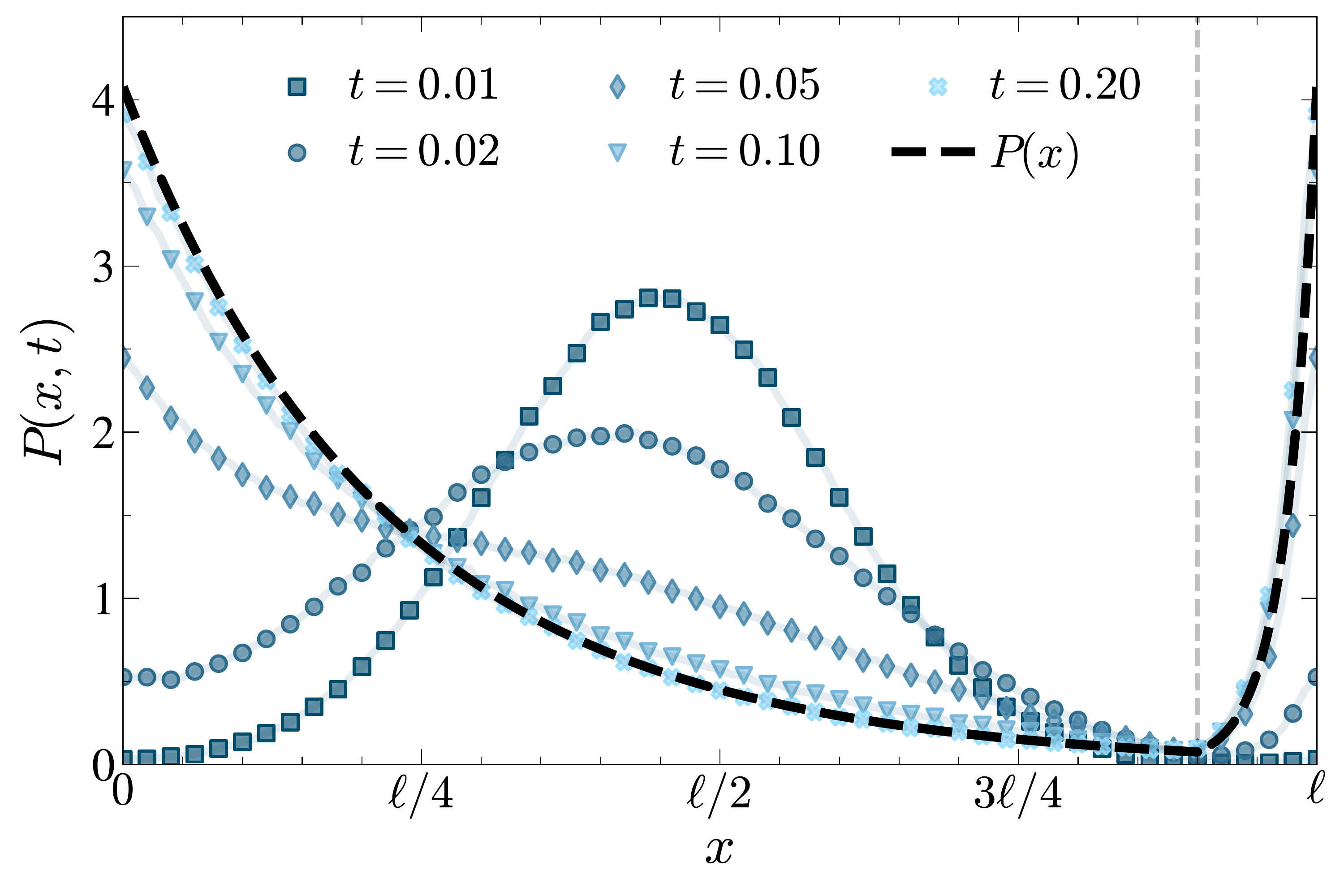}\caption{Time evolution of the overall probability density $P(x,t) = P_{R}(x,t) + P_{L}(x,t)$ obtained from averaging $10^{6}$ stochastic trajectories of an RnT particle subject to a ratchet potential $U(x)$, Eq.~(\ref{eq:RatchetPotential}). In each realisation, a particle is initialised with an equal probability of being a right mover or a left mover at $x=\ell/2$ and $t=0$, i.e.\ $P_{R}(x,0)$ = $P_{L}(x,0) = \delta(x-\ell/2)/2$, before it is evolved in time by numerically integrating Eq.~(\ref{eq:Langevin}) in timesteps of $\mathrm{d}t = 10^{-5}$. At large times $t \rightarrow \infty$, the simulated densities converge to the theoretical steady-state density $P(x) = P_{R}(x) + P_{L}(x)$, Eq.~(\ref{eq:ParticleDensityGeneralSoln}). The vertical dashed line indicates the position of the ratchet apex, $x = a$.}\label{fig:ConvergenceSteadyStateDensity}
    \end{figure}

The solutions for the steady-state probability densities, Eq.~(\ref{eq:ParticleDensityGeneralSoln}), are plotted in Fig.~\ref{fig:DensityCurrentPotential}(a). The solutions for the densities were confirmed by Monte Carlo simulations of the Langevin equation (\ref{eq:Langevin}) by demonstrating the average density of many ($10^{6}$) realisations of the stochastic dynamics converges to the theoretical steady-state density $P(x)$ as $t \rightarrow \infty$, Fig.~\ref{fig:ConvergenceSteadyStateDensity}.

In Fig.~\ref{fig:DensityCurrentPotential}(a), particles are mostly aggregated around the minimum of the ratchet and their density decays further up either slope. Left movers are more likely to be found close to $x = 0$ than right movers as their preferred direction of motion is against the steeper slope of the potential resulting in greater confinement. The strength of the confinement can be adjusted by varying $h$. For $h \rightarrow 0$, the densities tend to uniform distributions $P_{R}(x) = P_{L}(x) = 1/(2\ell)$, since particles are unconfined. Uniform densities are similarly obtained for $v \rightarrow \infty$, $f \rightarrow \infty$ or $D \rightarrow \infty$ as these cases result in the particle decoupling from the potential. For $h \rightarrow \infty$, particles are confined to the point $x = 0$ and so $P_{R,L}(x) \rightarrow \delta(x)/2$. In the equilibrium limit, $v\rightarrow 0$ or $\gamma \rightarrow \infty$, the density tends to a Boltzmann distribution $P^{[i]}(x) \propto \exp({-(U'^{[i]}+f)x/D})$. In the limit $\gamma \rightarrow 0$, right movers and left movers decouple and the solution reduces to that of two drift-diffusive species in the ratchet potential $U(x)$. Finally, taking $D \rightarrow 0$ with $f=0$ yields agreement with the $D=0$ case studied in Ref.~\cite{angelani2011active}.



\section{Currents}\label{sec:Currents}

The currents for right movers and left movers are, respectively,
\begin{subequations}\label{eq:ParticleCurrents}
\begin{alignat}{2}
J_{R}^{\ABindex{i}}(x,t) &= \left(v-f-U'^{[i]}\right)P^{[i]}_R(x,t) - D\frac{\partial P^{[i]}_R(x,t)}{\partial x}, \label{eq:RightParticleCurrent} \\
J_{L}^{\ABindex{i}}(x,t) &= \left(-v-f-U'^{[i]}\right)P^{[i]}_L(x,t) - D\frac{\partial P^{[i]}_L(x,t)}{\partial x}, \label{eq:LeftParticleCurrent}
\end{alignat}
\end{subequations}
where, again, we define separate contributions for the different regions of the ratchet (denoted by the superscript $[i]$) and we recall the $U'^{[i]}$ are constant, see Sec.~\ref{sec:Model}. The Fokker-Planck equation (\ref{eq:FokkerPlanckEquation}) can be rewritten as a continuity equation in terms of the currents defined in Eq.~(\ref{eq:ParticleCurrents}),
\begin{subequations}\label{eq:ContinuityEquations}
\begin{alignat}{2}
\frac{\partial P^{[i]}_R(x,t)}{\partial t}+ \frac{\partial J^{[i]}_R(x,t)}{\partial x} - \mathcal{J}_{L \rightarrow R}^{[i]}(x,t) &= 0, \label{eq:RightContinuityEquation} \\
\frac{\partial P^{[i]}_L(x,t)}{\partial t} + \frac{\partial J^{[i]}_L(x,t)}{\partial x} + \mathcal{J}_{L \rightarrow R}^{[i]}(x,t) &= 0, \label{eq:LeftContinuityEquation}
\end{alignat}
\end{subequations}
where we have also defined the flux density of transitions from the left-moving state to the right-moving state as
\begin{equation}\label{eq:FluxDensityTransitions}
\mathcal{J}_{L \rightarrow R}^{[i]}(x,t) \equiv \gamma\left(P^{[i]}_L(x,t)-P^{[i]}_R(x,t)\right) = -\mathcal{J}_{R \rightarrow L}^{[i]}(x,t).
\end{equation}
The overall current is given by $J(x,t) = J_{R}(x,t) + J_{L}(x,t)$. By summing Eqs.~(\ref{eq:RightContinuityEquation}) and (\ref{eq:LeftContinuityEquation}), it can be seen that the stationary condition $\partial_{t}P^{[i]}(x) = 0$ implies the overall current at stationarity is constant across space since this leads to $\partial_{x}J^{[i]} = 0$. Hence, $J$ can be calculated from the probability densities on just a single section $[i]$ of the ratchet since the other section must have the same current. If we choose $i=1$, corresponding to $x \in [0,a)$, then the resulting expression for $J = J_{R}^{[1]}(x) + J_{L}^{[1]}(x)$ can be simplified significantly by setting $x=0$, yielding a closed expression in terms of the known coefficients $\mathcal{A}_{R}^{[1]}, \mathcal{A}_{L}^{[1]}, \mathcal{A}_{R}^{[2]}, \dots, \mathcal{D}_{L}^{[2]}$ from Eq.~(\ref{eq:ParticleDensityGeneralSoln}),
\begin{small}
\begin{equation}\label{eq:OverallCurrentExplicit}
    J = \sum_{\mathcal{Z}=\mathcal{A}}^{\mathcal{C}} \left[v\mathcal{Z}_{-}^{[1]} - \left(f+\frac{h}{a}+D\lambda_{\mathcal{Z}}^{[1]} \right)\mathcal{Z}_{+}^{[1]} \right] - \left(f+\frac{h}{a}\right)\mathcal{D}^{[i]},
\end{equation}
\end{small}where we have defined $\mathcal{Z}_{\pm}^{[i]} \equiv \mathcal{Z}_{R}^{[i]} \pm \mathcal{Z}_{L}^{[i]}$ and have used that $\mathcal{D}_{R}^{[i]} = \mathcal{D}_{L}^{[i]} \equiv \mathcal{D}^{[i]}$, see Appendix~\ref{app:DerivationParticleDensity}.

\begin{figure}
        \centering
       \includegraphics[width=0.4\textwidth, trim = 4cm 5.2cm 2.3cm 0.8cm, clip]{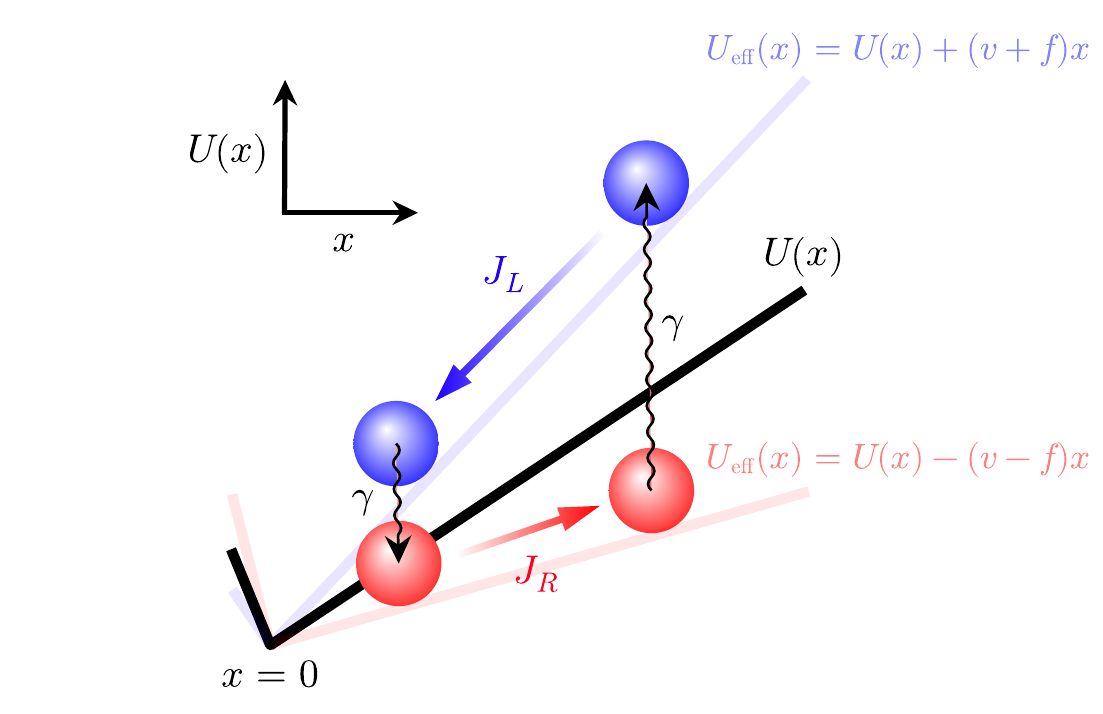}
       \caption{The currents for each species, $J_{R}(x)$ and $J_{L}(x)$, are greater in magnitude close to the potential minimum at $x = 0$ due to a recycling of particles that tumble before crossing the potential barrier. When in the right-moving state (red spheres, bottom), particles experience a weaker effective potential $U_{\text{eff}}(x) = U(x) -(v-f)x$ than when in the left-moving state (blue spheres, top), resulting in, on average, a greater number of right movers diffusing up the slope before tumbling and returning back down as left movers.}\label{fig:BoostedCurrent}
    \end{figure}

\begin{figure}
        \centering
       \includegraphics[width=0.45\textwidth, trim=0.3cm 0.5cm 0.3cm 0.3cm, clip]{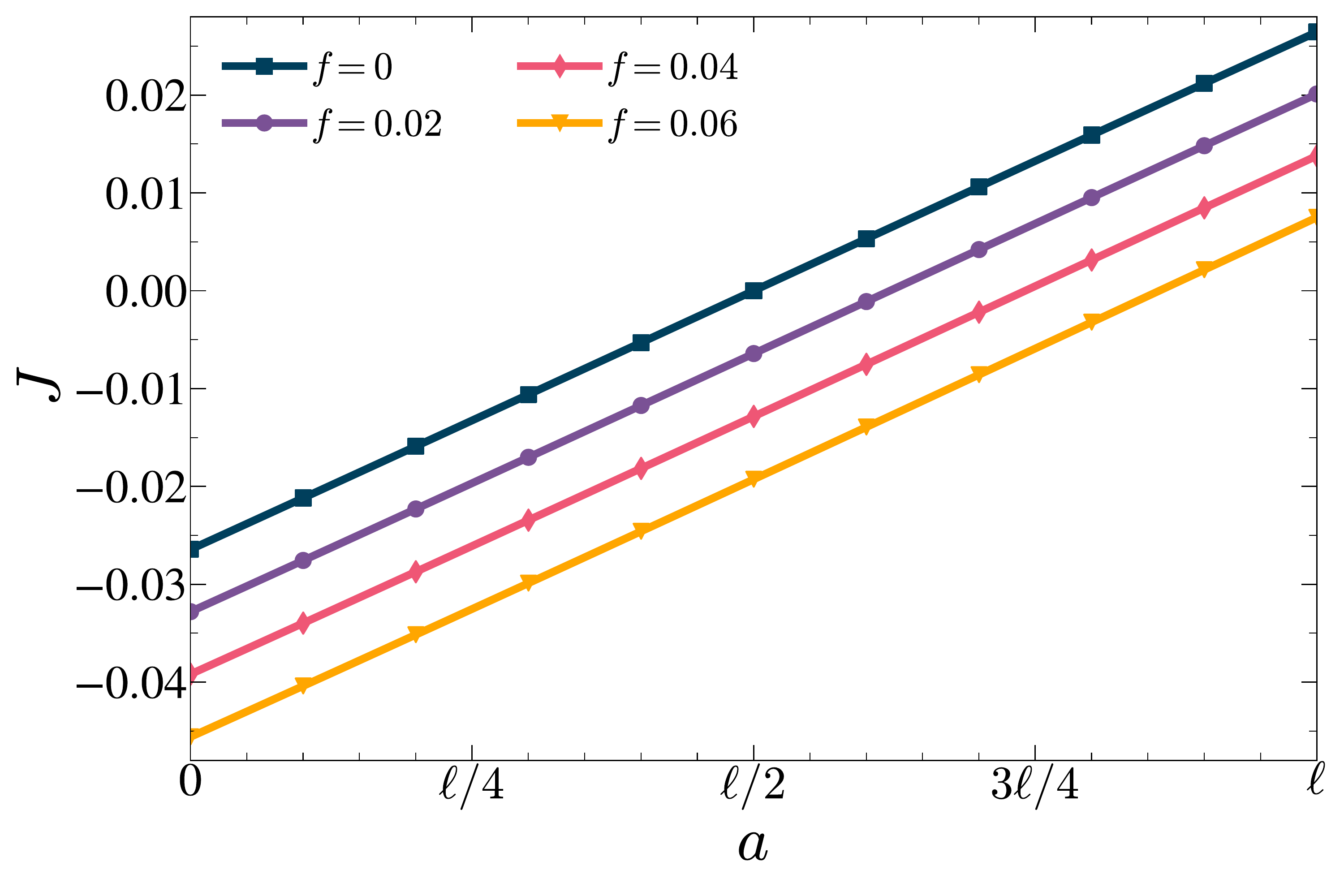}\caption{Steady-state current $J$, Eq.~(\ref{eq:OverallCurrentExplicit}), as a function of the position of the ratchet apex $a$ for various external forces $f$. Maximum current occurs for $a \rightarrow \ell^{-}$ using the default parameters defined in Sec.~\ref{sec:Model}.}\label{fig:CurrentVsPeakPosition}
    \end{figure}

The currents, Eqs.~(\ref{eq:ParticleCurrents}) and (\ref{eq:OverallCurrentExplicit}), and flux density of transitions, Eq.~(\ref{eq:FluxDensityTransitions}), are plotted in Fig.~\ref{fig:DensityCurrentPotential}(b). The currents for each species, $J_{R}(x)$ and $J_{L}(x)$, are greater in magnitude close to the potential minimum at $x = 0$. This results from particles that tumble before crossing the top of the potential barrier at $x=a$ and so are recycled to the bottom of the potential, e.g.\ right movers travel part way up the slope $U'^{[1]}$, spanning $0 \leq x < a$, before they tumble and return down the slope as left movers (and vice versa for particles on the opposite slope $U'^{[2]}$). This phenomenon is still present when particles are confined, e.g.\ we can still have $J_{R}(x) > 0$ on $U'^{[1]}$ despite $v - f < U'^{[1]}$. As illustrated in Fig.~\ref{fig:BoostedCurrent}, this is because particles experience different ``effective" potentials $U_{\text{eff}}(x)$ depending on their internal state. Specifically, a right-mover can be viewed as a Brownian particle moving in an effective potential $U_{\text{eff}}(x) = U(x) - (v-f)x$. Assuming $v - f>0$, right movers experience a smaller $U'^{[1]}_{\text{eff}}$ than left movers. This asymmetry in $U'^{[1]}_{\text{eff}}$ means, on average, a greater number of right movers than left movers will diffuse up the slope $U'^{[1]}$. When these additional right movers tumble, they will return down the slope as left movers, resulting in a current for each particle species just as in the unconfined case. As $D$ is decreased, these confined currents become more localised around $x = 0$ as the chance of a particle crossing the potential barrier becomes increasingly unlikely.

\begin{figure}
        \centering
       \includegraphics[width=0.45\textwidth, trim=0.2cm 0.4cm 0.2cm 0.2cm, clip]{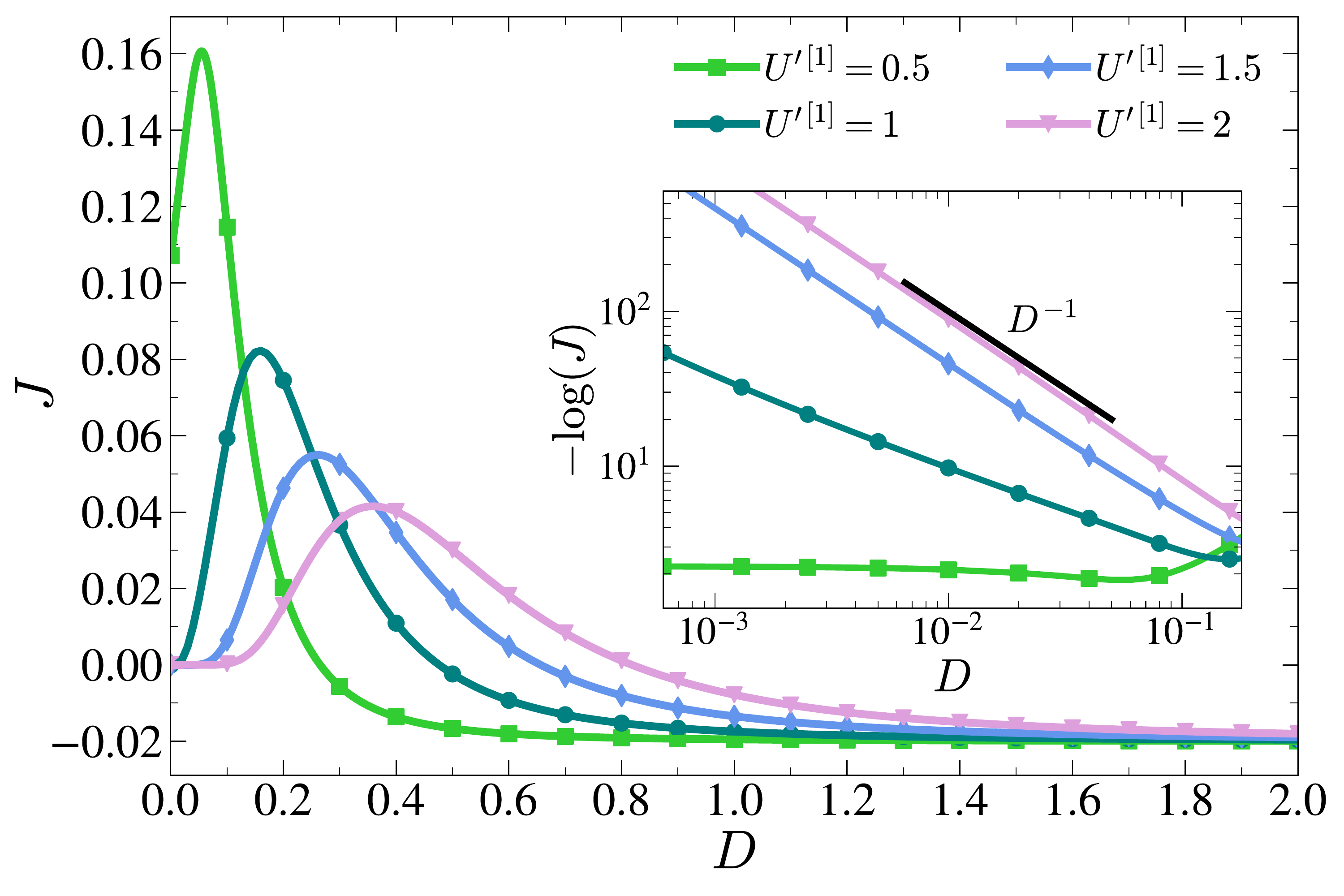}\caption{Steady-state current $J$, Eq.~(\ref{eq:OverallCurrentExplicit}), as a function of diffusion $D$ for various magnitudes of the slope $U'^{[1]} = h/a$ for fixed $a = 0.9$ and $f = 0.02$. For $D \rightarrow \infty$, the current asymptotically approaches $J = -f/\ell$. For $D \rightarrow 0$, the current vanishes if the particle is confined, i.e.\ $v - f < U'^{[1]}$ and $v + f < |U'^{[2]}|$. For $v - f > U'^{[1]}$, positive current is possible for $D = 0$, as illustrated by the line for $U'^{[1]} = 0.5$. The inset log-log plot demonstrates that agreement with the Arrhenius equation $J \sim \exp(-H/D)$ for small $D$ improves with increasing confinement.}\label{fig:CurrentVsDiffusion}
    \end{figure}

Based on the mechanism described above and in Fig.~\ref{fig:BoostedCurrent}, we expect each particle to contribute $\sim\gamma$ to the currents $J_{R,L}(x)$. As will become apparent for the entropy production in Sec.~\ref{sec:EntropyProduction}, a particularly relevant case is the form of the confined currents as $D \rightarrow 0$. In this equilibrium-like case, the current obeys the Arrhenius law $J \sim \exp(-H/D) \approx 0$, see inset of Fig.~\ref{fig:CurrentVsDiffusion}, where $H$ is the ``effective" height of the potential adjusted for particle activity \cite{Woillez2019Jun}. Hence, we can assume particles almost never overcome the potential barrier as $D \to 0$, and can thus approximate the barrier height $h$ to be effectively infinite. Taking the slope $U'^{[1]}$ as an example, the largest contribution to the left-moving current $J_{L}(x)$ at a position $x$ will come from right movers that tumble higher up the slope, and so $|J_{L}^{[1]}(x)| \approx \gamma \int_{x}^{\infty}\mathrm{d}x'~P_{R}^{[1]}(x')$. Since the confined case is effectively equilibrium, then the right-moving probability density is given by a Boltzmann distribution $P_{R}^{[1]}(x) \sim \exp(-x/l_{D})/l_{D}$, assuming relaxation to the steady state occurs much faster than transmutation between species and where $l_{D} = D/(U'^{[1]}-v + f) > 0$ is the characteristic length scale of diffusive spreading. Hence, when particles are confined, $0< v-f < U'^{[1]}$ and $v+f < |U'^{[2]}|$, the current in the limit of vanishing diffusion is
\begin{equation}\label{eq:CirculatingCurrent}
    \lim_{D \rightarrow 0} |J_{L}^{[1]}(x)| \sim \gamma e^{-\frac{x}{l_{D}}}.
\end{equation}
Since $J \rightarrow 0$ for $D \rightarrow 0$ in the confined case, then $|J_{R}(x)| = |J_{L}(x)|$ and so $|J_{R}^{[1]}(x)|$ is also equal to the right-hand side of Eq.~(\ref{eq:CirculatingCurrent}). Similar arguments can be used to calculate the contributions $J_{R,L}^{[2]}(x)$ to the confined currents on the other slope $U'^{[2]}$ for $D \rightarrow 0$.

\begin{figure}
    \centering
    \includegraphics[width=0.45\textwidth, trim=0.2cm 0.4cm 0.2cm 0.2cm, clip]{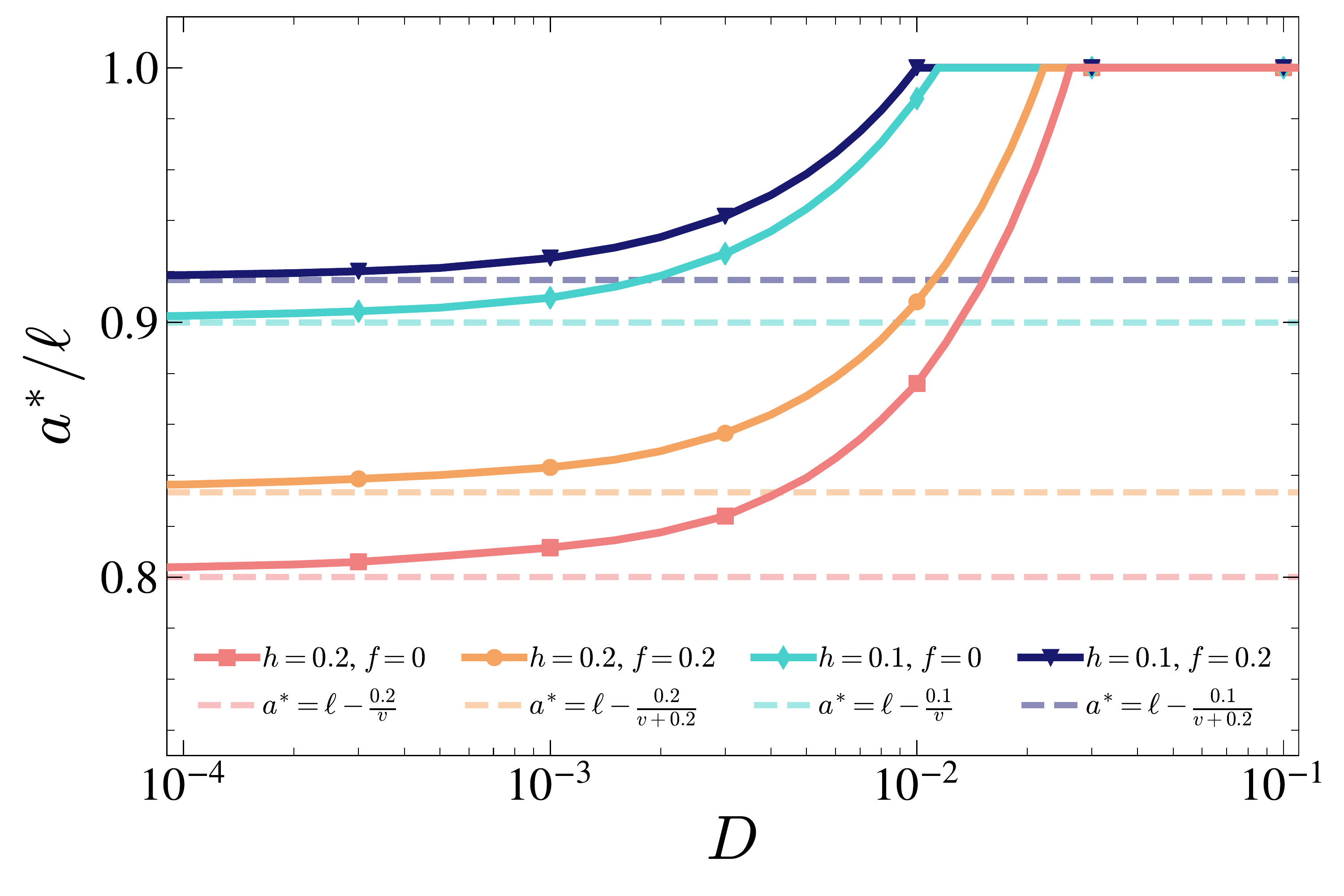}\caption{Position of the ratchet apex $a^{*}$ resulting in maximum current $J$, Eq.~(\ref{eq:OverallCurrentExplicit}), as a function of diffusion strength $D$. Maximum current occurs for $a^{*} = \ell - h/(v+f)$ for $D \rightarrow 0$ and $a^{*} \rightarrow \ell^{-}$ for large $D$.}\label{fig:MaxCurrentPositionVsDiffusion}
\end{figure}

In Fig.~\ref{fig:CurrentVsPeakPosition}, it can be seen that motion is rectified to the right ($J > 0$) for $a > \ell/2$ and is rectified to the left ($J < 0$) for $a < \ell/2$ when $f=0$. Contrary to that of the $D=0$ case studied in Ref.~\cite{angelani2011active}, where maximum positive current occurs for $a = \ell - h/(v+f)$, we find for the parameters used in Fig.~\ref{fig:CurrentVsPeakPosition} that maximum current occurs for $a \rightarrow \ell^{-}$, i.e.\ when there is an infinitely steep slope $U'^{[2]}$ opposing the motion of left movers. While taking $D \rightarrow 0$ in our model recovers the result of Ref.~\cite{angelani2011active}, we observe in Fig.~\ref{fig:MaxCurrentPositionVsDiffusion} that the position $a^{*}$ of the ratchet apex resulting in maximum current occurs at some nontrivial $\ell - h/(v+f) < a^{*} < \ell $ for small but finite diffusion, $D > 0$. To see why, consider first $a^{*}$ for the $D=0$ case, which arises by choosing the exact $a$ at which left movers are \textit{only just} confined by the potential, i.e.\ $v+f = h/(\ell - a)$, in order to not increase the uphill distance the unconfined right movers must traverse. However, for finite diffusion, $D>0$, left movers are able to overcome the barrier to the left, even for $v+f < h/(\ell - a)$. Hence, to achieve maximum current $J$ for $D>0$, one must increase $a$ to hinder the left movers' ability to overcome the barrier, while still bearing in mind the compromise that applies to right movers in the $D=0$ case. As $D$ is increased further, $a^{*}$ also increases up until the ``best'' one can do to minimise the left-moving current $J_{L}(x)$ is to set $a \rightarrow \ell^{-}$.

\begin{figure}
    \centering
    \includegraphics[width=0.45\textwidth, trim=0.2cm 0.4cm 0.2cm 0.2cm, clip]{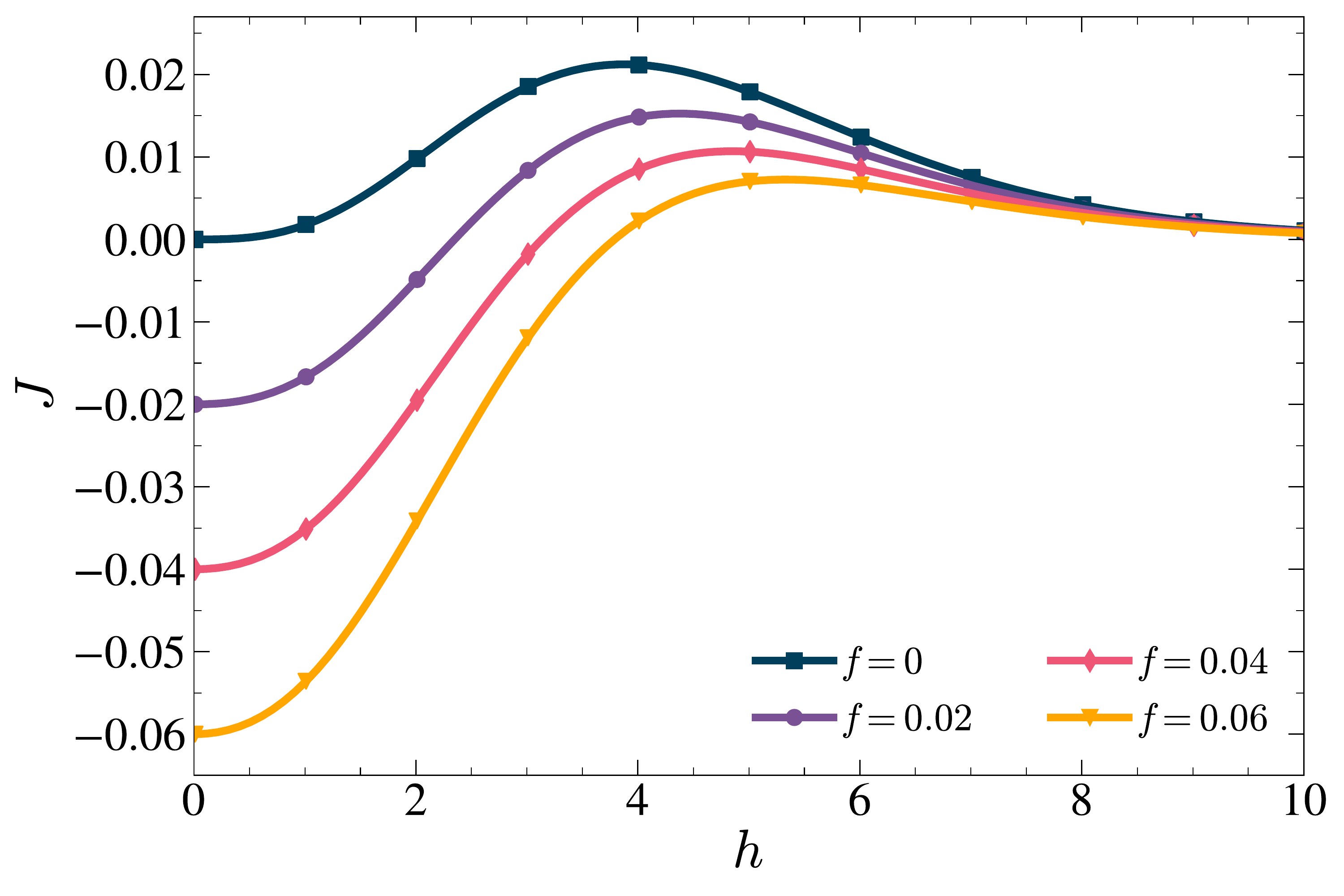}\caption{Steady-state current $J$, Eq.~(\ref{eq:OverallCurrentExplicit}), as a function of the ratchet height $h$ for various external forces $f$. For $h = 0$, the current is given by $J = -f/\ell$. The current vanishes for all $f$ as $h \rightarrow \infty$, due to the confinement at $x=0$ being infinitely strong. The current peaks around $h\approx 4$, reflecting a compromise in choosing a potential $U(x)$, Eq.~(\ref{eq:RatchetPotential}), that rectifies the symmetric RnT motion while not confining particles too strongly.}\label{fig:CurrentVsHeight}
\end{figure}

For completeness, we provide a detailed derivation of the $D = 0$ current $j$ in Appendix \ref{app:DerivationParticleDensity_D=0}, resulting in closed-form expressions that are compact enough to be written down, Eqs.~(\ref{app:eq:D=0_CurrentExactExpression_Unconfined})-(\ref{app:eq:app:eq:D=0_current_leftconfined_parameters}). This derivation generalises the results of Ref.~\cite{angelani2011active} to include an additional linear external force $f \neq 0$ or, equivalently, to allow for asymmetric self-propulsion speeds between right movers and left movers.
    
The effect of varying the ratchet height $h$ on the current is plotted in Fig.~\ref{fig:CurrentVsHeight}. For vanishing height $h = 0$, equivalent to the absence of a potential $U(x)$, Eq.~(\ref{eq:RatchetPotential}), the current is given by $J = -f/\ell$. As $h \rightarrow \infty$, the current vanishes as particles become confined by infinitely steep slopes at both sides of $x=0$. In between these extremes, $J$ peaks at a finite value of $h$. This reflects that the potential $U(x)$ that optimises the current must be suitably strong to rectify the symmetric RnT motion while not confining particles too strongly.

The current $J$ similarly peaks at finite values of $v$ and $D$, the latter of which is plotted in Fig.~\ref{fig:CurrentVsDiffusion}. As $v \rightarrow \infty$ or $D \rightarrow \infty$, the particle decouples from the potential such that $J \rightarrow -f/\ell$. This is also the case for when the external force dominates, i.e.\ $|f| \rightarrow \infty$. The behaviour of $J$ for $D \rightarrow 0$ depends on whether the particle is confined. If the particle's self-propulsive force is not large enough to penetrate the potential barrier, i.e.\ $v - f < U'^{[1]}$ and $v + f < |U'^{[2]}|$, then the particle is confined to the potential well at $x = 0$ and thus $J = 0$, see Fig.~\ref{fig:CurrentVsDiffusion}. If the particle is unconfined, $v - f > U'^{[1]}$ \textit{or} $v + f > |U'^{[2]}|$, then there is a nonzero current $J >0$ as $D \rightarrow 0$.

For $\gamma = 0$, the overall current $J$ is the average for that of uncoupled drift-diffusive right movers and left movers. For $\gamma \rightarrow \infty$, the current tends to $J \rightarrow -f/\ell$ since particles are no longer persistent, and the dynamics reduce to that of a purely diffusive particle moving in the potential $U(x) + fx$.
    


\section{Entropy production}\label{sec:EntropyProduction}

The entropy production rate $\dot{S}_{\mathrm{int}}$ quantifies the degree to which time-reversal symmetry is broken and therefore distinguishes between nonequilibrium, $\dot{S}_{\mathrm{int}} > 0$, and equilibrium, $\dot{S}_{\mathrm{int}} = 0$, systems. In the present system, there are several sources of symmetry breaking, such as asymmetric overall drift speeds due to the external force $f$, and that particles tend to run faster down the sides of the ratchet potential $U(x)$, Eq.~(\ref{eq:RatchetPotential}), when the force due to the potential is aligned with the particle's own self-propulsion speed $v$, similarly to RnT particles in a harmonic potential \cite{garcia2021run}. Starting from the Gibbs-Shannon entropy \cite{shannon1948mathematical}, we take a well-established approach \cite{seifert2005entropy, alston2022non, razin2020entropy} to obtain an exact expression for the entropy production rate at stationarity, Eq.~(\ref{eq:EPR_FinalExpression}).

The Gibbs-Shannon entropy for the present system is defined as
\begin{equation}\label{eq:GibbsShannonEntropy}
    S(t) = -\sum_{k=R,L}\int_{0}^{\ell}\mathrm{d}x~P_{k}(x,t)\log\left(\frac{P_{k}(x,t)}{\bar{P}}\right),
\end{equation}
where $\bar{P}$, with $\left[ \bar{P} \right] = 1/\ell$, is for dimensional consistency in the logarithm but will drop out of the final expression for $\dot{S}_{\mathrm{int}}$. The time derivative of Eq.~(\ref{eq:GibbsShannonEntropy}) is  
\begin{equation}\label{eq:TimeDerivativeGibbsShannonEntropy}
    \dot{S}(t) = -\sum_{k=R,L}\int_{0}^{\ell}\mathrm{d}x~\frac{\partial P_{k}(x,t)}{\partial t}\left(\log\left(\frac{P_{k}(x,t)}{\bar{P}}\right)+1\right).
\end{equation}
After relating $\partial_{t}P_{k}(x,t)$ in Eq.~(\ref{eq:TimeDerivativeGibbsShannonEntropy}) to the currents $J_{k}(x,t)$ and $\mathcal{J}_{L \rightarrow R}(x,t)$ through the continuity equations, Eq.~(\ref{eq:ContinuityEquations}), integrating by parts, and dropping boundary terms due to the periodic boundary conditions, we eventually obtain
\begin{equation}\label{eq:TimeDerivativeEntropy1}
\begin{split}
    \dot{S}(t) = -\int_{0}^{\ell}\mathrm{d}x\Bigg(&\sum_{k=R,L}\frac{J_{k}(x,t)}{P_{k}(x,t)}\frac{\partial P_{k}(x,t)}{\partial x}\\
    &+ \mathcal{J}_{L\rightarrow R}(x,t)\log\left(\frac{P_{R}(x,t)}{P_{L}(x,t)}\right)\Bigg).
\end{split}
\end{equation}
Using the definition of the currents, Eq.~(\ref{eq:ParticleCurrents}), $\partial_{x}P_{k}(x,t)$ in Eq.~(\ref{eq:TimeDerivativeEntropy1}) can be rewritten to yield
\begin{equation}\label{eq:TimeDerivativeEntropy2}
\begin{split}
    &\dot{S}(t) = \int_{0}^{\ell}\mathrm{d}x\Bigg[\sum_{k=R,L}\frac{J_{k}^{2}(x,t)}{D P_{k}(x,t)} + \frac{f + U'(x)}{D}J(x,t)\\
    &- \frac{v}{D}\left(J_{R}(x,t)-J_{L}(x,t) \right) + \mathcal{J}_{L\rightarrow R}(x,t)\log\left(\frac{P_{L}(x,t)}{P_{R}(x,t)}\right)\Bigg],
\end{split}
\end{equation}
where we have used that $J_{R}(x,t) + J_{L}(x,t) = J(x,t)$.

At this point, we separate the time derivative of the entropy into the contributions $\dot{S}(t) = \dot{S}_{\mathrm{int}}(t) - \dot{S}_{\mathrm{ext}}(t)$ and identify the positive-definite terms in Eq.~(\ref{eq:TimeDerivativeEntropy2}) as those corresponding to the internal entropy production rate of the system $\dot{S}_{\mathrm{int}}(t)$ \cite{cocconi2020entropy, alston2022non, razin2020entropy}. We thus have
\begin{equation}\label{eq:EntropyProductionRateGeneral}
    \dot{S}_{\mathrm{int}}(t) = \dot{S}_{R}(t) + \dot{S}_{L}(t) + \dot{S}_{R \leftrightarrow L}(t)
\end{equation}
where
\begin{equation}\label{eq:EPR_Density_RightMovers}
    \dot{S}_{R/L}(t) \equiv \int_{0}^{\ell}\mathrm{d}x~\dot{s}_{R/L}(x,t) \equiv \int_{0}^{\ell}\mathrm{d}x~\frac{J_{R/L}^{2}(x,t)}{D P_{R/L}(x,t)}
\end{equation}
is the entropy production of right/left movers and $\dot{s}_{R/L}(x,t) \equiv J_{R/L}^{2}(x,t)/(DP_{R/L}(x,t))$ is the local entropy production \emph{density} of right/left movers at position $x$ and time $t$ \cite{razin2020entropy, garcia2021run}. Similarly, 
\begin{equation}\label{eq:EPR_DensityTransitions}
\begin{split}
    \dot{S}_{R \leftrightarrow L}(t) &\equiv \int_{0}^{\ell}\mathrm{d}x~\dot{s}_{R \leftrightarrow L}(x,t)\\
    &\equiv \int_{0}^{\ell}\mathrm{d}x~\mathcal{J}_{L\rightarrow R}(x,t)\log\left(\frac{P_{L}(x,t)}{P_{R}(x,t)}\right)
\end{split}
\end{equation}
is the entropy production due to transitions between right movers and left movers, with $\dot{s}_{R \leftrightarrow L}(x,t) = \mathcal{J}_{L\rightarrow R}(x,t)\log\left(P_{L}(x,t)/P_{R}(x,t)\right)$ its local density. Note that $\dot{S}_{R \leftrightarrow L}(t) \geq 0$ for all $t$ since $\mathcal{J}_{L\rightarrow R}(x,t)$, Eq.~(\ref{eq:FluxDensityTransitions}), and $\log\left(P_{L}(x,t)/P_{R}(x,t)\right)$ are both positive when $P_{L}(x,t) > P_{R}(x,t)$ and both negative when $P_{L}(x,t) < P_{R}(x,t)$.

\begin{figure}
    \centering
    \includegraphics[width=0.45\textwidth, trim=0.3cm 0.5cm 0.3cm 0.3cm, clip]{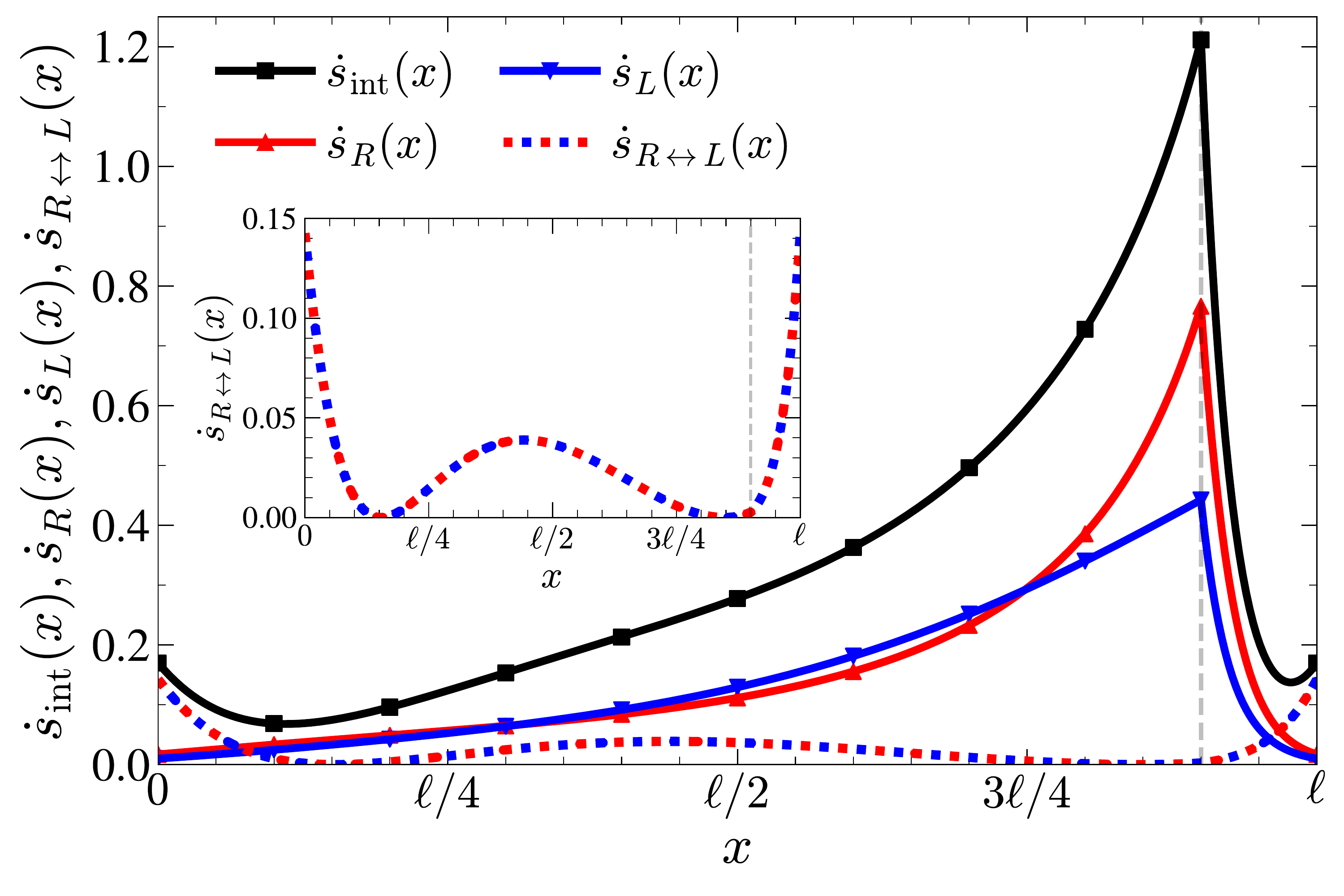}\caption{Steady-state entropy production densities for right movers $\dot{s}_{R}(x)$, Eq.~(\ref{eq:EPR_Density_RightMovers}), left movers $\dot{s}_{L}(x)$, Eq.~(\ref{eq:EPR_Density_RightMovers}), transitions between states $\dot{s}_{R \leftrightarrow L}(x)$, Eq.~(\ref{eq:EPR_DensityTransitions}) (inset), and their sum $\dot{s}_{\mathrm{int}}(x) = \dot{s}_{R}(x) + \dot{s}_{L}(x) + \dot{s}_{R \leftrightarrow L}(x)$. The vertical dashed lines indicate the position of the ratchet apex, $x=a$.}\label{fig:EntropyProductionDensity}
\end{figure}

The steady-state entropy production densities, as defined by $\lim\limits_{t\rightarrow\infty}\dot{s}_{R/L}(x,t) \equiv \dot{s}_{R/L}(x)$ and $\lim\limits_{t\rightarrow\infty}\dot{s}_{R \leftrightarrow L}(x,t) \equiv \dot{s}_{R \leftrightarrow L}(x)$ in Eqs.~(\ref{eq:EPR_Density_RightMovers}) and (\ref{eq:EPR_DensityTransitions}) respectively, are plotted in Fig.~\ref{fig:EntropyProductionDensity} for the same parameters as in Fig.~\ref{fig:DensityCurrentPotential}. We find the overall contribution from right movers to the steady-state entropy production $\dot{s}_{R}(x)$ is larger than that from left movers $\dot{s}_{L}(x)$ in the case where motion is rectified to the right, i.e.\ $J>0$ as a result of $a>\ell/2$. This is expected since left movers are more confined than right movers for $a>\ell/2$ and $f=0$. The entropy production densities for both states, $\dot{s}_{R}(x)$ and $\dot{s}_{L}(x)$, are most concentrated around $x = a$, where the magnitude of the current per particle $J/P(x)$, i.e.\ the mean local velocity, is greatest. The total entropy production density $\dot{s}_{\mathrm{int}}(x)$ has a local maximum at $x=0$ as a result of the entropy production from transitions between states $\dot{s}_{R \leftrightarrow L}(x)$, Eq.~(\ref{eq:EPR_DensityTransitions}). The confinement due to the ratchet results in a higher concentration of particles around $x=0$, see Fig.~\ref{fig:DensityCurrentPotential}(a), and thus a greater number of symmetry-breaking tumble events, which produces a local peak in the entropy production density. The density for transitions between states $\dot{s}_{R \leftrightarrow L}(x)$, essentially the Kullback-Leibler divergence of $P_{R}(x)$ and $P_{L}(x)$ \cite{kullback1951information}, has another local maximum in the bulk. In the case of the default parameters used in Fig.~\ref{fig:EntropyProductionDensity}, this local maximum occurs between $x=\ell/4$ and $x=\ell/2$, coinciding with a large local difference between the probability densities for each species, $P_{R}(x)$ and $P_{L}(x)$, as seen in Fig.~\ref{fig:DensityCurrentPotential}.

\begin{figure}
    \centering
    \includegraphics[width=0.45\textwidth, trim=0.3cm 0.5cm 0.3cm 0.3cm, clip]{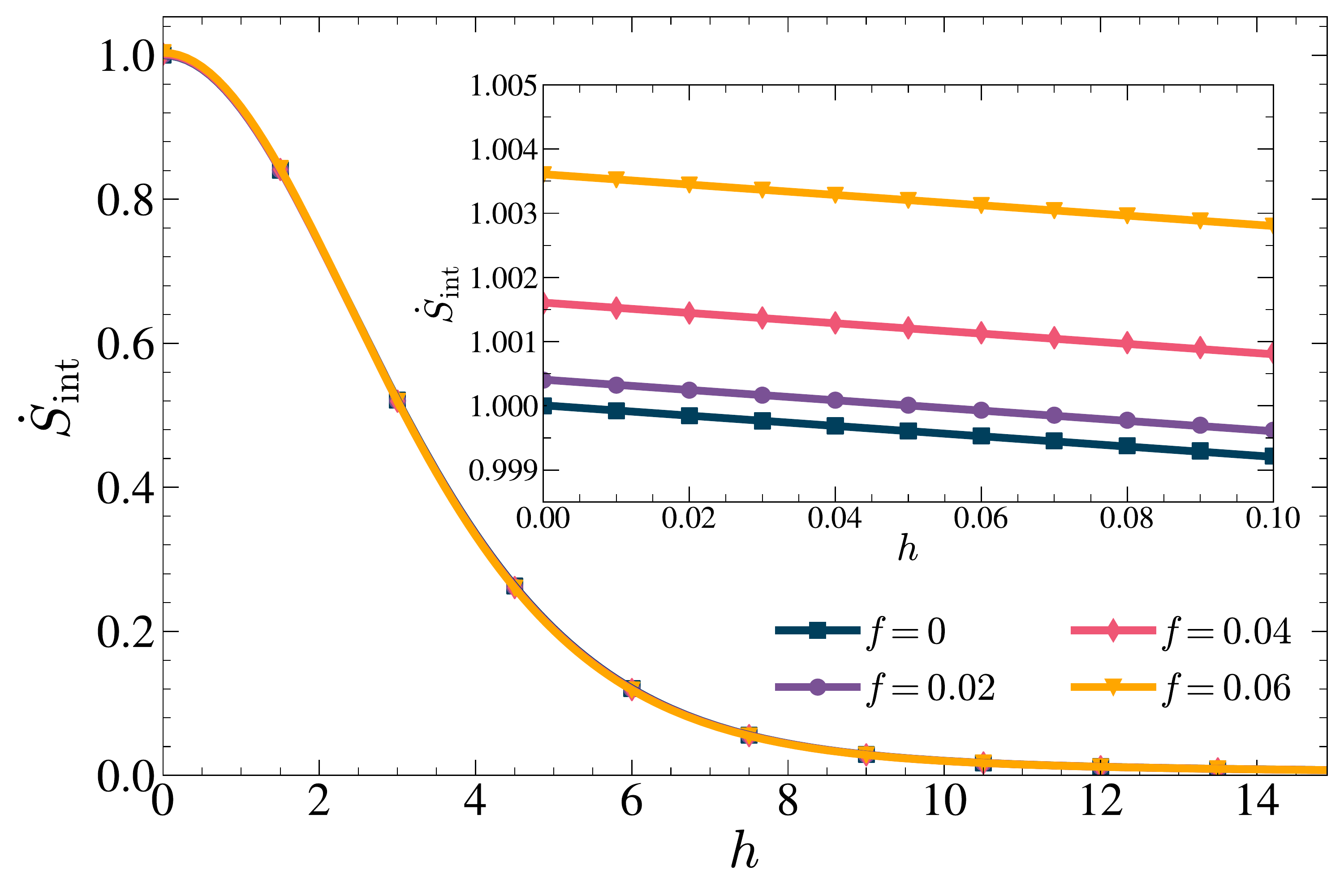}\caption{Steady-state entropy production rate $\dot{S}_{\mathrm{int}}$, Eq.~(\ref{eq:EPR_FinalExpression}), as a function of the ratchet height $h$ for various external forces $f$. In the absence of a ratchet, $h = 0$, the entropy production is given by $\dot{S}_{\mathrm{int}} = (v^{2}+f^{2})/D$ (inset), whilst $\dot{S}_{\text{int}} \rightarrow 0$ as $h \rightarrow \infty$ for all $f$.}\label{fig:entropy_against_h}
\end{figure}

After identifying the terms contributing to the internal entropy production rate $\dot{S}_{\text{int}}(t)$, Eq.~(\ref{eq:EntropyProductionRateGeneral}), the remaining terms in $\dot{S}(t)$, Eq.~(\ref{eq:TimeDerivativeEntropy2}), are identified as the entropy flux lost from the system to the environment, i.e.\
\begin{equation}\label{eq:EPR_FluxToEnvironment}
\begin{split}
    \dot{S}_{\mathrm{ext}}(t) = \frac{1}{D}\int_{0}^{\ell}\mathrm{d}x~\big[&v(J_{R}(x,t)-J_{L}(x,t))\\
    & - \left(f + U'(x)\right)J(x,t)\big].
\end{split}
\end{equation}

At stationarity, $\dot{S} = 0$ implies $\dot{S}_{\mathrm{int}} = \dot{S}_{\mathrm{ext}}$, where $\dot{S} \equiv \lim\limits_{t\rightarrow\infty}\dot{S}(t)$ etc. Thus, it is convenient to evaluate the steady-state entropy production $\dot{S}_{\mathrm{int}}$ through the simpler expression for the steady-state entropy flux $\dot{S}_{\mathrm{ext}}$, Eq.~(\ref{eq:EPR_FluxToEnvironment}), whence we obtain
\begin{equation}\label{eq:EPR1}
    \dot{S}_{\mathrm{int}} = \frac{1}{D}\int_{0}^{\ell}\mathrm{d}x~\left[v\left(J_{R}(x)-J_{L}(x)\right) -(f+U'(x))J\right].
\end{equation}
Then, using that $J$ is constant across space, Eq.~(\ref{eq:OverallCurrentExplicit}), and $U(x)$ is periodic, Eq.~(\ref{eq:RatchetPotential}), we obtain
\begin{equation}\label{eq:EPR2}
    \dot{S}_{\mathrm{int}} = \frac{v}{D}\int_{0}^{\ell}\mathrm{d}x~\left(J_{R}(x)-J_{L}(x)\right) - \frac{f\ell J}{D}.
\end{equation}
After reinserting the expressions for the individual currents $J_{R,L}(x)$ from Eq.~(\ref{eq:ParticleCurrents}), and applying the normalisation $\int_{0}^{\ell}\mathrm{d}x~P(x) = 1$ and periodic boundary conditions $P_{R,L}(0) = P_{R,L}(\ell)$, Eq.~(\ref{eq:EPR2}) becomes
\begin{equation}\label{eq:EPR3}
    \dot{S}_{\mathrm{int}} = \frac{v^{2}}{D} - \frac{f\ell J}{D} - \frac{v}{D}\int_{0}^{\ell}\mathrm{d}x~U'(x)\left(P_{R}(x)-P_{L}(x)\right),
\end{equation}
which, in the absence of an external counterforce $f$ and ratchet potential $U(x)$, Eq.~(\ref{eq:RatchetPotential}), recovers the entropy production rate of a free particle $v^{2}/D$ (in the case where its internal state is known at all times) \cite{cocconi2020entropy}. Finally, splitting the integral in Eq.~(\ref{eq:EPR3}) into separate contributions for each region of the ratchet and inserting the expressions for the probability densities, Eq.~(\ref{eq:ParticleDensityGeneralSoln}), we obtain an expression in terms of the known coefficients $\mathcal{A}_{R}^{[1]}, \mathcal{A}_{L}^{[1]}, \mathcal{A}_{R}^{[2]}, \dots, \mathcal{D}_{L}^{[2]}$,
\begin{small}
\begin{equation}\label{eq:EPR_FinalExpression}
\begin{split}
    \dot{S}_{\mathrm{int}} &= \frac{v^{2}}{D} - \frac{f\ell J}{D} - \frac{v}{D}\sum_{\mathcal{Z}=\mathcal{A}}^{\mathcal{C}}\Bigg(\frac{h}{a}\mathcal{Z}_{-}^{[1]}\frac{e^{\lambda_{\mathcal{Z}}^{[1]}a}-1}{\lambda_{\mathcal{Z}}^{[1]}}\\
    &~~~~~~~~~~~~~~~~~~~~~~~~~~~-\frac{h}{\ell-a}\mathcal{Z}_{-}^{[2]}\frac{e^{\lambda_{\mathcal{Z}}^{[2]}\ell}-e^{\lambda_{\mathcal{Z}}^{[2]}a}}{\lambda_{\mathcal{Z}}^{[2]}}\Bigg),
\end{split}
\end{equation}
\end{small}where we have used $\mathcal{D}_{R}^{[i]} = \mathcal{D}_{L}^{[i]}$, see Appendix~\ref{app:DerivationParticleDensity}, and have adopted the same shorthand notation, i.e.\ $\mathcal{Z}_{-}^{[i]} \equiv \mathcal{Z}_{R}^{[i]} - \mathcal{Z}_{L}^{[i]}$, used in Eq.~(\ref{eq:OverallCurrentExplicit}).

The entropy production $\dot{S}_{\mathrm{int}}$, Eq.~(\ref{eq:EPR_FinalExpression}), is plotted in Fig.~\ref{fig:entropy_against_h} against the height $h$ of the ratchet. For $h = 0$, the entropy production reduces to $\dot{S}_{\mathrm{int}} = (v^{2} + f^{2})/D$, see inset of Fig.~\ref{fig:entropy_against_h}, which is that of a free RnT particle $v^{2}/D$ plus a positive semidefinite contribution $f^{2}/D$ for all $f$ due to the asymmetry in the overall drift speeds breaking time-reversal symmetry. The entropy production monotonically decreases with $h$, tending to $\dot{S}_{\mathrm{int}} \rightarrow 0$ as $h \rightarrow \infty$ since particles become confined to the potential minimum at $x=0$ and therefore the system contains only stationary particles that are trivially at equilibrium.

\begin{figure}
    \centering
    \includegraphics[width=0.45\textwidth, trim=0.3cm 0.5cm 0.3cm 0.3cm, clip]{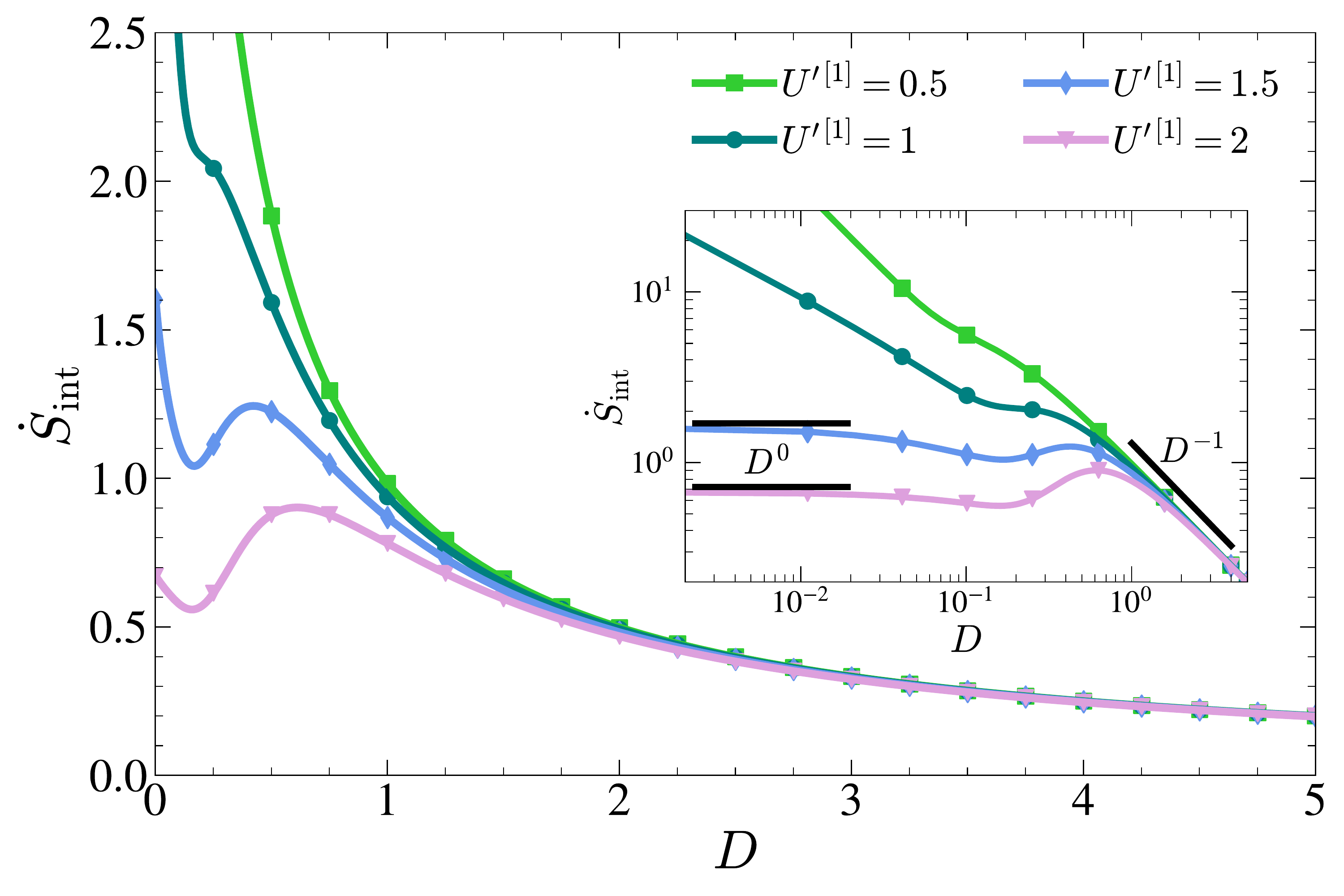}\caption{Steady-state entropy production rate $\dot{S}_{\mathrm{int}}$, Eq.~(\ref{eq:EPR_FinalExpression}), as a function of diffusion $D$ for various magnitudes of the slope $U'^{[1]} = h/a$ for fixed $a = 0.9$ and $f = 0$. The entropy production vanishes as $D \rightarrow \infty$, while for $D \rightarrow 0$, $\dot{S}_\text{int}$ diverges or remains finite depending on whether the particle is confined, i.e.\ whether $v - f < U'^{[1]}$ and $v + f < |U'^{[2]}|$.}\label{fig:EntropyVsDiffusion}
\end{figure}

As is typically the case for RnT particles \cite{razin2020entropy, garcia2021run}, $\dot{S}_{\text{int}}$ monotonically increases with $v$ and diverges as $v \rightarrow \infty$. For $D \rightarrow \infty$, the entropy production vanishes $\dot{S}_{\text{int}} \rightarrow 0$ since any differences between the deterministic components of the forward and time-reversed trajectories of the RnT particle are washed out by diffusive noise. For $v \rightarrow 0$, $\dot{S}_{\text{int}}$ tends to a positive value due to there still being a current from the remaining drift $-f$.

As for the current $J$ in Sec.~\ref{sec:Currents}, the $D \rightarrow 0$ limit of $\dot{S}_{\text{int}}$ depends on whether the particle is confined, i.e.\ whether $v-f<U'^{[1]}$ and $v+f<|U'^{[2]}|$. In typical systems of RnT particles \cite{razin2020entropy, garcia2021run}, the entropy production diverges as $D \rightarrow 0$, since the dynamics become more deterministic and it thus becomes easier to distinguish between forward and time-reversed trajectories. However, in the present case, this effect competes with the increasing confinement of the particles as $D \rightarrow 0$ for $v-f<U'^{[1]}$ and $v+f<|U'^{[2]}|$. This results in the entropy production $\dot{S}_{\text{int}}$ remaining finite in the confined case, but diverging in the unconfined case, as $D \rightarrow 0$, see Fig.~\ref{fig:EntropyVsDiffusion}. It is somewhat remarkable that the confining case results in $\dot{S}_{\text{int}} > 0$ for $D \rightarrow 0$ as one may na\"{i}vely expect $\dot{S}_{\text{int}}$ to vanish if the particles are confined. However, the nonvanishing entropy production as $D \rightarrow 0$ is a result of the confined currents, $J_{R}(x)$ and $J_{L}(x)$, discussed in Sec.~\ref{sec:Currents}. There, it was argued the magnitude of these currents scales approximately as $|J_{R,L}^{[1]}(x)| \sim \gamma\exp(-x/l_{D})$, Eq.~(\ref{eq:CirculatingCurrent}), for $D \rightarrow 0$ on the slope $U'^{[1]}$, where $l_{D} = D/(U'^{[1]}-v + f)$. Inserting Eq.~(\ref{eq:CirculatingCurrent}) into Eq.~(\ref{eq:EPR2}), we obtain for the entropy production in the confined case,
\begin{equation}\label{eq:EPR_D=0}
\begin{split}
    \lim_{D \rightarrow 0} \dot{S}_{\mathrm{int}} &\sim \lim_{D \rightarrow 0}\gamma~\frac{v}{D} \int_{0}^{a} \mathrm{d}x~e^{-\frac{x}{l_{D}}} + \int_{a}^{\ell}\mathrm{d}x~\dots\\
    &\sim \gamma~ \frac{v}{U'^{[1]}-v+f} + \dots~,
\end{split}
\end{equation}
where $\dots$ indicates additional nondiverging contributions from the confined currents on the $i=2$ section of the ratchet, spanning $a \leq x < \ell$. Eq.~(\ref{eq:EPR_D=0}) demonstrates that $\dot{S}_{\text{int}}$ remains finite as $D \rightarrow 0$ in the confined case, see also Fig.~\ref{fig:EntropyVsDiffusion}. 

In Fig.~\ref{fig:EntropyVsDiffusion}, we also observe that $\dot{S}_{\mathrm{int}}$ has two local extrema in its dependence on $D$ in the confined case. As an explanation, we first note in Fig.~\ref{fig:CurrentVsDiffusion} that $J \sim \exp(-H/D) \approx 0$ for small $D$ in the confined case and picks up abruptly as $D$ is increased. Thus, it is expected that $\dot{S}_{\text{int}}$ should initially decrease with $D$ in the confined case since the particle dynamics become less deterministic with no significant gain in the overall current $J$ until diffusion-mediated barrier crossings significantly increase in frequency. As $D$ is increased further, the onset of the rapid increase in $J$ results in a similar increase in $\partial\dot{S}_{\text{int}}/\partial D$, as can be seen by comparing Figs.~\ref{fig:CurrentVsDiffusion} and \ref{fig:EntropyVsDiffusion} (allowing for the minimal effect of the slight difference in external forces $f$ used in each figure). The entropy production $\dot{S}_{\mathrm{int}}$ thus attains a local minimum as $D$ is increased as there becomes an appreciable rise in diffusion-mediated barrier crossings that generate a current $J$. However, as $D$ is increased further, the particle eventually decouples from the potential $U(x)$, Eq.~(\ref{eq:RatchetPotential}), and so $\dot{S}_{\text{int}}$ must decrease with $D$ again as $D$ becomes large. As a result, the entropy production also attains a local maximum in the confined case before eventually decaying like that of a free particle, i.e.\ $\dot{S}_{\mathrm{int}} \sim 1/D$, see inset of Fig.~\ref{fig:EntropyVsDiffusion}.

For $a < \ell/2$, we found the entropy production $\dot{S}_{\text{int}}$ is a monotonically increasing function of the external force $f$ since increasing the external force enhances the existing bias of the system to rectify motion to the left which, in turn, increases the degree to which time-reversal symmetry is broken. In the confined case, $v-f<U'^{[1]}$ and $v+f< |U'^{[2]}|$, for $a > \ell/2$, increasing the external force $f$ initially decreases the overall bias $J>0$ of the system, and so the entropy production $\dot{S}_{\text{int}}$ attains a global minimum at finite $f$ before increasing with $f$ again. As $|f| \rightarrow \infty$, the entropy production tends to $\dot{S}_{\text{int}} \rightarrow f^{2}/D$ as the external force dominates over the ratchet force $U'(x)$ and the self-propulsion $v$.

For $\gamma \rightarrow 0$, the entropy production $\dot{S}_{\text{int}}$ reduces to the average for that of uncoupled drift-diffusive right movers and left movers moving in a potential $U(x) + fx$. The entropy production also monotonically increases with $\gamma$ and asymptotically approaches a value $\dot{S}_{\text{int}} \leq (v^{2} + f^{2})/D$ as $\gamma \rightarrow \infty$.



\section{Power output and efficiency}\label{sec:OptimalWork}

\begin{figure}
    \centering
    \includegraphics[width=0.45\textwidth, trim=0.3cm 0.5cm 0.3cm 0.3cm, clip]{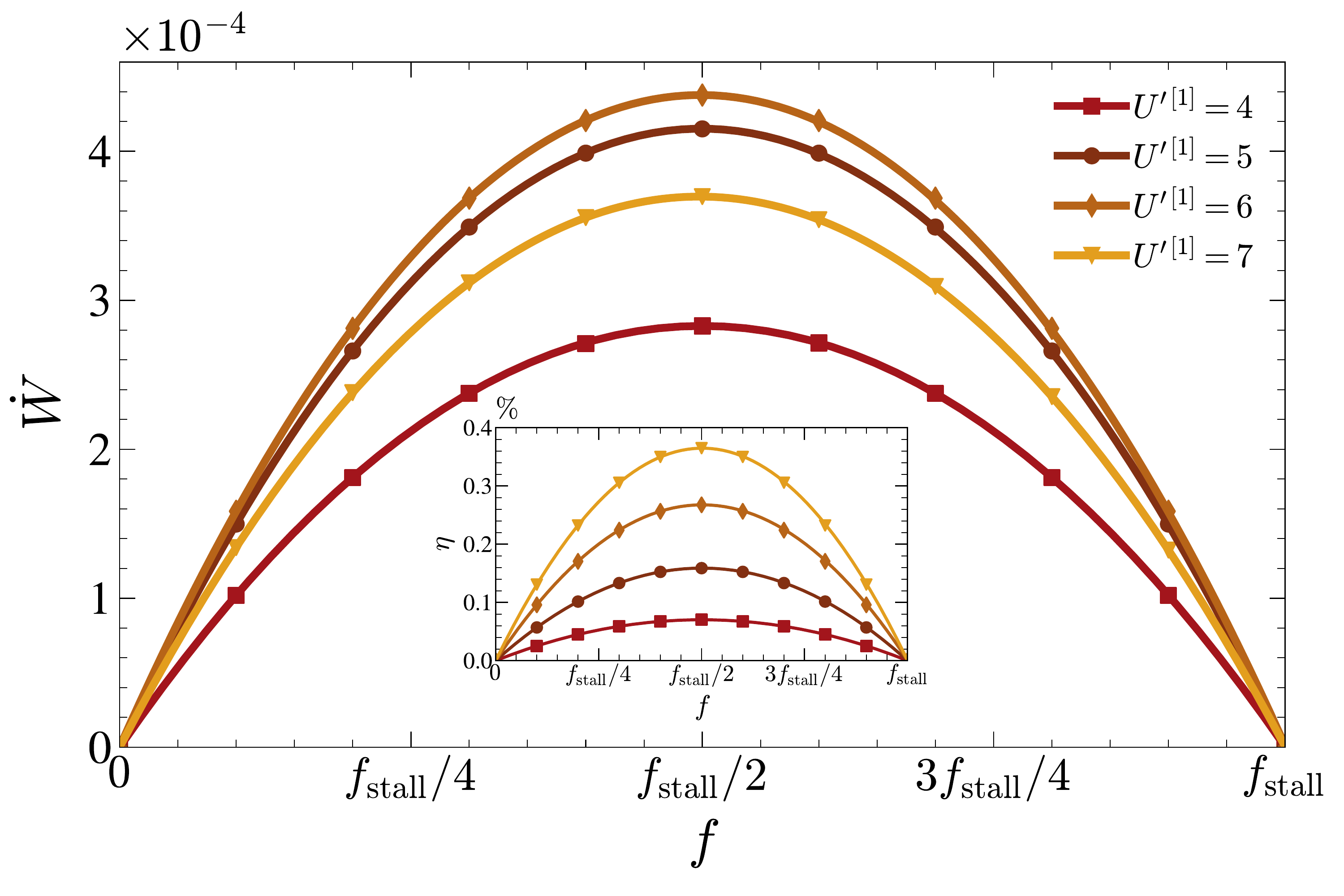}\caption{Steady-state power output $\dot{W}$, Eq.~(\ref{eq:PowerOutput}), as a function of the external force $f$ for various magnitudes of the slope $U'^{[1]} = h/a$ for fixed $a = 0.9$. Maximum power output occurs for a slope $U'^{[1]} \approx 6$ for the default parameters defined in Sec.~\ref{sec:Model}. The power output $\dot{W}$ and thermodynamic efficiency $\eta$ (inset) are nonzero for external forces $0 < f < f_{\mathrm{stall}}$. The stall forces in this case are $f_{\mathrm{stall}} \approx 0.054, 0.083, 0.112, 0.138$ for each $U'^{[1]} = 4,5,6,7$ respectively.}\label{fig:PowerVsForce}
\end{figure}

By applying an external force $f > 0$ (in the negative $x$-direction, see Sec.~\ref{sec:Model}), work $W$ can be extracted from an RnT particle provided $J > 0$. The average rate of work, i.e.\ power, extracted from the particle in the steady state is
\begin{equation}\label{eq:PowerOutput}
     \dot{W} = J\ell f,
\end{equation}
where $\dot{W} > 0$ indicates work is being done \emph{by} the particle. The dependence of $\dot{W}$ on the system parameters is thus the same as for that of the current $J$, Eq.~(\ref{eq:OverallCurrentExplicit}), for constant $f$ and $\ell$. Hence, much of the asymptotic behaviour of the extractable power, Eq.~(\ref{eq:PowerOutput}), can be readily obtained from that of the current $J$ in Sec.~\ref{sec:Currents}, and we therefore omit this repetitive discussion. The only qualitatively different behaviour for $\dot{W}$ is obtained by varying the external force $f$, which is plotted in Fig.~\ref{fig:PowerVsForce} alongside the thermodynamic efficiency $\eta \equiv \dot{W}/P_{\mathrm{in}} = \dot{W}/(\dot{W} + D \dot{S}_{\mathrm{int}})$ \cite{pietzonka2019autonomous} for gradients $U'^{[1]} = h/a$ close to that achieving maximum extractable power for the default parameters defined in Sec.~\ref{sec:Model}. For these particular parameters, the current is approximated by $J \approx J_{0}(1 - f/f_{\mathrm{stall}})$ for $f_{\mathrm{stall}} \neq 0$, where $f_{\mathrm{stall}}$ is the stall force at which the current vanishes, $J = 0$, i.e.\ the point at which the force from the external load exactly cancels the current generated by the rectified motion. Hence, for these parameters where the approximation $J \approx J_{0}(1 - f/f_{\mathrm{stall}})$ holds, the maximum extractable power $\dot{W}$, Eq.~(\ref{eq:PowerOutput}), and in turn the maximum efficiency $\eta$, occur for $f \approx f_{\mathrm{stall}}/2$, as seen in Fig.~\ref{fig:PowerVsForce}. In the absence of an external force $f=0$, no work $W$ is produced and the efficiency of the power extraction is thus also $\eta = 0$. For $f < 0$, work is done \textit{on} the particle and so $\dot{W} < 0$ and $\eta < 0$. For $f = f_{\mathrm{stall}}$, we have $\dot{W} = 0$ and $\eta = 0$ because of the absence of a current $J =0$, and for forces greater than the stall force $f > f_{\mathrm{stall}}$, work is done \textit{on} the particle because the current generated by the force $f$ outweighs that generated by the rectification, causing the particle to be ``dragged backwards''. As $|f| \rightarrow \infty$, the power $\dot{W} \rightarrow -f^{2}$ and thus the efficiency $\eta \rightarrow -\infty$ since $D\dot{S}_{\text{int}} \rightarrow f^{2}$, see Sec.~\ref{sec:EntropyProduction}.

\begin{figure}
    \centering
    \includegraphics[width=0.45\textwidth, trim=0.3cm 0.5cm 0.3cm 0.3cm, clip]{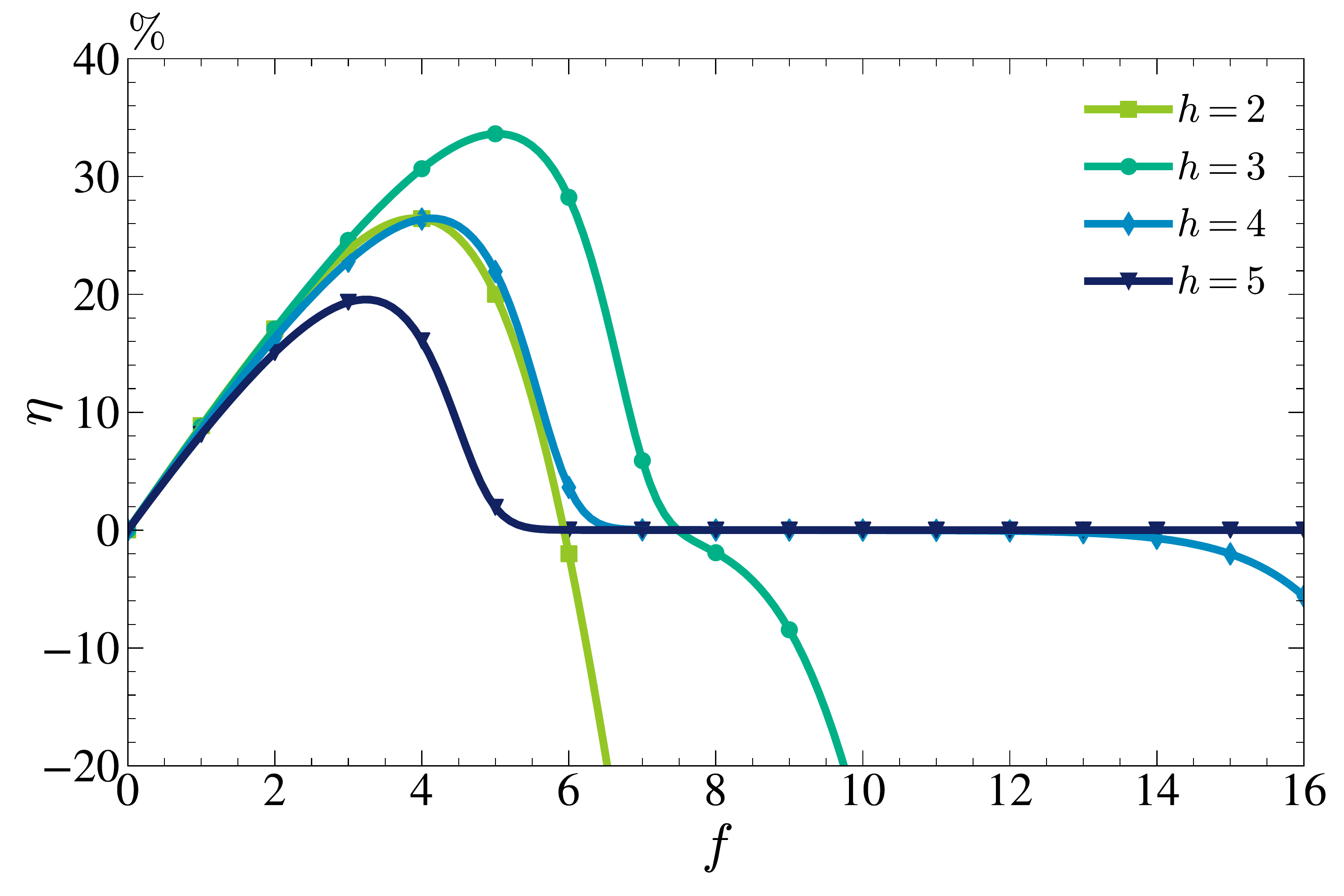}\caption{Steady-state efficiency $\eta$ as a function of the external force $f$ for various heights $h$ of the ratchet and for particle parameters $v=10$, $\gamma = 1$, and $D = 0.1$. For $h \gtrsim 4$, there is a range of external forces for which $\eta \approx 0$.}\label{fig:HighEfficiencyVsForce}
\end{figure}

In Fig.~\ref{fig:PowerVsForce}, the efficiency is small $\eta < 1\%$ for the default parameter values defined in Sec.~\ref{sec:Model} since, in this confined case, a current $J$ can be generated only by rare diffusion-mediated barrier crossings. However, the efficiency $\eta$ is vastly improved in the case where the particle is not confined by the potential $U(x)$, Eq.~(\ref{eq:RatchetPotential}), and for small diffusion $D$. Using our default system parameters $L=1$ and $a=0.9$, this is illustrated in Fig.~\ref{fig:HighEfficiencyVsForce} for particle parameters $v=10$, $\gamma = 1$, and $D = 0.1$, corresponding to a P\'{e}clet number $\mathrm{Pe} = v^{2}/(D\gamma) = 1000$ on the order of that of an \textit{E.\ coli} bacterium \footnote{The typical run speed and tumble rate of an \textit{E.\ coli} are $v \approx 15~\mu\mathrm{m}~\mathrm{s}^{-1}$ and $\gamma \approx 1~\mathrm{s}^{-1}$ respectively \cite{molaei2014failed}. The \emph{thermal} diffusion constant $D$ of an \textit{E.\ coli} can be calculated from the Stokes-Einstein equation $D = k_{B}T/(6\pi\eta_{\text{vis}} r)$. By approximating an \textit{E.\ coli} as a sphere of radius $r = 1~\mu\mathrm{m}$ immersed in water (of dynamic viscosity $\eta_{\text{vis}} \approx 10^{-3}~\mathrm{Pa}~\mathrm{s}$), this gives $D \approx 0.2~\mu\mathrm{m}^{2}~\mathrm{s}^{-1}$ at room temperature $T = 300~\mathrm{K}$. Thus, the P\'{e}clet number of an \textit{E.\ coli} is on the order of $\mathrm{Pe} = v^{2}/(D\gamma) \approx 1000$}. These parameters are seen to result in a much higher efficiency of $\eta \approx 35\%$ for $h = 3$ and $f \approx 5$. However, these parameters are by no means those that result in the maximum possible efficiency $\eta$ for this system. In fact, arbitrarily high efficiencies can be obtained by taking $D \to 0$ and $\gamma \to 0$ while choosing the ratchet parameters in such a way that left movers are confined, $v+f < |U'^{[2]}|$, but right movers are not, $v-f > U'^{[1]}$, and choosing an external force $f$ slightly smaller than the stall force $f_{\mathrm{stall}}$ to minimise dissipation $D\dot{S}_{\mathrm{int}}$ while maintaining positive power output $\dot{W}$. In this case, there is a ``ratchet-and-pawl" effect where, in the absence of tumbling $\gamma = 0$, right movers produce useful work $W$ against the external force $f$, while left movers are confined to $x=0$ and therefore do not contribute negatively to the efficiency $\eta$. This highlights, however, the limitations of so-called ``dry" active matter considered here, for which hydrodynamic interactions are neglected and thus particles do not dissipate energy while stuck at walls \cite{Marchetti2013Jul}. We expect the efficiency $\eta$ of this system would be lower for the more realistic case of ``wet" active matter where momentum is conserved through hydrodynamic interactions.

In Fig.~\ref{fig:HighEfficiencyVsForce}, for $h = 4$ and $h=5$, we observe a range of external forces $f$ for which the efficiency is $\eta \approx 0$. This is possible for external forces that satisfy $f_{\mathrm{stall}} < f < h/(\ell -a) - v$, as these are larger than the stall force, resulting in motion to the right being suppressed, but still small enough that the combination of the external force $f$ and self-propulsion $v$ is not enough to overcome the potential barrier to the left, and so motion to the left is also suppressed. In the case of small but nonvanishing diffusion $D$ in Fig.~\ref{fig:HighEfficiencyVsForce}, there is still an appreciable chance of the particle overcoming the barrier as a result of diffusive fluctuations and so the range of forces $f$ for which $\eta \approx 0$ does not extend the \textit{entire} way to $f = h/(\ell -a) - v$.

\begin{figure}
        \centering
       \includegraphics[width=0.4\textwidth, trim=4.4cm 1cm 5.4cm 0cm, clip]{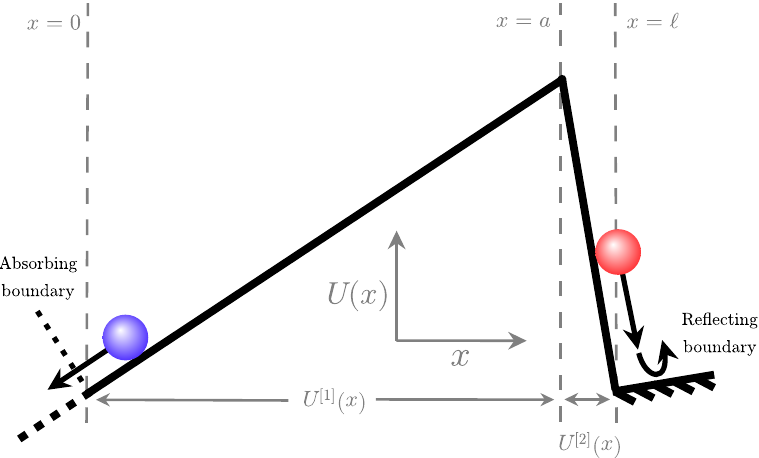}\caption{Schematic representation of configuration (ii) in Sec.~\ref{sec:FirstPassage}, used to study the mean first-passage time of RnT particles in a ratchet potential with a reflecting boundary at $x=\ell$ and an absorbing boundary at $x=0$. Configuration (i) is the same as configuration (ii) but with the reflecting boundary at $x=\ell$ replaced by an absorbing boundary.}\label{fig:MFPTSchematic}
    \end{figure}

Regarding the maximum power $\dot{W}$, Eq.~(\ref{eq:PowerOutput}), that can be extracted from this system, it is known in the case of $D=0$ that maximum current $j$, and hence maximum power $\dot{W}$, is achieved by taking $h \rightarrow 0$ and $\ell \rightarrow 0$ for fixed $a = \ell - h/(v + f) > \ell/2$ \cite{angelani2011active}, corresponding to an optimal ratchet on a ring of vanishing circumference that a particle circumnavigates infinitely quickly. However, this diverging $D=0$ current $j$ fails to take into account the finite size of the particles which would restrict the ratchet height $h$ and length $\ell$ from becoming too small. Instead, for given particle parameters $v$ and $\gamma$, and physically appropriate values of $h$ and $\ell$, one could find the optimal power $\dot{W}$ in the $D = 0$ case by using that maximum current occurs for an apex position $a = \ell - h/(v+f)$, which leaves the external force $f$ as the only remaining parameter to be optimised.



\section{First-passage properties}\label{sec:FirstPassage}

In this section, we calculate the mean first-passage times $\tau_{R,L}(x)$ and splitting probabilities $\Pi_{R,L}(x)$ for an RnT particle in a ratchet potential $U(x)$, Eq.~(\ref{eq:RatchetPotential}), on a bounded interval $x \in [0,\ell]$. Firstly, the mean first-passage times can be calculated for the cases of (i) two absorbing boundary conditions at $x = 0$ and $x = \ell$, and (ii) a single reflecting boundary at $x = \ell$ and an absorbing boundary at $x = 0$, Fig.~\ref{fig:MFPTSchematic}. We define the mean first-passage times $\tau_{R/L}(x)$ as the average time it takes the RnT particle to be absorbed at a boundary, and thus leave the system, given the particle was initialised in the right/left-moving state at position $x \in [0,\ell]$. The mean first-passage times can be shown to satisfy the backward equations \cite{van1992stochastic}
\begin{widetext}
\begin{subequations}\label{eq:MFPT_DifferentialEquation}
\begin{alignat}{2}
-1 &= D\frac{\mathrm{d}^{2} \tau_R^{[i]}(x)}{\mathrm{d} x^{2}}+\left(v-f-U'^{[i]}\right)\frac{\mathrm{d}\tau_R^{[i]}(x)}{\mathrm{d} x}-\gamma\left(\tau_R^{[i]}(x)-\tau_L^{[i]}(x)\right), \label{eq:MFPT_DifferentialEquationRightMovers}\\
-1 &= D \frac{\mathrm{d}^{2}\tau_L^{[i]}(x)}{\mathrm{d} x^{2}} +\left(-v-f-U'^{[i]}\right)\frac{\mathrm{d}\tau_L^{[i]}(x)}{\mathrm{d} x}-\gamma\left(\tau_L^{[i]}(x)-\tau_R^{[i]}(x)\right), \label{eq:MFPT_DifferentialEquationLeftMovers}
\end{alignat}
\end{subequations}
with boundary conditions (i) $\tau_{R,L}^{[1]}(0) = \tau_{R,L}^{[2]}(\ell) = 0$, and (ii) $\tau_{R,L}^{[1]}(0) = 0$ and $\partial_{x}\tau_{R,L}^{[2]}(x)|_{x=\ell} = 0$, for each system configuration. The solutions for the mean first-passage time have the form
\begin{equation}\label{eq:MeanFirstPassageGeneralSoln}
\tau_{R,L}^{[i]}(x)=
 \tilde{\mathcal A}_{R,L}^{\ABindex{i}}e^{\tilde\lambda_{\mathcal{A}}^{\ABindex{i}}x}
+ \tilde{\mathcal B}_{R,L}^{\ABindex{i}}e^{\tilde\lambda_{\mathcal{B}}^{\ABindex{i}}x}
+ \tilde{\mathcal C}_{R,L}^{\ABindex{i}}e^{\tilde\lambda_{\mathcal{C}}^{\ABindex{i}}x}
+ \tilde{\mathcal D}_{R,L}^{\ABindex{i}}+\frac{x}{f+U'^{[i]}}~,
\end{equation}
\end{widetext}
where, compared to the general solution for the probability densities $P_{R,L}^{[i]}(x)$ in Eq.~(\ref{eq:ParticleDensityGeneralSoln}), an additional term that is linear in $x$ arises from the particular solution to the nonhomogeneous Eq.~(\ref{eq:MFPT_DifferentialEquation}), see Appendix~\ref{app:DerivationMeanFirstPassageSplitProb}. Again, as for the probability densities $P_{R,L}^{[i]}(x)$ in Sec.~\ref{sec:ParticleDensities}, we fix the 16 coefficients $\tilde{\mathcal{A}}_{R}^{[1]}, \tilde{\mathcal{A}}_{L}^{[1]}, \tilde{\mathcal{A}}_{R}^{[2]}, \dots, \tilde{\mathcal{D}}_{L}^{[2]}$ by substituting each linearly independent term in Eq.~(\ref{eq:MeanFirstPassageGeneralSoln}) into Eq.~(\ref{eq:MFPT_DifferentialEquation}), before applying boundary conditions and continuity conditions, see Appendix~\ref{app:DerivationMeanFirstPassageSplitProb}.

For configuration (i), the presence of two absorbing boundaries allows us to also calculate the splitting probabilities $\Pi_{R,L}(x)$ for the system. The splitting probabilities are the probabilities of a particle being absorbed at the left-hand boundary $x = 0$ (before being absorbed at the right-hand boundary $x=\ell$) given it was initialised as a right mover $\Pi_{R}(x)$ or as a left mover $\Pi_{L}(x)$ at position $x \in [0,\ell]$. The splitting probabilities at the opposite boundary $x=\ell$ are readily obtained from $1 - \Pi_{R,L}(x)$ and their discussion is therefore omitted. Similarly to the mean first-passage times $\tau_{R,L}(x)$, Eq.~(\ref{eq:MeanFirstPassageGeneralSoln}), the splitting probabilities obey the backward equations \cite{van1992stochastic}
\begin{widetext}
\begin{subequations}\label{eq:SplittingProb_DifferentialEquation}
\begin{alignat}{2}
0 &= D\frac{\mathrm{d}^{2} \Pi^{[i]}_R(x)}{\mathrm{d} x^{2}}+\left(v-f-U'^{[i]}\right)\frac{\mathrm{d}\Pi^{[i]}_R(x)}{\mathrm{d} x}-\gamma\left(\Pi^{[i]}_R(x)-\Pi^{[i]}_L(x)\right), \label{eq:SplittingProb_DifferentialEquationRightMovers}\\
 0&= D \frac{\mathrm{d}^{2}\Pi^{[i]}_L(x)}{\mathrm{d} x^{2}} +\left(-v-f-U'^{[i]}\right)\frac{\mathrm{d}\Pi^{[i]}_L(x)}{\mathrm{d} x}-\gamma\left(\Pi^{[i]}_L(x)-\Pi^{[i]}_R(x)\right), \label{eq:SplittingProb_DifferentialEquationLeftMovers}
\end{alignat}
\end{subequations}
\end{widetext}
which are homogeneous, meaning the solutions $\Pi^{[i]}_{R,L}(x)$ to Eq.~(\ref{eq:SplittingProb_DifferentialEquation}) can be obtained in a similar manner to that of the homogeneous Eq.~(\ref{eq:FokkerPlanckEquation}), i.e\
\begin{equation}\label{eq:SplittingProbabilitiesGeneralSoln}
\Pi_{R,L}^{[i]}(x)=
 \check{\mathcal A}_{R,L}^{\ABindex{i}}e^{\tilde\lambda_{\mathcal{A}}^{\ABindex{i}}x}
+ \check{\mathcal B}_{R,L}^{\ABindex{i}}e^{\tilde\lambda_{\mathcal{B}}^{\ABindex{i}}x}
+ \check{\mathcal C}_{R,L}^{\ABindex{i}}e^{\tilde\lambda_{\mathcal{C}}^{\ABindex{i}}x}
+ \check{\mathcal D}_{R,L}^{\ABindex{i}},
\end{equation}
where expressions for the 16 coefficients $\check{\mathcal{A}}_{R}^{[1]}, \check{\mathcal{A}}_{L}^{[1]}, \check{\mathcal{A}}_{R}^{[2]}, \dots, \check{\mathcal{D}}_{L}^{[2]}$ are found by applying the boundary conditions $\Pi_{R,L}^{[1]}(0) = 1$ and $\Pi_{R,L}^{[2]}(\ell) = 0$, as well as analogous continuity conditions to that of the mean first-passage times $\tau_{R,L}^{[i]}(x)$, see Appendix~\ref{app:DerivationSplitProb}.
\begin{figure*}
\subfloat{\includegraphics[width=0.66\columnwidth, trim=0cm 0.7cm 0.4cm 0.15cm, clip]{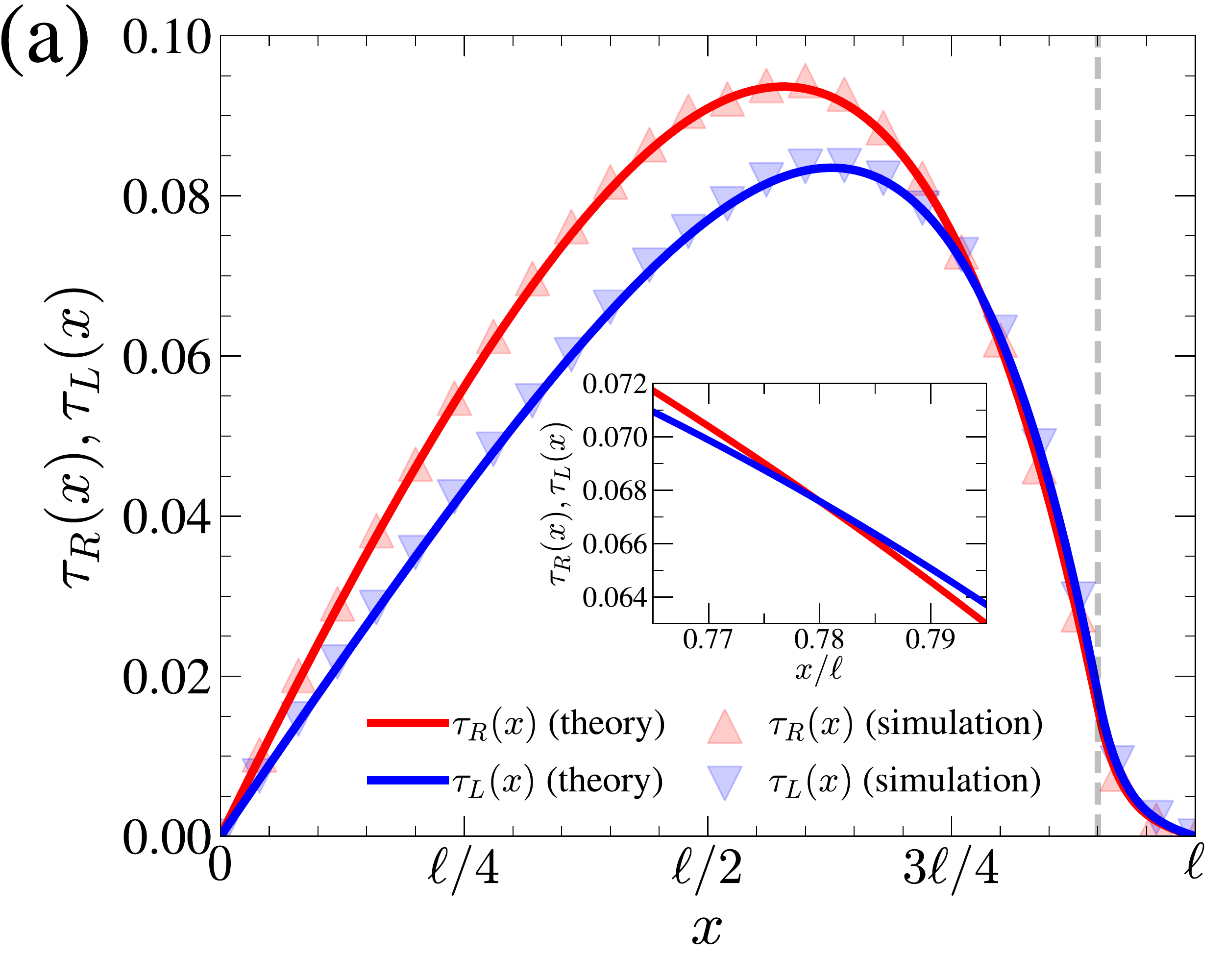}
}\hfill
\subfloat{\includegraphics[width=0.66\columnwidth, trim=0.15cm 0.7cm 0.4cm 0.1cm, clip]{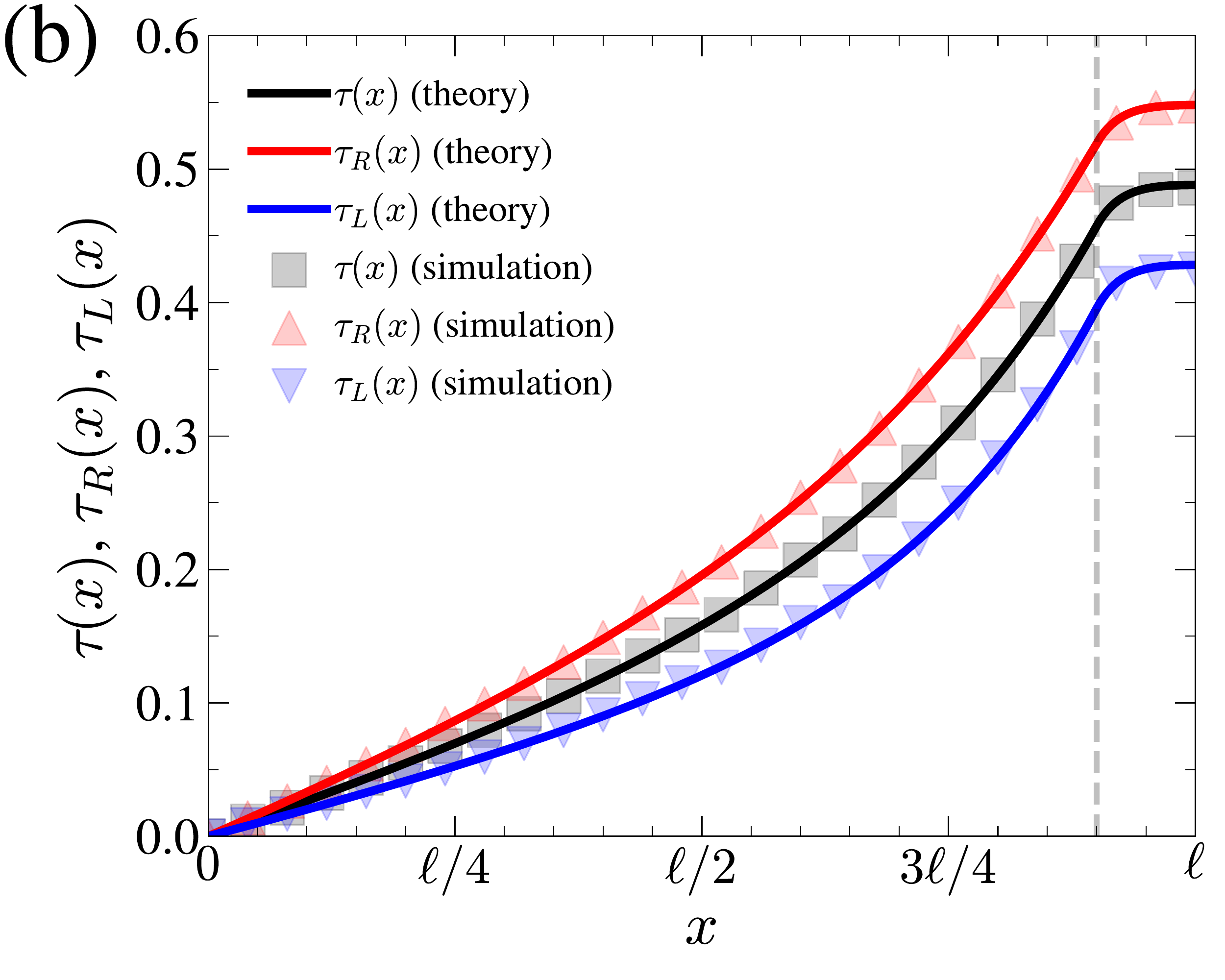}
}\hfill
\subfloat{\includegraphics[width=0.66\columnwidth, trim=0.15cm 0.7cm 0.4cm 0.15cm, clip]{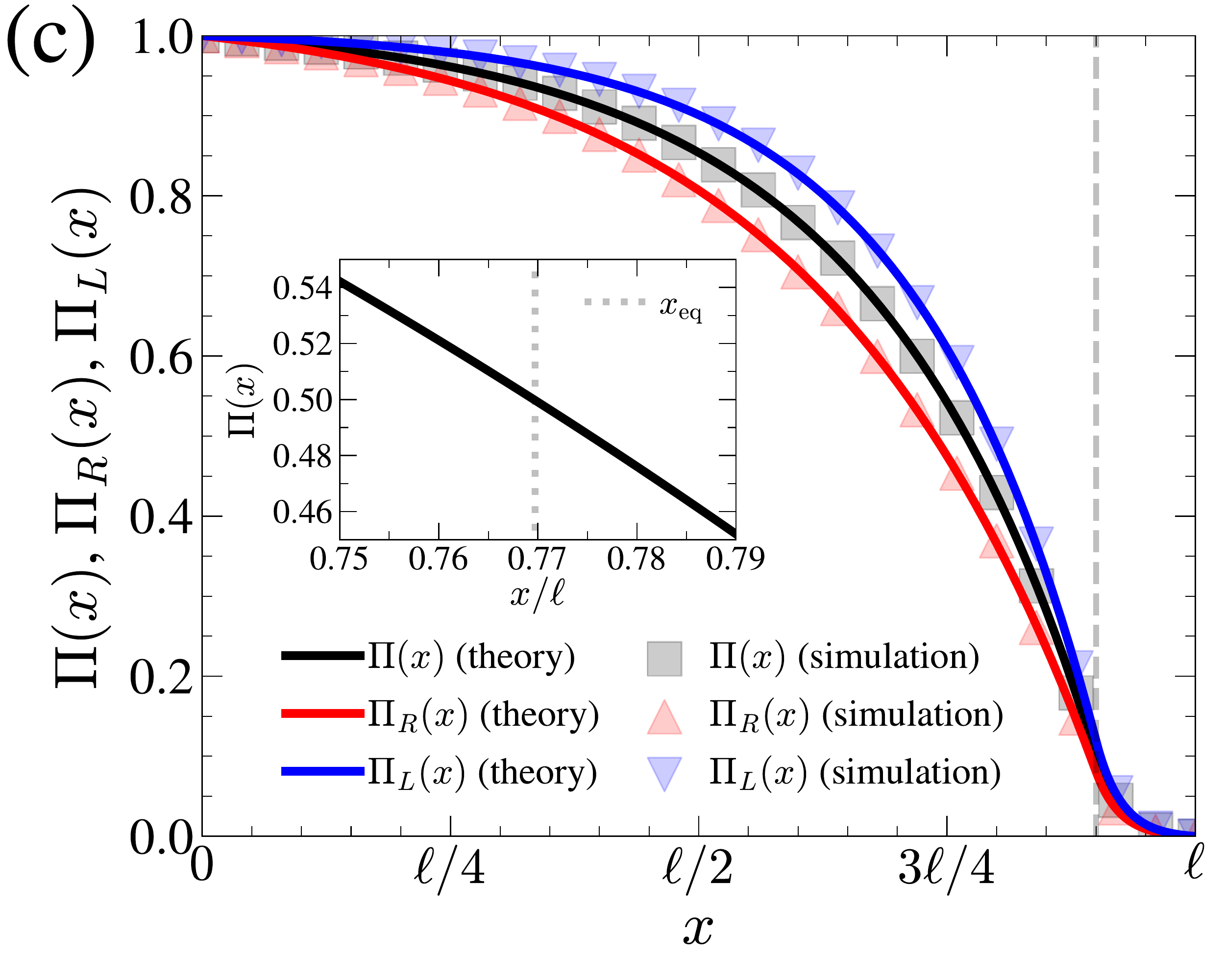}
}
\caption{Comparison between simulations (markers) and theoretical results (lines) for mean first-passage times $\tau_{R/L}(x)$, Eq.~(\ref{eq:MeanFirstPassageGeneralSoln}), in the case of (a) absorbing boundaries at $x=0$ and $x=\ell$, and (b) an absorbing boundary at $x=0$ and a reflecting boundary at $x = \ell$, and for (c) splitting probabilities $\Pi_{R/L}(x)$, Eq.~(\ref{eq:SplittingProb_DifferentialEquation}), for initialisation of an RnT particle as a right/left mover. As in Fig.~\ref{fig:ConvergenceSteadyStateDensity}, simulations were performed by numerically integrating the Langevin equation (\ref{eq:Langevin}) in timesteps of $\mathrm{d}t = 10^{-5}$. The simulation data were obtained from the average time it takes the particle to escape the interval, in the case of (a) and (b), and the proportion of times it escapes through $x=0$, in the case of (c), for $10^{5}$ realisations at each $x/\ell = 0, 0.01, \dots, 1$. The vertical dashed lines indicate the position of the ratchet apex, $x=a$. The theoretical results, Eqs.~(\ref{eq:MeanFirstPassageGeneralSoln}) and (\ref{eq:SplittingProbabilitiesGeneralSoln}), are in good agreement with simulations.}
\label{fig:MFPTandSplittingProbability}
\end{figure*}
\begin{figure*}
\subfloat{%
  \includegraphics[width=0.66\columnwidth, trim=0.5cm 0.7cm 0.4cm 0.1cm, clip]{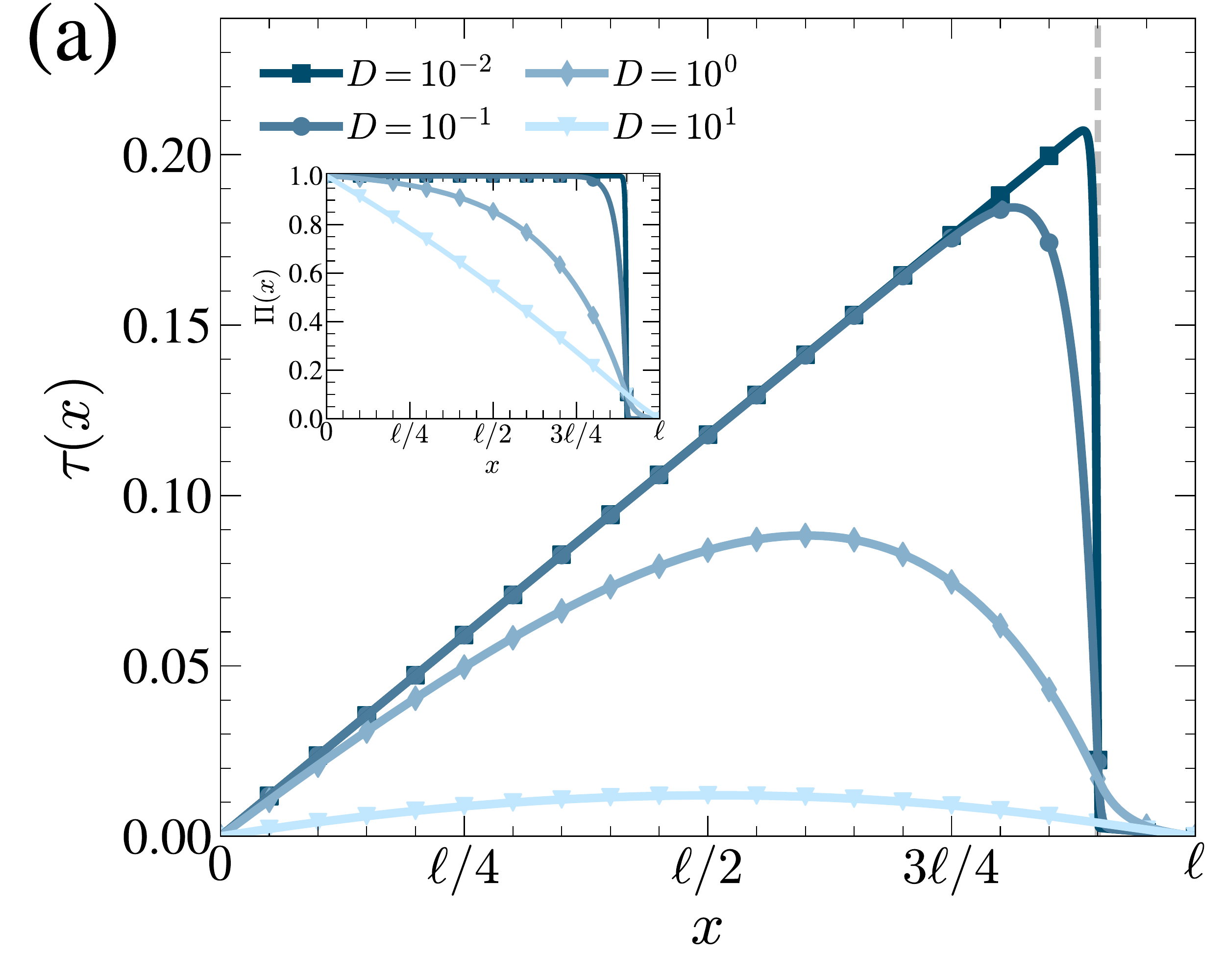}%
}\hfill
\subfloat{%
  \includegraphics[width=0.66\columnwidth, trim=0.5cm 0.7cm 0.4cm 0.1cm, clip]{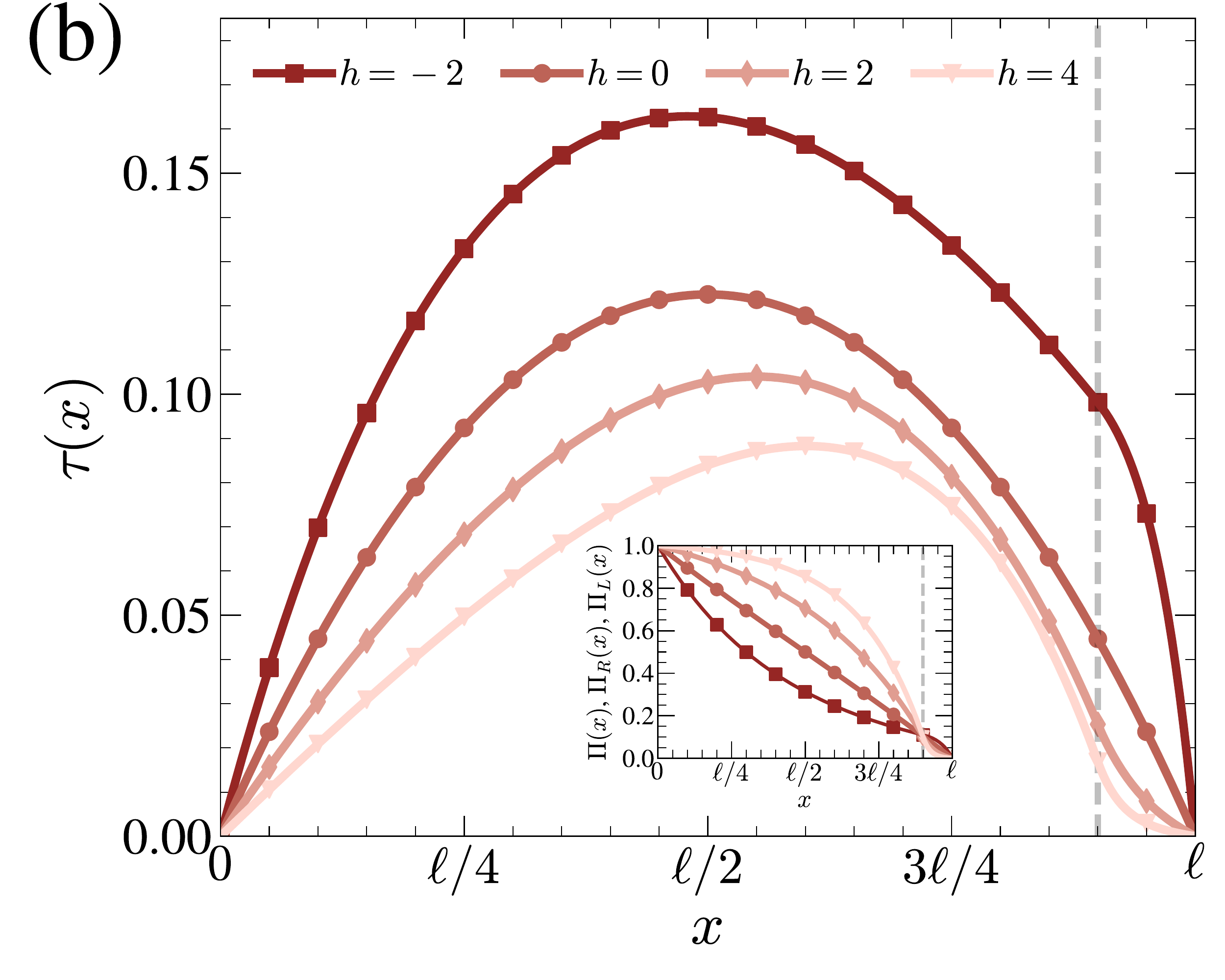}%
}\hfill
\subfloat{%
  \includegraphics[width=0.66\columnwidth, trim=0.5cm 0.7cm 0.4cm 0.1cm, clip]{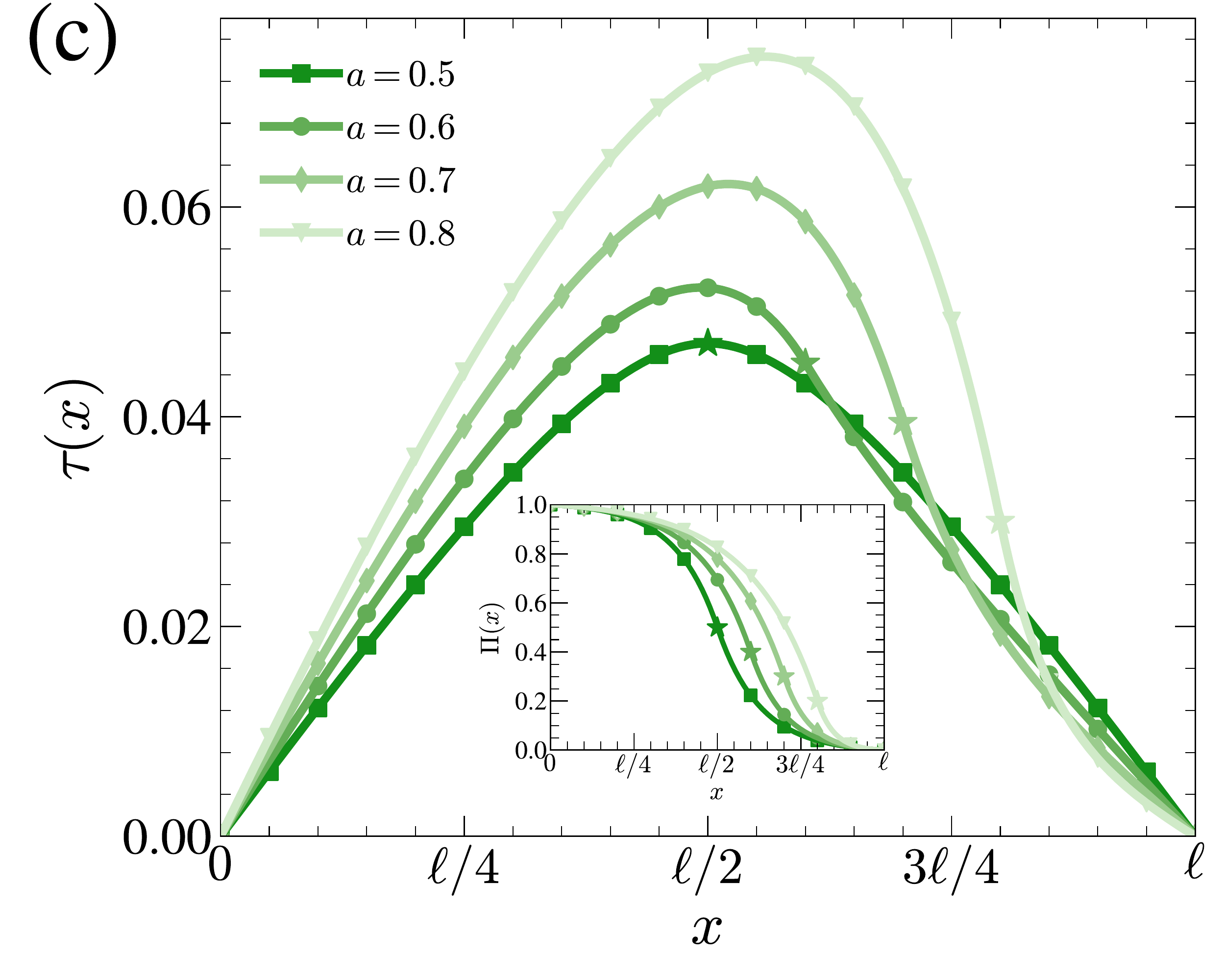}%
}
\caption{Mean first-passage times $\tau(x)$, Eq.~(\ref{eq:MeanFirstPassageGeneralSoln}), and corresponding splitting probabilities $\Pi(x)$ (insets), Eq.~(\ref{eq:SplittingProbabilitiesGeneralSoln}), for varying (a) the diffusion constant $D$, (b) the height of the ratchet $h$, and (c) the position of the ratchet peak $a$, in the case of absorbing boundaries at $x=0$ and $x=\ell$. In (a) and (b), vertical dashed lines indicate the position of the ratchet apex, $x=a$. In (c), the apexes are indicated by star data points for each curve.}
\label{fig:MFPTsVaryingParameters}
\end{figure*}

In Fig.~\ref{fig:MFPTandSplittingProbability}(a), we plot the mean first-passage times $\tau_{R,L}(x)$, Eq.~(\ref{eq:MeanFirstPassageGeneralSoln}), using our default parameter values, Sec.~\ref{sec:Model}, for absorbing boundaries at $x=0$ and $x=\ell$. Both $\tau_{R}(x)$ and $\tau_{L}(x)$ are skewed to the right compared to a symmetric distribution, which is the case for a Brownian particle in free space, i.e.\ the maximum mean first-passage times both occur for $x > \ell/2$, due to the asymmetry $a > \ell/2$ of the system. This asymmetry becomes more pronounced as the distance $|a - \ell/2|$ from the symmetric potential is increased, see Fig.~\ref{fig:MFPTsVaryingParameters}(c). The ratchet breaks the symmetry in the mean first-passage times $\tau_{R,L}(x)$ of an RnT particle in the absence of a potential, which satisfies $\tau_{R}(x) = \tau_{L}(\ell-x)$ \cite{Malakar2018Apr}, but is restored upon taking $h \rightarrow 0$, Fig.~\ref{fig:MFPTsVaryingParameters}(b). As a result of $a > \ell/2$, a particle initialised on the slope $U'^{[1]}$, i.e.\ at $x \in [0,a)$, experiences an average force to the left, assuming $v-f < U'^{[1]}$, and thus has a higher chance to escape through the left-hand boundary $x=0$ than for the free case. This is seen in the splitting probabilities $\Pi(x)$ plotted in Fig.~\ref{fig:MFPTandSplittingProbability}(c), where an equal chance of escaping through either boundary, i.e.\ $\Pi(x) = 0.5$, occurs at $x = x_{\text{eq}} \approx 3\ell/4$. For our default parameters, Sec.~\ref{sec:Model}, this point $x_{\text{eq}}$ also approximately coincides with the point where the mean first-passage times for right movers and left movers are the same, i.e.\ $\tau_{R}(x_{\text{eq}}) \approx \tau_{L}(x_{\text{eq}})$, Fig.~\ref{fig:MFPTandSplittingProbability}(a). As $D$ is increased, the effect of the ratchet on the mean first-passage times $\tau_{R,L}(x)$ is diminished, resulting in an average time to exit the interval that is symmetric about $x=\ell/2$, Fig.~\ref{fig:MFPTsVaryingParameters}(a). Similarly, the large diffusion limit $D \to \infty$ results in the familiar form $\Pi(x) = 1 - x/\ell$ for the splitting probability, see inset of Fig.~\ref{fig:MFPTsVaryingParameters}(a).

The bounded interval also allows us to explore the first-passage properties for the case $h<0$, which is no longer degenerate, as in Secs.~\ref{sec:Model}-\ref{sec:OptimalWork}, since the boundaries break translational symmetry. The resulting inverted potential for $h < 0$ has the character of a confining potential well, rather than a potential barrier, similarly to the potential $U(x) \sim |x|$ considered in Refs.~\cite{dhar2019run, demaerel2018active}. However, in the present case, our framework allows us to explore a more general asymmetric variation of this piecewise-linear potential well. As seen in Fig.~\ref{fig:MFPTsVaryingParameters}(b), $h<0$ results in an overall increase in the mean first-passage time $\tau(x)$, as well as concave distributions in the region $a \leq x \leq \ell$ for both $\tau(x)$ and the splitting probability $\Pi(x)$, compared to the convex curves that occur in this region when $h>0$ due to the favourable gradient towards the right-hand boundary.

For configuration (ii), where the absorbing boundary at $x=\ell$ is replaced with a reflecting boundary, the mean first-passage time of a right mover is trivially greater than that of a left mover for any initial position due to the initial self-propulsion of a right mover always being directed away from the only available exit to the interval, Fig.~\ref{fig:MFPTandSplittingProbability}(b). This is also reflected in the splitting probabilities $\Pi_{R,L}(x)$ in Fig.~\ref{fig:MFPTandSplittingProbability}(c), where the probability for the particle to exit at the left-hand boundary at $x=0$ is always greater for an initial left mover than an initial right mover.

Finally, for completeness, in Appendix~\ref{app:ConditionalSPderivation} we demonstrate how to derive the ``conditional splitting probabilities'' $\pi_{X|Y}(x_{0})$ for a particle initialised in state $Y \in \{R,L\}$ at $x = x_{0}$ to leave through the left-hand boundary at $x=0$ in the self-propulsion state $X \in \{R,L\}$. This generalises what was first derived in Ref.~\cite{Cocconi2022Nov} to incorporate piecewise-linear potentials $U(x)$, Eq.~(\ref{eq:RatchetPotential}). In the free case \cite{Cocconi2022Nov}, conditional splitting probabilities allow for the study of particles that exit an interval via diffusive fluctuations even when the particle's self-propulsion favours escape at the opposite boundary, e.g.\ when a right mover exits through the left-hand boundary at $x=0$. The derivation of the conditional splitting probabilities here can be used to study the combined role that diffusion and a potential play in this phenomenon.

\section{Conclusion}\label{sec:Conclusion}

We have studied in detail the motion of run-and-tumble particles in a piecewise-linear ``ratchet" potential $U(x)$, Eq.~(\ref{eq:RatchetPotential}), that are also subject to an external counterforce $f$ for the purpose of extracting useful work. We derived the stationary probability densities $P_{R,L}(x)$, Eq.~(\ref{eq:ParticleDensityGeneralSoln}), in Sec.~\ref{sec:ParticleDensities} and Appendix~\ref{app:DerivationParticleDensity} for both right-moving and left-moving particle species, Fig.~\ref{fig:DensityCurrentPotential}(a), and confirmed the numerically simulated time-dependent densities converge to these expressions as $t \rightarrow \infty$, Fig.~\ref{fig:ConvergenceSteadyStateDensity}.

From the densities $P_{R,L}(x)$, we calculated the corresponding steady-state currents $J_{R}(x)$, $J_{L}(x)$, and $J = J_{R}(x) + J_{L}(x)$, Eq.~(\ref{eq:OverallCurrentExplicit}), in Sec.~\ref{sec:Currents} and analysed how they depend on the system's parameters. We found the current $J$ peaks at a finite value of the ratchet potential height $h$ and decays to zero as $h \to \infty$, Fig.~\ref{fig:CurrentVsHeight}. The current $J$ displays similar nonmonotonic dependencies on the magnitude of the particle's self-propulsion speed $v$ and the translational diffusion $D$. Depending on the strength of the diffusion $D$, the position of the apex $a$ resulting in maximum current $J$ occurs between $a = \ell - h/(v+f)$ for $D = 0$ and $a \rightarrow \ell^{-}$ for large $D$, Fig.~\ref{fig:MaxCurrentPositionVsDiffusion}. As $D \rightarrow 0$, the current $J$ either remains finite or vanishes depending on whether the particle is confined, Fig.~\ref{fig:CurrentVsDiffusion}, i.e.\ whether $v-f < h/a$ and $v+f < h/(\ell -a)$, representing confinement conditions for right movers and left movers respectively. Simple closed-form expressions were obtained for the $D=0$ current in both the unconfined, Eq.~(\ref{app:eq:D=0_CurrentExactExpression_Unconfined}), and confined, Eq.~(\ref{app:eq:D=0_current_leftconfined}), cases in Appendix~\ref{app:DerivationParticleDensity_D=0}. The currents for each particle species, $J_{R}(x)$ and $J_{L}(x)$, are greater in magnitude close to the potential minimum at $x=0$ due to particles that tumble before crossing the potential barrier, see Fig.~\ref{fig:DensityCurrentPotential}(b). This phenomenon is still present even when particles are confined because of weaker ``effective" potentials that particles experience as a result of their self-propulsion, Fig.~\ref{fig:BoostedCurrent}. While many ratchet models exhibit current reversals \cite{Reimann1996Dec}, where particles preferentially flow against the steeper slope of the potential, we do not observe any such phenomenon in the present system. This is in agreement with the findings of Ref.~\cite{Ai2017Mar}, where there is no current reversal for overdamped active particles moving in a piecewise-linear potential, but there is for underdamped active particles. Hence, introducing inertia into the present system could result in a current reversal and thus allow power extraction from a ``negative" load, $f<0$, i.e.\ an external force applied in the opposite direction to that considered here, which could be explored in future work.

In Sec.~\ref{sec:EntropyProduction}, we derived the steady-state entropy production rate $\dot{S}_{\mathrm{int}}$, Eq.~(\ref{eq:EPR_FinalExpression}), and entropy production densities $\dot{s}_{R}(x)$, $\dot{s}_{L}(x)$, and $\dot{s}_{R \leftrightarrow L}(x)$, Eqs.~(\ref{eq:EPR_Density_RightMovers}) and (\ref{eq:EPR_DensityTransitions}), for each particle species. The total entropy production density $\dot{s}_{\mathrm{int}}(x) = \dot{s}_{R}(x) + \dot{s}_{L}(x) + \dot{s}_{R \leftrightarrow L}(x)$ is greatest at the apex $x=a$ of the ratchet where the mean local velocity of particles is greatest, Fig.~\ref{fig:EntropyProductionDensity}. The steady-state entropy production $\dot{S}_{\mathrm{int}}$ was found to decrease monotonically with the height $h$ of the potential, Fig.~\ref{fig:entropy_against_h}. Remarkably, the entropy production $\dot{S}_{\mathrm{int}}$ can have a nonmonotonic dependence on $D$ and, in the limit $D \rightarrow 0$, remains finite when the particle is confined due to the resulting scaling of the currents for each particle species, Fig.~\ref{fig:EntropyVsDiffusion}. We would expect to see similar behaviour for the entropy production of run-and-tumble particles in confining potentials of the form $U_{p}(x) \sim |x|^{p}$ for $p = 1$, studied in Ref.~\cite{dhar2019run} for the case $D = 0$. Therefore, studying the entropy production of run-and-tumble particles subject to potentials of the form $U_{p}(x)\sim |x|^{p}$ for $D > 0$ would be an interesting avenue for further research. It would also be interesting to consider smooth ratchet potentials, where particles can become confined to a finite region rather than a single point, to verify whether entropy production diverges in the confined case as $D \rightarrow 0$. Furthermore, throughout this paper we have calculated average values for our observables of interest. A study of higher-order moments, such as fluctuations of the entropy production, would be interesting to consider for future work.

The system studied in this work can be interpreted as an autonomous engine extracting work $W$ from a run-and-tumble particle. For such an ``active engine'' we studied the extractable power $\dot{W}$, Eq.~(\ref{eq:PowerOutput}), and the thermodynamic efficiency $\eta$ in Sec.~\ref{sec:OptimalWork}. Poor efficiencies $\eta < 1\%$ are observed in the case where the particle is confined by the potential $U(x)$, Eq.~(\ref{eq:RatchetPotential}), and therefore able to generate a current $J$ only via diffusive fluctuations, Fig.~\ref{fig:PowerVsForce}. However, efficiencies $\eta$ arbitrarily close to $100\%$ are possible for $\gamma \rightarrow 0$ and $D \rightarrow 0$ in the case where right movers are unconfined, $v-f > U'^{[1]}$, but left movers are confined, $v+f < |U'^{[2]}|$, as this prevents left movers from dissipating wasted energy (at least when particles are modelled as ``dry" active matter). It would be interesting to calculate the maximum efficiency $\eta$ of this system in the case of ``wet" active matter where momentum is conserved through hydrodynamic interactions.

Finally, in Sec.~\ref{sec:FirstPassage} and Appendix~\ref{app:DerivationFirstPassageProperties}, we derived the mean first-passage times $\tau_{R,L}(x)$, Eq.~(\ref{eq:MeanFirstPassageGeneralSoln}), and splitting probabilities $\Pi_{R,L}(x)$, Eq.~(\ref{eq:SplittingProbabilitiesGeneralSoln}), for this system under different boundary conditions and explored how they depend on the system's parameters, Figs.~\ref{fig:MFPTandSplittingProbability} and \ref{fig:MFPTsVaryingParameters}.

\begin{acknowledgements}
We are grateful to Gunnar Pruessner, Thibault Bertrand, Ziluo Zhang, Jacob Knight, Luca Cocconi, Henry Alston, and Sinead O'Brien for useful discussions. C.R.\ acknowledges support from the Engineering and Physical Sciences Research Council (Grant No.\ 2478322). Z.Z.\ acknowledges support from the Department of Mathematics, Imperial College London.

C.R.\ and Z.Z.\ contributed equally to this work.
\end{acknowledgements}



\appendix
\onecolumngrid

\section{Derivation of steady-state probability densities $P_{R,L}^{[i]}(x)$}\label{app:DerivationParticleDensity}

To solve the coupled Fokker-Planck equations,
Eq.~(\ref{eq:FokkerPlanckEquation}), for the steady-state probability densities $P_{R,L}^{[i]}(x)$, we rewrite Eq.~(\ref{eq:FokkerPlanckEquation}) as a four-dimensional first-order ODE system for both $i=1$ and $i=2$. Introducing $\eta^{[i]}_{R,L}(x) \equiv P'^{[i]}_{R,L}(x)$, the steady-state Fokker-Planck equations, i.e.\ Eq.~(\ref{eq:FokkerPlanckEquation}) with $\partial_{t}P_{R,L}^{[i]} = 0$, can be rewritten as
\begin{equation}
    \mathbf{y}'(x)=\mathsf{A}\mathbf{y}(x),\label{eq:density_matrix_form}
    \end{equation}
where $\mathbf{y}'$ is the derivative of $\mathbf{y}$ with respect to $x$, and we have defined
\begin{equation}
    \mathbf{y}(x)=\begin{pmatrix}
    P_R^{[i]}(x)\\
    \eta_R^{[i]}(x)\\
    P_L^{[i]}(x)\\
    \eta_L^{[i]}(x)
    \end{pmatrix} \quad \text{and} \quad \mathsf{A} = \begin{pmatrix}
    0&1&0&0\\
    \frac{\gamma}{D}&\frac{v-f-U'^{[i]}}{D}&-\frac{\gamma}{D}&0\\
    0&0&0&1\\
    -\frac{\gamma}{D}&0&\frac{\gamma}{D}&\frac{-v-f-U'^{[i]}}{D}
    \end{pmatrix}.
\end{equation}

As Eq.~(\ref{eq:density_matrix_form}) is linear and homogeneous for both $i=1$ and $i=2$, we make the ansatz $P_{R,L}^{[i]}(x) = \mathcal{Z}_{R,L}^{[i]} \exp( \lambda^{[i]}x)$, where the $\lambda^{[i]}$ are the eigenvalues of $\mathsf{A}$. After substituting this ansatz into Eq.~(\ref{eq:density_matrix_form}), the eigenvalues $\lambda^{[i]}$ are found to satisfy a quartic characteristic equation (arising from two coupled quadratic equations),
\begin{equation}
    D^2 (\lambda^{[i]})^4 + (\lambda^{[i]})^3  \left(2 D f+2 D U'^{[i]}\right) + (\lambda^{[i]})^2  \left(-2 \gamma  D+f^2+2 f U'^{[i]}+\left(U'^{[i]}\right)^2-v^2\right)-\lambda^{[i]}  \left(2 \gamma  f+2 \gamma  U'^{[i]}\right)=0.\label{eq:eigenvalue_density}
\end{equation}
The solutions $\lambda^{[i]}$ to Eq.~(\ref{eq:eigenvalue_density}) are too long to display conveniently as equations. However, one of the solutions to Eq.~(\ref{eq:eigenvalue_density}) is $\lambda^{\ABindex{i}} = 0$ for both $i \in \{1,2\}$, and so the general solution to the matrix differential equation (\ref{eq:density_matrix_form}) is 
\begin{equation}\label{app:eq:ParticleDensityGeneralSoln}
P_{R,L}^{\ABindex{i}}(x)=
 \mathcal A_{R,L}^{\ABindex{i}}e^{\lambda_{\mathcal{A}}^{\ABindex{i}}x}
+ \mathcal B_{R,L}^{\ABindex{i}}e^{\lambda_{\mathcal{B}}^{\ABindex{i}}x}
+ \mathcal C_{R,L}^{\ABindex{i}}e^{\lambda_{\mathcal{C}}^{\ABindex{i}}x}
+ \mathcal D_{R,L}^{\ABindex{i}},
\end{equation}
where $\lambda_{\mathcal{A}}^{\ABindex{i}},\lambda_{\mathcal{B}}^{\ABindex{i}},\lambda_{\mathcal{C}}^{\ABindex{i}}$ are the other three nonzero solutions (for a given $[i]$) to the quartic equation (\ref{eq:eigenvalue_density}).

It remains to calculate the 16 coefficients $\mathcal{A}_{R}^{[1]}, \mathcal{A}_{L}^{[1]}, \mathcal{A}_{R}^{[2]}, \dots, \mathcal{D}_{L}^{[2]}$ in Eq.~(\ref{app:eq:ParticleDensityGeneralSoln}). First, the ratios of the coefficients $\mathcal{A}_{R}^{[i]}/\mathcal{A}_{L}^{[i]}$, representing half of the degrees of freedom, are fixed from substituting Eq.~(\ref{app:eq:ParticleDensityGeneralSoln}) into Eq.~(\ref{eq:density_matrix_form}) and requiring the resulting prefactor of each linearly independent $\exp(\lambda^{[i]}x)$ term to vanish. This results in
\begin{equation}\label{app:eq:CoefficientRatios}
\frac{\mathcal{Z}_{R}^{[i]}}{\mathcal{Z}_{L}^{[i]}} = \frac{\gamma}{\gamma + \left(v-f-U'^{[i]}\right)\lambda_{\mathcal{Z}}^{[i]} - D(\lambda_{\mathcal{Z}}^{[i]})^{2}}, \quad \text{and} \quad \mathcal{D}_{R}^{[i]} = \mathcal{D}_{L}^{[i]},
\end{equation}
where $\mathcal{Z} \in \{ \mathcal{A}, \mathcal{B}, \mathcal{C}$\}. The remaining eight degrees of freedom are fixed through normalisation as well as continuity of the probability densities $P_{R,L}^{[i]}(x)$ and currents $J_{R,L}^{[i]}(x)$, Eq.~(\ref{eq:ParticleCurrents}), for each species across $x=0$ and $x=a$. Specifically, the continuity conditions are
\begin{subequations}\label{app:eq:ContinuityConditions}
\begin{alignat}{4}
P_{R,L}^{[1]}(0) &= P_{R,L}^{[2]}(\ell),\\
J_{R,L}^{[1]}(0) &= J_{R,L}^{[2]}(\ell),\\
P_{R,L}^{[1]}(a) &= P_{R,L}^{[2]}(a),\\
J_{R,L}^{[1]}(a) &= J_{R,L}^{[2]}(a).
\end{alignat}
\end{subequations}
Finally, we also have the normalisation condition
\begin{equation}\label{app:eq:NormalisationCondition}
\int_{0}^{a}\mathrm{d}x\left(P_{R}^{[1]}(x)+P_{L}^{[1]}(x)\right) + \int_{a}^{\ell}\mathrm{d}x\left(P_{R}^{[2]}(x)+P_{L}^{[2]}(x)\right) = 1.
\end{equation}

Only 15 of the resulting 16 equations from Eqs.~(\ref{app:eq:CoefficientRatios}) and (\ref{app:eq:ContinuityConditions}) are linearly independent. After removing any one of these equations, the steady-state densities $P_{R,L}^{[i]}(x)$ for RnT particles moving in a piecewise-linear ratchet potential $U(x)$, Eq.~(\ref{eq:RatchetPotential}), can be obtained in closed form by solving the remaining Eqs.~(\ref{app:eq:CoefficientRatios}), (\ref{app:eq:ContinuityConditions}) and (\ref{app:eq:NormalisationCondition}) for the coefficients $\mathcal{A}_{R}^{[1]}, \mathcal{A}_{L}^{[1]}, \mathcal{A}_{R}^{[2]}, \dots, \mathcal{D}_{L}^{[2]}$. 


\section{Derivation of $D=0$ steady-state probability densities $p_{R,L}^{[i]}(x)$ and currents $j_{R,L}^{[i]}(x)$}\label{app:DerivationParticleDensity_D=0}
Here, we derive the steady-state probability densities $p_{R,L}^{[i]}(x)$ and currents $j_{R,L}^{[i]}(x)$ for $D=0$. We set $D=0$ in Eq.~(\ref{eq:FokkerPlanckEquation}) so that the coupled Fokker-Planck equations in the absence of diffusion become
\begin{subequations}\label{eq:FokkerPlanckEquation_D=0}
\begin{alignat}{2}
\frac{\partial p^{[i]}_R(x,t)}{\partial t} &= -\left(v-f-U'^{[i]}\right)\frac{\partial p^{[i]}_R(x,t)}{\partial x}-\gamma\left(p^{[i]}_R(x,t)-p^{[i]}_L(x,t)\right),\label{eq:FokkerPlanckEquationRightMover_D=0} \\
\frac{\partial p^{[i]}_L(x,t)}{\partial t} &=  -\left(-v-f-U'^{[i]}\right)\frac{\partial p^{[i]}_L(x,t)}{\partial x}-\gamma\left(p^{[i]}_L(x,t)-p^{[i]}_R(x,t)\right).
\label{eq:FokkerPlanckEquationLeftMoverD=0}
\end{alignat}
\end{subequations}
The currents for right movers and left movers then become, respectively,
\begin{subequations}\label{eq:ParticleCurrents_D=0}
\begin{alignat}{2}
j_{R}^{\ABindex{i}}(x,t) &= \left(v-f-U'^{[i]}\right)p^{[i]}_R(x,t), \label{eq:RightParticleCurrent_D=0} \\
j_{L}^{\ABindex{i}}(x,t) &= \left(-v-f-U'^{[i]}\right)p^{[i]}_L(x,t). \label{eq:LeftParticleCurrent_D=0}
\end{alignat}
\end{subequations}

To solve the coupled Fokker-Planck equations (\ref{eq:FokkerPlanckEquation_D=0}) in the steady state, $\partial_{t}p_{R,L}^{[i]} = 0$, we again rewrite them in a matrix form as a two-dimensional system of first-order ODEs for both $i=1$ and $i=2$, i.e.\
\begin{equation}
    \hat{\mathbf{y}}'(x)=\hat{\mathsf{A}}\hat{\mathbf{y}}(x),\label{eq:density_matrix_form_D=0}
\end{equation}
where
\begin{equation}
\hat{\mathbf{y}}(x)=
\begin{pmatrix}
p_R^{[i]}(x)\\
p_L^{[i]}(x)
\end{pmatrix}
\quad 
\text{and}\quad
\hat{\mathsf{A}}=
\begin{pmatrix}
-\frac{\gamma}{v-f-U'^{[i]}(x)}&\frac{\gamma}{v-f-U'^{[i]}(x)}\\
\frac{\gamma}{-v-f-U'^{[i]}(x)}&-\frac{\gamma}{-v-f-U'^{[i]}(x)}
\end{pmatrix}.
\end{equation}
As Eq.~(\ref{eq:density_matrix_form_D=0}) is homogenous, we again make the ansatz $p_{R,L}^{[i]}(x) = \hat{\mathcal{Z}}_{R,L}^{[i]} \exp( \hat{\lambda}^{[i]}x)$, where $\hat{\lambda}^{[i]}$ are the eigenvalues of $\hat{\mathsf{A}}$. The two eigenvalues are found to be
\begin{equation}
    \hat{\lambda}_{\mathcal{A}}^{[i]}=\frac{2\gamma~(f+U'^{[i]})}{(-v+f+U'^{[i]})(v+f+U'^{[i]})} \quad \text{and} \quad \hat{\lambda}_{\mathcal{B}}^{[i]}=0.
\end{equation}
The general solutions to the matrix ODE (\ref{eq:density_matrix_form_D=0}) are then
\begin{subequations}\label{app:eq:ParticleDensityGeneralSoln_D=0}
\begin{alignat}{2}
p_{R,L}^{\ABindex{1}}(x) &=
 \hat{\mathcal A}_{R,L}^{\ABindex{1}}e^{\hat{\lambda}_{\mathcal{A}}^{\ABindex{1}}x}
+ \hat{\mathcal B}_{R,L}^{\ABindex{1}}+\hat{\mathcal{C}}_{R,L}
\delta(x),\\
p_{R,L}^{\ABindex{2}}(x) &=
 \hat{\mathcal A}_{R,L}^{\ABindex{2}}e^{\hat{\lambda}_{\mathcal{A}}^{\ABindex{2}}x}
+ \hat{\mathcal B}_{R,L}^{\ABindex{2}},
\end{alignat}
\end{subequations}
where the $\delta(x)$ term is due to the possibility of particles becoming trapped at $x=0$ in the absence of diffusion. There are 10 coefficients $\hat{\mathcal{A}}_{R}^{[1]}, \hat{\mathcal{A}}_{L}^{[1]}, \hat{\mathcal{A}}_{R}^{[2]}, \dots, \hat{\mathcal{B}}_{L}^{[2]}, \hat{\mathcal{C}}_{R}, \hat{\mathcal{C}}_{L}$ in Eq.~(\ref{app:eq:ParticleDensityGeneralSoln_D=0}), but only 9 of these need to be fixed at any given time since $\hat{\mathcal{C}}_{R}$ and $\hat{\mathcal{C}}_{L}$ are both nonzero \textit{only} in the case where the particle is completely confined, i.e.\ $v-f < U'^{[1]}$ \textit{and} $v + f < |U'^{[2]}|$. In this case, the solutions are trivially given by $p_{R,L}(x) = \delta(x)/2$. Therefore, to fix the 9 coefficients that are nonzero, we first substitute Eq.~(\ref{app:eq:ParticleDensityGeneralSoln_D=0}) into the Fokker-Planck equation (\ref{eq:FokkerPlanckEquation_D=0}), and require the resulting prefactor of each linearly independent $\exp(\hat{\lambda}^{[i]}x)$ to vanish, resulting in
\begin{equation}
\label{app:eq:CoefficientRatios_D=0}
\frac{\hat{\mathcal{A}}_{R}^{[i]}}{\hat{\mathcal{A}}_{L}^{[i]}}=\frac{\gamma }{\gamma+(v-f-U'^{[i]}(x))\hat{\lambda}_{\mathcal{A}}^{[i]}} \quad \text{and} \quad \hat{\mathcal{B}}_{R}^{[i]}=\hat{\mathcal{B}}_{L}^{[i]},
\end{equation}
and leaving 5 degrees of freedom to fix.

Continuity in the densities $p_{R,L}^{[i]}(x)$ is no longer a viable condition to fix the degrees of freedom since the absence of diffusion ceases to smooth out the solutions to Eq.~(\ref{eq:density_matrix_form_D=0}), resulting in the densities becoming discontinuous at $x=0$ and $x=a$. The form of the solutions depends on the relations between the self-propulsion $v$ and the slopes of the potential, $U'^{[1]}=h/a$ and $U'^{[2]}=-h/(\ell-a)$, in particular whether or not the particle is confined. For instance, the particle could climb over both slopes of the potential, i.e.\ $v-f > U'^{[1]}$ \textit{and} $v + f > |U'^{[2]}|$, only one slope, $v-f > U'^{[1]}$ \textit{or} $v + f > |U'^{[2]}|$, or be completely confined, $v-f < U'^{[1]}$ \textit{and} $v + f < |U'^{[2]}|$. At $x=0$, if the particle is confined by the barrier to the left, $v + f < |U'^{[2]}|$, then there is a $\delta(x)$ contribution to the left-moving density $p_{L}^{[1]}(x)$ with prefactor $\hat{\mathcal{C}}_{L}$. A tumbling event from a left mover to a right mover thus results in an instantaneous contribution $\gamma\hat{\mathcal{C}}_{L}$ to the right-moving current $j^{[1]}_{R}(x)$ at $x=0$. Similarly, if the particle were instead confined to the right, $v - f < U'^{[1]}$, then there would be a $\delta(x)$ contribution to the right-moving density $p_{R}^{[1]}(x)$ with prefactor $\hat{\mathcal{C}}_{R}$, and therefore a contribution $-\gamma\hat{\mathcal{C}}_{R}$ to the left-moving current $j^{[1]}_{L}(x)$ at $x=0$. Hence, in the case where the particle is confined only to the left, $v + f < |U'^{[2]}|$ \textit{and} $v - f > U'^{[1]}$, the continuity conditions are
\begin{subequations}
\begin{align}
    j^{[1]}_{R}(0) &= j^{[2]}_{R}(\ell)+\gamma\hat{\mathcal{C}}_{L},\\
    j_{R}^{[1]}(a) &= j_{R}^{[2]}(a),\\
    p_{L}^{[1]}(a) &= 0,\\
    p_{L}^{[2]}(a) &= 0,
\end{align}
\end{subequations}
where, since the particle cannot cross the left barrier, there is no contribution to the left-moving probability density $p_{L}(x)$ at $x=a$. Similarly, in the case where the particle is instead confined only to the right, $v + f > |U'^{[2]}|$ \textit{and} $v - f < U'^{[1]}$, the continuity conditions are
\begin{subequations}
\begin{align}
    j^{[1]}_{L}(0) &= j^{[2]}_{L}(\ell)-\gamma\hat{\mathcal{C}}_{R},\\
    j_{L}^{[1]}(a) &= j_{L}^{[2]}(a),\\
    p_{R}^{[1]}(a) &= 0,\\
    p_{R}^{[2]}(a) &= 0.
\end{align}
\end{subequations}
Finally, if the particle can overcome both barriers, $v-f > U'^{[1]}$ \textit{and} $v + f > |U'^{[2]}|$, then the continuity conditions are
\begin{subequations}
\begin{align}
    j^{[1]}_{R,L}(0) &= j^{[2]}_{R,L}(\ell),\\
    j_{R,L}^{[1]}(a) &= j_{R,L}^{[2]}(a).
\end{align}
\end{subequations}
The remaining degree of freedom is then fixed through the normalisation condition,
\begin{equation}\label{app:eq:NormalisationConditionD=0}
\int_{0}^{a}\mathrm{d}x\left(p_{R}^{[1]}(x)+p_{L}^{[1]}(x)\right) + \int_{a}^{\ell}\mathrm{d}x\left(p_{R}^{[2]}(x)+p_{L}^{[2]}(x)\right) = 1.
\end{equation}

The overall current $j = j_{R}(x) + j_{L}(x)$ has a closed-form expression that is compact enough to be written down. In the case where the particle is unconfined and therefore able to cross both barriers, $v + f > |U'^{[2]}|$ \textit{and} $v - f > U'^{[1]}$, the current is given by
\begin{equation}\label{app:eq:D=0_CurrentExactExpression_Unconfined}
    j_{\leftrightarrow} = \frac{2\gamma(a f+h)^2 (a f-f \ell+h)^2 \left(e^{\kappa + \delta \ell}-e^{ \delta a}\right)}{\theta (e^{ \delta a } - e^{\kappa + \delta \ell}) + h^2 \ell^{2} v^2( e^{\kappa}-1)(e^{\delta a}-e^{\delta \ell})},
\end{equation}
where 
\begin{subequations}
\begin{align}
\delta &=\frac{2 \gamma  (a-\ell) (a f-f \ell+h)}{(a f-f \ell+h)^2-v^2 (a-\ell)^2},\\
\kappa &=\frac{2 \gamma a^2 (a f+h)}{(a f+h)^2-a^2 v^2},\\
\theta &= 2 \gamma \ell  (a f+h) (a f-f \ell+h) (a (a f+2 h)-\ell (a f+h)).
\end{align}
\end{subequations} 
However, when the particle is confined only to the left, $v + f < |U'^{[2]}|$ \textit{and} $v - f > U'^{[1]}$, the current is given by 
\begin{equation}\label{app:eq:D=0_current_leftconfined}
    j_{\not\leftarrow} =\frac{2\gamma(a f+h)^2 (a f-f \ell+h)^2}{\tilde\theta_1 e^{-\kappa}+\tilde\theta_2 e^{\delta (\ell-a)}+\tilde\theta_3 - \theta},
\end{equation}
where 
\begin{subequations}\label{app:eq:app:eq:D=0_current_leftconfined_parameters}
\begin{align}
    \tilde\theta_1 &= (af-f \ell+h)^2 (a (f+v)+h)^2,\\
    \tilde\theta_2 &= -(a f + h)^2 (h + (a - \ell) (f + v))^2,\\
    \tilde\theta_3 &= h \ell^{2} v^{2} (2 a f + h)  -2 h \ell v (a f + h) (a f + h - f \ell) - 
      2 h \ell a v^{2}(a f + h).
\end{align}
\end{subequations}

Finally, the overall current $j_{\not\rightarrow}$ in the case where the particle is confined only to the right, $v + f > |U'^{[2]}|$ \textit{and} $v - f < U'^{[1]}$, can be found by inverting the overall sign in Eq.~(\ref{app:eq:D=0_current_leftconfined}) and replacing $a \rightarrow \ell - a$ in Eqs.~(\ref{app:eq:D=0_current_leftconfined}) and (\ref{app:eq:app:eq:D=0_current_leftconfined_parameters}) above.


\section{Derivation of first-passage properties}\label{app:DerivationFirstPassageProperties}

\subsection{Mean first-passage times $\tau_{R,L}^{[i]}(x)$}\label{app:DerivationMeanFirstPassageSplitProb}

As the coupled ODEs for the mean first-passage time $\tau_{R,L}^{[i]}(x)$, Eq.~(\ref{eq:MFPT_DifferentialEquation}), are similar to the coupled Fokker-Planck equations (\ref{eq:FokkerPlanckEquation}) for the steady-state probability density $P_{R,L}^{[i]}(x)$, the procedure for solving Eq.~(\ref{eq:MFPT_DifferentialEquation}) follows similarly to solving Eq.~(\ref{eq:FokkerPlanckEquation}) in Appendix~\ref{app:DerivationParticleDensity}. However, Eq.~(\ref{eq:MFPT_DifferentialEquation}) is nonhomogeneous and so we must also find a particular solution satisfying the equation. Again, Eq.~(\ref{eq:MFPT_DifferentialEquation}) can be rewritten as a four-dimensional system of first-order ODEs for both $i=1$ and $i=2$. As in the previous appendices, introducing $\tilde\eta^{[i]}_{R,L}(x)=\tau_{R,L}'^{[i]}(x)$ allows
Eq.~(\ref{eq:MFPT_DifferentialEquation}) to be written as
\begin{equation}
    \tilde{\mathbf{y}}'(x)=\tilde{\mathsf{A}}\tilde{\mathbf{y}}(x)+\tilde{\mathbf{b}}\label{eq:MFPT_matrix_form}
    \end{equation}
where
\begin{equation}\label{eq:MFPT_matrix_form_components}
    \tilde{\mathbf{y}}(x)=\begin{pmatrix}
    \tau_{R}^{[i]}(x)\\
    \tilde\eta_R^{[i]}(x)\\
    \tau_{L}^{[i]}(x)\\
    \tilde\eta_L^{[i]}(x)
    \end{pmatrix}, \quad  \tilde{\mathsf{A}}=\begin{pmatrix}
    0&1&0&0\\
    \frac{\gamma}{D}&\frac{-v+f+U'^{[i]}}{D}&-\frac{\gamma}{D}&0\\
    0&0&0&1\\
    -\frac{\gamma}{D}&0&\frac{\gamma}{D}&\frac{v+f+U'^{[i]}}{D}
    \end{pmatrix}\quad
    \text{and} \quad
    \tilde{\mathbf{b}}=\begin{pmatrix}
    -1\\0\\-1\\0
    \end{pmatrix}.
\end{equation}

First, the solution to the homogeneous equation is found by substituting the ansatz $\tau_{R,L}^{[i]}(x) = \tilde{\mathcal{Z}}_{R,L}^{[i]} \exp( \tilde{\lambda}^{[i]}x)$ into Eq.~(\ref{eq:MFPT_matrix_form}) to obtain a quartic equation for the eigenvalues $\tilde{\lambda}^{[i]}$,
\begin{equation}
    D^2 (\tilde{\lambda}^{[i]})^4 -(\tilde{\lambda}^{[i]})^3  \left(2 D f+2 D U'^{[i]}\right)+(\tilde{\lambda}^{[i]})^2  \left(-2 \gamma  D+f^2+2 f U'^{[i]}+\left(U'^{[i]}\right)^2-v^2\right)+\tilde{\lambda}^{[i]}  \left(2 \gamma  f+2 \gamma  U'^{[i]}\right)=0,
    \label{eq:eigenvalue_MFPT}
\end{equation}
which differs to Eq.~(\ref{eq:eigenvalue_density}) only in the sign of the eigenvalues, i.e.\ $\lambda^{[i]} \rightarrow -\tilde{\lambda}^{[i]}$. As in Appendix~\ref{app:DerivationParticleDensity}, one eigenvalue for each of $i=1$ and $i=2$ vanishes and so the solution to the homogeneous part of Eq.~(\ref{eq:MFPT_matrix_form}) has the same form as for $P_{R,L}^{[i]}(x)$ in Eq.~(\ref{app:eq:ParticleDensityGeneralSoln}), i.e.\ 
\begin{equation}\label{app:eq:MeanFirstPassageGeneralSoln}
\tau_{R,L}^{[i]}\Big|_{\text{H}}(x)=
 \tilde{\mathcal A}_{R,L}^{\ABindex{i}}e^{\tilde\lambda_{\mathcal{A}}^{\ABindex{i}}x}
+ \tilde{\mathcal B}_{R,L}^{\ABindex{i}}e^{\tilde\lambda_{\mathcal{B}}^{\ABindex{i}}x}
+ \tilde{\mathcal C}_{R,L}^{\ABindex{i}}e^{\tilde\lambda_{\mathcal{C}}^{\ABindex{i}}x}
+ \tilde{\mathcal D}_{R,L}^{\ABindex{i}}.
\end{equation}
Next, it is easily verified that the particular solution satisfies
\begin{equation}
    \tau_{R,L}^{[i]}\Big|_{\text{P}}(x)=\tilde{\mathcal E}_{R,L}^{\ABindex{i}}x+\tilde{\mathcal F}_{R,L}^{\ABindex{i}}\label{eq:MPFT_particular_integral},
\end{equation}
and after substituting Eq.~(\ref{eq:MPFT_particular_integral}) into Eq.~(\ref{eq:MFPT_matrix_form}), we obtain 
\begin{subequations}
\begin{alignat}{2}
\label{MPFT_equation1}
    (v-f-U'^{[i]})\tilde{\mathcal E}_{R}^{\ABindex{i}}-\gamma(\tilde{\mathcal E}_{R}^{\ABindex{i}}x+\tilde{\mathcal F}_{R}^{\ABindex{i}}-\tilde{\mathcal E}_{L}^{\ABindex{i}}x-\tilde{\mathcal F}_{L}^{\ABindex{i}}) &= 0,\\
    \label{MPFT_equation2}
    (-v-f-U'^{[i]})\tilde{\mathcal E}_{L}^{\ABindex{i}}-\gamma({\mathcal E}_{L}^{\ABindex{i}}x+\tilde{\mathcal F}_{L}^{\ABindex{i}}-\tilde{\mathcal E}_{R}^{\ABindex{i}}x-\tilde{\mathcal F}_{R}^{\ABindex{i}}) &= 0.
\end{alignat}
\end{subequations}
As Eqs.~(\ref{MPFT_equation1}) and (\ref{MPFT_equation2}) are valid for all $x \in [0,\ell]$, then we have that
\begin{equation}\label{app:eq:ParticularCoefficientE}
    \tilde{\mathcal E}_{R}^{\ABindex{i}}=\tilde{\mathcal E}_{L}^{\ABindex{i}}
\end{equation}
and
\begin{subequations}
\begin{alignat}{2}
\label{MPFT_equation3}
        (v-f-U'^{[i]})\tilde{\mathcal E}_{R}^{\ABindex{i}}-\gamma(\tilde{\mathcal F}_{R}^{\ABindex{i}}-\tilde{\mathcal F}_{L}^{\ABindex{i}}) &= 0,\\
    (-v-f-U'^{[i]})\tilde{\mathcal E}_{L}^{\ABindex{i}}-\gamma({\mathcal F}_{L}^{\ABindex{i}}-\tilde{\mathcal F}_{R}^{\ABindex{i}}) &= 0.
    \label{MPFT_equation4}
\end{alignat}
\end{subequations}
Adding Eqs.~(\ref{MPFT_equation3}) and (\ref{MPFT_equation4}), then combining with Eq.~(\ref{app:eq:ParticularCoefficientE}), yields
\begin{equation}
   \tilde{\mathcal E}_{R}^{\ABindex{i}} = \tilde{\mathcal E}_{L}^{\ABindex{i}} = \frac{1}{f+U'^{[i]}}.
\end{equation}
Absorbing the constant $\tilde{\mathcal F}_{R,L}^{\ABindex{i}}$ into $\tilde{\mathcal D}_{R,L}^{\ABindex{i}}$, the full solution to Eq.~(\ref{eq:MFPT_matrix_form}) is therefore
\begin{equation}
    \tau_{R,L}^{[i]}(x) = \tau_{R,L}^{[i]}\Big|_{\text{H}}(x) + \tau_{R,L}^{[i]}\Big|_{\text{P}}(x) = 
 \tilde{\mathcal A}_{R,L}^{\ABindex{i}}e^{\tilde\lambda_{\mathcal{A}}^{\ABindex{i}}x}
+ \tilde{\mathcal B}_{R,L}^{\ABindex{i}}e^{\tilde\lambda_{\mathcal{B}}^{\ABindex{i}}x}
+ \tilde{\mathcal C}_{R,L}^{\ABindex{i}}e^{\tilde\lambda_{\mathcal{C}}^{\ABindex{i}}x}
+ \tilde{\mathcal D}_{R,L}^{\ABindex{i}}+\frac{x}{f+U'^{[i]}}\label{eq:MFPT_general_solution}.
\end{equation}

As for the probability densities $P_{R,L}^{[i]}(x)$ in Appendix~\ref{app:DerivationParticleDensity}, the ratios $\tilde{\mathcal{Z}}_{R}^{[i]}/\tilde{\mathcal{Z}}_{L}^{[i]}$, where $\tilde{\mathcal{Z}} \in \{\tilde{\mathcal{A}},\tilde{\mathcal{B}},\tilde{\mathcal{C}},\tilde{\mathcal{D}}\}$, are fixed by requiring each linearly independent term in Eq.~(\ref{eq:MFPT_general_solution}) to satisfy  Eq.~(\ref{eq:MFPT_matrix_form}), resulting in
\begin{equation}\label{app:eq:MFPT_CoefficientRatios}
\frac{\tilde{\mathcal{Z}}_{R}^{[i]}}{\tilde{\mathcal{Z}}_{L}^{[i]}} = \frac{\gamma}{\gamma - \left(v-f-U'^{[i]}\right)\tilde{\lambda}_{\mathcal{Z}}^{[i]} - D(\tilde{\lambda}_{\mathcal{Z}}^{[i]})^{2}}, \quad \text{and} \quad \tilde{\mathcal{D}}_{R}^{[i]} = \tilde{\mathcal{D}}_{L}^{[i]}.
\end{equation}
The remaining degrees of freedom are then fixed from the boundary conditions and continuity conditions,
\begin{subequations}
\begin{alignat}{3}
    \tau_{R,L}^{[1]}(0) &= 0,\\
    \tau_{R,L}^{[1]}(a) &= \tau_{R,L}^{[2]}(a),\\
    \frac{\mathrm{d}\tau_{R,L}^{[1]}(x)}{\mathrm{d}x}\Big|_{x=a} &= \frac{\mathrm{d}\tau_{R,L}^{[2]}(x)}{\mathrm{d}x}\Big|_{x=a},
\end{alignat}
and
\begin{equation}
    \text{(i)} \quad \tau_{R,L}^{[2]}(\ell) = 0, \quad \text{or} \quad \text{(ii)} \quad \frac{\mathrm{d}\tau_{R,L}^{[2]}(x)}{\mathrm{d}x}\Big|_{x=\ell} = 0,
\end{equation}
\end{subequations}
depending on whether the system has (i) an absorbing boundary or (ii) a reflecting boundary at $x=\ell$ (see, for instance, Ref.~\cite{ahmad2022first}).


\subsection{Splitting probabilities $\Pi_{R,L}^{[i]}(x)$}\label{app:DerivationSplitProb}
To derive the splitting probabilities $\Pi_{R,L}^{[i]}(x)$, we define $\check\eta^{[i]}_{R,L}(x)=\Pi_{R,L}'^{[i]}(x)$ and $\check{\mathbf{y}}(x)= (\pi_{R}^{[i]}(x), \check\eta_{R}^{[i]}(x), \pi_{L}^{[i]}(x), \check\eta_{L}^{[i]}(x))$ such that the governing equations (\ref{eq:SplittingProb_DifferentialEquation}) for the splitting probabilities $\Pi_{R,L}^{[i]}(x)$ can be written as a four-dimensional system of first-order ODEs, $\check{\mathbf{y}}'(x)=\tilde{\mathsf{A}}\breve{\mathbf{y}}(x)$, where $\tilde{\mathsf{A}}$ is the same operator acting on the mean first-passage times $\tau_{R,L}^{[i]}(x)$ above, Eq.~(\ref{eq:MFPT_matrix_form_components}). Thus, the splitting probabilities $\Pi_{R,L}^{[i]}(x)$ obey the same (backward) equations as for that of the mean first-passage times $\tau_{R,L}^{[i]}(x)$, Eqs.~(\ref{eq:MFPT_matrix_form})-(\ref{eq:MFPT_matrix_form_components}), aside from being homogeneous, i.e.\ $\tilde{\mathbf{b}} = \mathbf{0}$. Hence, the splitting probabilities have the same form of general solution as for the homogeneous part to the mean first-passage times, Eq.~(\ref{app:eq:MeanFirstPassageGeneralSoln}), but with different coefficients, i.e.\
\begin{equation}\label{app:eq:SplittingProbabilitiesSoln}
\Pi_{R,L}^{[i]}(x)=
 \check{\mathcal A}_{R,L}^{\ABindex{i}}e^{\tilde\lambda_{\mathcal{A}}^{\ABindex{i}}x}
+ \check{\mathcal B}_{R,L}^{\ABindex{i}}e^{\tilde\lambda_{\mathcal{B}}^{\ABindex{i}}x}
+ \check{\mathcal C}_{R,L}^{\ABindex{i}}e^{\tilde\lambda_{\mathcal{C}}^{\ABindex{i}}x}
+ \check{\mathcal D}_{R,L}^{\ABindex{i}},
\end{equation}
where the eigenvalues $\tilde{\lambda}^{[i]}$ are the same as those obtained through solving Eq.~(\ref{eq:eigenvalue_MFPT}) and the ratios $\check{\mathcal{Z}}_{R}^{[i]}/\check{\mathcal{Z}}_{L}^{[i]}$, where $\check{\mathcal{Z}} \in \{\check{\mathcal{A}},\check{\mathcal{B}},\check{\mathcal{C}},\check{\mathcal{D}}\}$, also satisfy equations analogous to that of the mean first-passage times $\tau_{R,L}^{[i]}(x)$, i.e.\ Eq.~(\ref{app:eq:MFPT_CoefficientRatios}). The remaining degrees of freedom in the general solution for the splitting probabilities $\Pi_{R,L}^{[i]}(x)$, Eq.~(\ref{app:eq:SplittingProbabilitiesSoln}), are fixed from the boundary conditions and continuity conditions,
\begin{subequations}\label{app:eq:SplittingProbabilityBoundaryContinuityConditions}
\begin{alignat}{4}
    \Pi_{R,L}^{[1]}(0) &= 1,\\
    \Pi_{R,L}^{[2]}(\ell) &= 0,\\
    \Pi_{R,L}^{[1]}(a) &= \Pi_{R,L}^{[2]}(a),\\
    \frac{\mathrm{d}\Pi_{R,L}^{[1]}(x)}{\mathrm{d}x}\Big|_{x=a} &= \frac{\mathrm{d}\Pi_{R,L}^{[2]}(x)}{\mathrm{d}x}\Big|_{x=a},
\end{alignat}
\end{subequations}
recalling that we define $\Pi_{R/L}(x)$ to be the probabilities for a right/left mover to exit at the \textit{left}-hand boundary of the system at $x=0$.

\subsection{Conditional splitting probabilities $\pi_{X|Y}^{[i]}(x)$}\label{app:ConditionalSPderivation}
The splitting probabilities $\Pi_{R,L}^{[i]}(x)$ calculated above can be generalised to be conditioned on the particle's self-propulsion state as it exits the interval. Following the calculation in Ref.~\cite{Cocconi2022Nov}, we define $\pi_{X|Y}(x_{0})$ as the ``conditional splitting probability'' for a particle initialised in state $Y \in \{R,L\}$ at $x = x_{0}$ to leave through the left-hand boundary at $x=0$ in the self-propulsion state $X \in \{R,L\}$. Hence, there are four conditional splitting probabilities that exist for each boundary, obtained from every possible permutation of $X,Y \in \{R,L\}$. Introducing $\breve{\eta}^{[i]}_{X|Y}(x)=\pi_{X|Y}'^{[i]}(x)$, the conditional splitting probabilities satisfy $\breve{\mathbf{y}}'_{R,L}(x)=\tilde{\mathsf{A}}\breve{\mathbf{y}}_{R,L}(x)$ where $\tilde{\mathsf{A}}$ is the same operator, defined in Eq.~(\ref{eq:MFPT_matrix_form_components}), acting on the ``unconditional'' splitting probabilities $\Pi_{R,L}^{[i]}(x)$ and mean first-passage times $\tau_{R,L}^{[i]}(x)$ above, and $\breve{\mathbf{y}}_{R,L}(x)= (\pi_{R,L|R}^{[i]}(x), \breve\eta_{R,L|R}^{[i]}(x), \pi_{R,L|L}^{[i]}(x), \breve\eta_{R,L|L}^{[i]}(x))$. Hence, the conditional splitting probabilities satisfy two different sets of coupled ODEs, one for each conditional exit state. As these ODEs are the same as for that of the unconditional splitting probabilities, Eq.~(\ref{eq:SplittingProb_DifferentialEquation}), they thus have the same general solution given in Eq.~(\ref{app:eq:SplittingProbabilitiesSoln}), i.e.\
\begin{equation}\label{app:eq:ConditionalSPsSoln}
\pi_{X|Y}^{[i]}(x)=
 \breve{\mathcal A}_{X|Y}^{\ABindex{i}}e^{\tilde\lambda_{\mathcal{A}}^{\ABindex{i}}x}
+ \breve{\mathcal B}_{X|Y}^{\ABindex{i}}e^{\tilde\lambda_{\mathcal{B}}^{\ABindex{i}}x}
+ \breve{\mathcal C}_{X|Y}^{\ABindex{i}}e^{\tilde\lambda_{\mathcal{C}}^{\ABindex{i}}x}
+ \breve{\mathcal D}_{X|Y}^{\ABindex{i}},
\end{equation}
where $X, Y \in \{R,L\}$ and the eigenvalues $\tilde{\lambda}^{[i]}$ are found through solving Eq.~(\ref{eq:eigenvalue_MFPT}). The coefficients $\breve{\mathcal{A}}_{X|Y}^{[1]}, \breve{\mathcal{A}}_{X|Y}^{[2]}, \dots, \breve{\mathcal{D}}_{X|Y}^{[2]}$ are fixed in the same manner as for the mean first-passage times $\tau_{R,L}^{[i]}(x)$ and splitting probabilities $\Pi_{R,L}^{[i]}(x)$ above, by requiring each linearly independent term to satisfy the ODEs before applying boundary conditions and continuity conditions. The boundary conditions of the ODEs are
\begin{subequations}\label{eq:CSP_boundary_conditions}
\begin{alignat}{4}
    \pi^{[1]}_{R|R}(0) = \pi^{[1]}_{L|L}(0) &=1,\\
    \pi^{[1]}_{R|L}(0)=\pi^{[1]}_{L|R}(0) &=0,\\
   \pi^{[2]}_{R|R}(\ell)=\pi^{[2]}_{L|L}(\ell) &=0,\\
   \pi^{[2]}_{R|L}(\ell)=\pi^{[2]}_{L|R}(\ell) &=0,
\end{alignat}
\end{subequations}
and the continuity conditions are analogous to that of the unconditional splitting probabilities in Eq.~(\ref{app:eq:SplittingProbabilityBoundaryContinuityConditions}), i.e.\
\begin{subequations}\label{eq:CSP_continuity_conditions}
\begin{alignat}{4}
    \pi^{[1]}_{R,L|R}(a) &= \pi^{[2]}_{R,L|R}(a),\\
    \pi^{[1]}_{R,L|L}(a) &= \pi^{[2]}_{R,L|L}(a),\\
    \frac{\mathrm{d}\pi_{R,L|R}^{[1]}(x)}{\mathrm{d}x}\Big|_{x=a} &= \frac{\mathrm{d}\pi_{R,L|R}^{[2]}(x)}{\mathrm{d}x}\Big|_{x=a},\\
    \frac{\mathrm{d}\pi_{R,L|L}^{[1]}(x)}{\mathrm{d}x}\Big|_{x=a} &= \frac{\mathrm{d}\pi_{R,L|L}^{[2]}(x)}{\mathrm{d}x}\Big|_{x=a}.
\end{alignat}
\end{subequations}



\bibliography{references}

\end{document}